\documentclass[fleqn,usenatbib]{mnras}

\usepackage[T1]{fontenc}
\usepackage{ae,aecompl}

\usepackage{graphicx}   
\usepackage{amsmath}    
\usepackage{amssymb}    
\usepackage{pdflscape}
\usepackage{amsbsy}                      
\usepackage[normalem]{ulem}              
\usepackage{epstopdf}
\usepackage[caption=false]{subfig}
\usepackage{placeins}
\usepackage{color}
\usepackage{url}
\newcommand\MyHead[2]{%
 \multicolumn{1}{|c|}{\parbox{#1}{\centering #2}}
}

\usepackage[bordercolor=white,backgroundcolor=gray!30,linecolor=black,colorinlistoftodos]{todonotes}

\title[Galaxy Zoo: Mergers]{Galaxy Zoo: Mergers - Dynamical Models of Interacting Galaxies\thanks{We would like to thank the thousands of members of {\it Galaxy Zoo:~Mergers} for their help on the project. A full list of those who contributed is available at {\url{http://data.galaxyzoo.org/galaxy-zoo-mergers/authors.html}}}}

\author[Holincheck et al.]{Anthony J. Holincheck,$^1$\thanks{E-mail: aholinch@masonlive.gmu.edu}
John F. Wallin,$^2$\thanks{E-mail: jwallin@mtsu.edu} \thanks{Faculty Affiliate School of Physics, Astronomy \& Computational Science, George Mason University}
Kirk Borne,$^1$
Lucy Fortson,$^3$
\newauthor
Chris Lintott,$^4$
Arfon M. Smith,$^5$ \thanks{Currently a staff member at GitHub, Inc.} 
Steven Bamford,$^6$
William C. Keel,$^7$
\newauthor
and Michael Parrish$^5$ \\
$^1$School of Physics, Astronomy \& Computational Sciences, George Mason University, 4400 University Drive, Fairfax, VA, 22030, USA\\
$^2$Center for Computational Science \& Department of Physics and Astronomy, Middle Tennessee State University, \\ 
         1301 East Main Street, Murfreesboro, TN, 37132, USA\\
$^3$School of Physics and Astronomy, University of Minnesota, 116 Church St. SE, Minneapolis, MN 55455, USA\\
$^4$Oxford Astrophysics, Denys Wilkinson Building, Keble Road, Oxford OX1 3RH, UK\\
$^5$The Adler Planetarium, 1300 S. Lake Shore Drive, Chicago, IL 60605, USA\\
$^6$School of Physics \& Astronomy, The University of Nottingham, University Park, Nottingham, NG7 2RD, UK\\
$^7$Department of Physics and Astronomy, University of Alabama, Box 870324, Tuscaloosa, AL  35487, USA}

\date{Accepted XXX. Received YYY; in original form ZZZ}

\pubyear{2015}

\begin{document}
\label{firstpage}
\pagerange{\pageref{firstpage}--\pageref{lastpage}}
\maketitle

\begin{abstract}
The dynamical history of most merging galaxies is not well understood. Correlations between galaxy interaction and star formation have been found in previous studies, but require the context of the physical history of merging systems for full insight into the processes that lead to enhanced star formation. We present the results of simulations that reconstruct the orbit trajectories and disturbed morphologies of pairs of interacting galaxies. With the use of a restricted three-body simulation code and the help of Citizen Scientists, we sample $10^{5}$ points in parameter space for each system. We demonstrate a successful recreation of the morphologies of 62 pairs of interacting galaxies through the review of more than 3 million simulations. We examine the level of convergence and uniqueness of the dynamical properties of each system. These simulations represent the largest collection of models of interacting galaxies to date, providing a valuable resource for the investigation of mergers. This paper presents the simulation parameters generated by the project. They are now publicly available in electronic format at {\url{http://data.galaxyzoo.org/mergers.html}}. Though our best-fit model parameters are not an exact match to previously published models, our method for determining uncertainty measurements will aid future comparisons between models.  The dynamical clocks from our models agree with previous results of the time since the onset of star formation from star burst models in interacting systems and suggests that tidally induced star formation is triggered very soon after closest approach.
\end{abstract}


\section{Introduction}
One of the major processes affecting the formation and evolution of galaxies is mutual interaction. These encounters can include gravitational tidal distortion, mass transfer, and even mergers. In any hierarchical model, mergers are the key mechanism in galaxy formation and evolution. Galaxy interactions take place on timescales of a billion years or more. Even though we are able to look back through time to earlier epochs and see galaxies at many stages of interaction, we cannot hope to observe any particular system for more than just a single instant in time. Because of this static view provided by observations, researchers have turned to simulations of interacting systems. The assumption in modelling interacting systems is that distorted morphological features (tidal bridges, tails, etc.) are tied to the dynamical history of the system \citep{tt72}. By matching tidal features in models, we are matching the overall dynamical history of the systems.

\subsection{Modelling Populations of Interacting Systems}
\label{previous}

Previous progress in developing detailed models of specific, observed systems has been ad hoc. Since 2000, a number of researchers have developed semi-automated methods for trying to speed the process, \citep{theis01, wahde01, smith10}. These methods have seen success in matching simulated systems used as truth data, but their application to real sets of interacting galaxies usually requires detailed observational data beyond a simple image as well as customized fitness functions. For example, one additional piece of data critical for verifying these models is the velocity fields of the tidal features. To compare the velocity fields of simulations and models, we need to have relatively high velocity resolution (typically about $\sim1/10$ of the disc rotation velocity or $\sim30$ \ km~s$^{-1}$) and spatial resolution (about $\sim1/10$ of the disc sizes for the galaxies,  $\sim5$ arcmin for close galaxies)  for the systems being modelled. There are only a few systems where these kinematic data are available. Hence other methods that do not rely on kinematic data must be explored to obtain any systematic approach in learning models of interacting galaxies.

Determining the dynamical parameters for a model of a real system of interacting galaxies can be a time-consuming process. \cite{tt72} offered a series of coarse, yet revealing, parameter studies. For example they showed the different morphologies produced by varying the inclination angle while holding other values fixed. Using the physical intuition gained from such studies, researchers attempting to model a specific system can narrow the range of simulation parameters to be used. However, there is a tremendous amount of trial and error involved in finding a best-fit orbit. This is especially true if one is trying to match the kinematics data from the simulation to observations.\footnote{We discuss the limitations of modelling systems without their velocity field in section \ref{limitations}.}  \cite{hammer_hubble_2009} claims that `[t]he accurate modelling of both morphology and kinematics takes several months, from two to six months for a well-experimented user.'     The `Identikit' software \citep{barnes_identikit_2009} has made this process easier, but it still requires a great deal of effort to match real systems.

Several attempts at speeding and even automating this process have been published. One approach is to build a library of simulation results. Researchers would then browse the set of pre-computed results to look for simulated systems that matched observed ones. The search is conducted by examining the particle output at various time steps and rotating in three dimensions to attempt to locate the proper viewing angle. As simulation codes and computing power have evolved, more elaborate versions of libraries have been constructed (\citealt{kk_lib, howard_simulation_1993, galmer}). If the volume of potential parameter space used to describe a single pair of interacting galaxies is large, then the possible number of parameter sets to describe all interacting pairs must be even larger. Libraries of previous simulations will offer only rough matches at best. A further limitation is that the viewing angle parameters will be added to the list of parameters to be selected, further increasing the number of dimensions to search.

Another approach is to attempt automated optimization using fitness functions to match simulations to observed systems. \cite{wahde98} was one of the first to demonstrate the use of a genetic algorithm for optimizing models of interacting galaxies. A genetic algorithm (GA) uses the evolutionary processes of crossover and mutation to randomly assemble new offspring from an existing population of solutions. The parent solutions are chosen to generate offspring in proportion to their fitness. The more fit, or better matched, to the target system an individual model is, the more often it will contribute its genetic information to subsequent generations. The genes in this GA approach are simply the dynamical model parameters like inclination, mass ratio, disc orientations, etc. The fitness function to be evaluated and optimized needs to provide a meaningful quantitative value for how well a given simulation result matches the target system.

With a fitness function defined, a GA is seeded with an initial population and then set to evolve for some number of generations. Researchers trying to optimize galaxy models will use a population size of typically between 50 and 1000, and then will evolve the system for 50 generations. There is an extensive body of research on the convergence behaviour of GAs in terms of the nature of the fitness landscape being studied and the particular evolutionary mechanisms invoked \citep{dejong}.

At least three groups have published results of GA optimization of models of interacting galaxies: \cite{wahde01}, \cite{theis00}, and \cite{smith10}. They all demonstrate convergence to one or a few best-fit models for real systems based on matching morphological features. However, the convergence radius for these systems is not well documented. A large radius of convergence (perhaps even global in scale) is demonstrated by \cite{wahde98} and others when they are modelling\emph{ artificial systems}. These systems use the simulation code itself to generate a high-resolution simulated observation of a hypothetical system of interacting galaxies. The researchers are then able to use their GA to optimize and find a close fit to the known dynamical parameters. Additionally, to demonstrate convergence as well as some amount of uniqueness, it is customary to take the resulting best-fit models, apply a set of random alterations to the dynamical parameters, and then use these altered models as the initial population in a new GA run. If this population converges to the same best-fit model,  then some confidence in the local uniqueness of the model is gained. However, \cite{smith10} found four distinct best-fit models for the pair of galaxies NGC 7714/5. Even though \cite{smith10} used some kinematics data in their fitness calculation, this demonstrates the potential degeneracy within the models when the fitness is based in great part on the morphology. 

These automated matching systems have not yet been applied to large numbers of real galaxies. The largest number of models attempted in a single study for specific systems is 33 \citep{hammer_hubble_2009}. In this study, the `Identikit' software  \citep{barnes_identikit_2009} was used along with related simulation code to run full N-body simulations of just 48 different sets of initial conditions. The viewing angles of the results were then altered to find qualitative matches to the morphology and observed kinematics of the sample of 33 galaxy mergers. It is important to emphasize that a total of 48 simulation runs were used for all 33 mergers, not 48 simulations per merger. Thus \cite{hammer_hubble_2009} correctly indicate that uniqueness of the resulting solutions is not guaranteed. Another recent study \citep{scudder_galaxy_2015} looked at the interactions between 17 pairs of galaxies. The authors ran N-body simulations of binary galaxy mergers with 9 distinct values of gas fraction. These sophisticated simulations included radiative gas cooling, star formation and associated feedback, and chemical enrichment. However, the same set of 9 simulation runs, with the exact same orbits, mass ratios, and disc orientations, was used to generate simulated SFR for comparison with the observed values for all the pairs in their sample. The observed SFR was determined from HI content estimated from VLA observations. Their conclusion about the relative importance of initial gas fraction compared to interaction effects would be further strengthened by an attempt to model the specific dynamical history of each pair. 

There are several drawbacks to these current automated methods that hinder widespread application. The biggest problem seems to be the lack of a robust, quantifiable fitness function that can be used to model real interacting systems. Previous results using GA approaches have used simple boxes around tidal features and low resolution contour maps to compare the surface brightness of targeted systems to the N-body simulations. Using this approach, even a small angular displacement in a tidal feature causes a low fitness value. The higher surface brightness in the inside of the galaxy also influences the results. Fitting the inner contours of an interaction generally is not considered as important as matching the low surface brightness outer contours. Both of these effects cause unfit models to be selected more frequently than expected and fit models to be missed by GA codes. These problems lead to GA codes converging only on a few systems and not finding widespread application in modelling galaxy interactions. 

Recently, \cite{mortazavi_modeling_2014} have shown success with matching simulated data (both morphological and kinematic) with an automated score function applied to the `Identikit 2' software \citep{barnes_identikit_2011}. They also estimated systematic and random errors in matched orbit parameters by evaluating their fitness metric for several runs. However, their fitness metric scored only a subset of the available observation data by sampling small regions along tidal features. They recommended a total of 10 to 100 simulation runs where the selection of these regions is varied just to estimate the uncertainty in the value of their fitness metric. Their targets were chosen from a previously generated set of models with a narrow range of values for the various orbit parameters. With their approach it is unclear what is the total number of full N-body runs needed to converge on the orbit parameters when varied over a wide range.

At present, no general-purpose, automated fitness function with a wide convergence radius has been published in the literature.  Current functions have been demonstrated to be useful with simulated data and narrow ranges of parameter values.  We expect our results to aid in the development for a more broadly applicable fitness function.

\subsection{The Role of Citizen Scientists as Human Evaluators}  

A novel approach to the need to determine fit models is to employ human visual processing capabilities in a pipeline fashion. A human evaluator reviewing simulation output can be seen as applying a more robust fitness function. Our visual processing capability will allow us to see similar morphologies where a simple difference calculation will not. Also, by focusing on people's ability to match similar shapes, additional observational data are not needed to achieve initial convergence on morphologies. Parallelizing the human fitness function allows this method to be applied to a large number of systems. A single reviewer may lose interest after viewing 125,000 simulations of the same system looking for morphological matches. However, if the work is distributed across 1000 volunteers, each one would only need to review 125 simulations. This would achieve the same number of samples of parameter space as a 50/50/50 GA,\footnote{50 runs of a population of 50 models for 50 generations} though purely random samples would not necessarily guarantee convergence.

This paper describes the {\it Galaxy Zoo:~Mergers} project where we have applied this human visual processing capability to 62 pairs of interacting galaxies. With this methodology, volunteers can help explore parameter space and characterize the fitness of simulation outputs at each location in the space that is sampled. By combining the efforts of thousands of Citizen Scientists, detailed knowledge of the fitness landscape is gained. This knowledge is then used for the direct identification of the best-fit model for each of our 62 systems.  

The paper is organized as follows: in section \ref{sample}, we describe the sample selection for the interacting galaxy pairs (targets) presented to the public for comparison with simulations. In section \ref{methodology} we present the methodology starting with a discussion of the preparation of the target images followed by a description of the physical parameters used in the restricted three-body simulations to which the target images are compared. We then describe the simulations themselves in section \ref{sims}. Section \ref{model_selection} describes the model selection process and the process by which the initial model selections made by the public are further refined through three subsequent volunteer tasks. Results are presented in section \ref{results} where, for brevity, we  describe in detail results for one example system followed by summary information for the remaining 61 models (full information for all 62 models is presented in an online database.) Section \ref{analysis} describes our efforts to quantify the level of convergence of the models. We compare our results to previously published models for a number of systems. The section ends with a discussion of the distribution of the population of simulation parameters, the separation distance, interaction time and implications for triggered star formation. The paper finishes with concluding remarks in section \ref{conclusions}.


\section{The Sample for this Study}
\label{sample}

The sample used in this study was constructed from three existing catalogues of interacting galaxies. The criteria used were that the galaxies had to appear as a pair of interacting objects, had to have obvious tidal distortions and that the progenitor discs had to be at least minimally discernible in the image. The three catalogues are: (1) the Arp Catalog of Peculiar Galaxies \citep{arp_atlas_1966}; (2) the Sloan Digital Sky Survey Data Release 7 \citep{sdss_dr7_2009} and (3) a small set from the {\it Hubble Space Telescope} ({\it HST}). The {\it HST} images were part of an 18th anniversary press release and were primarily from the GOALS program \citep{2009PASP..121..559A}.  The HST images used in this project are single composite colour images created by the Space Telescope Science Institute Hubble Heritage team by combining three images using the F390W, F475X, and F600LP filters.  We studied 54 pairs of galaxies from the Arp catalogue that were observed for SDSS DR7. We used {\it HST} images for 8 additional pairs. Our sample does not exhaust any of these catalogues. The Arp catalogue contains a total of 338 targets. The catalogue presented in \cite{darg_galaxy_2010} contains over 3003 visually selected pairs of interacting galaxies at various stages of merging from SDSS images classified in {\it Galaxy Zoo} \citep{lintott_galaxy_2008}. Other researchers have identified additional mergers using the {\it Galaxy Zoo} and {\it Galaxy Zoo 2} morphologies \citep{casteels}. The set of {\it HST} images had a total of 59 images of interacting galaxies. Many of the images from these catalogues were excluded because either they were not from pairs of galaxies or we were unable to discern the progenitor discs. We estimate that on the order of an additional 100 pairs of galaxies from these catalogues could be studied using the method presented here; the 62 pairs presented here were the ones that were studied by the {\it Galaxy Zoo:~Mergers} volunteers during the active phase of the project.

One piece of data that is not used in matching models to interacting systems in this project is the velocity field measurements for our targeted system. In principle, a velocity field could be included in any citizen science interface. However these kinematic data are not available for most interacting galaxy systems. Even basic measurements of the centre of mass velocities of the galaxies in the SDSS catalogue lacks the resolution to make strong constraints on the kinematics in many interacting systems. Fiber collisions, spectroscopic measurements that are not aligned with the centres of galaxies, and relatively low spectral resolution in the survey create potential errors of hundreds of \ km~s$^{-1}$ in some systems. These errors are larger than the best case scenario suggested by \cite{keel_redshift} for paired systems with scatter between velocity indicators in the 20 to 54 \ km~s$^{-1}$ range.   Furthermore, velocity field measurements across the discs of interacting systems is extremely rare.   Therefore to model large samples of interacting systems, it is simply impractical to use velocity fields as a constraint. Instead, we can use morphologies as the primary constraint in our models and then tune them to match the velocity fields as they become available.

Table \ref{tartbl1} shows the target names (ranked by order in which the target was shown to the public), positions and cross identifications for galaxies from our sample. SDSS IDs are from DR7 with the exception of the Violin Clef galaxy from DR8. 

\begin{table*}
\caption[Target Objects]{\small The name, object ID, right ascension, declination, and cross identifications for our target galaxies. For the first 54 objects, SDSS images were used. The right ascension and declination are J2000. * indicates an {\it HST} image was used.}
\begin{center}
\scalebox{0.9}{
\begin{tabular}{| c | c | c | c | c | l |}
\hline
Display Order & Short Name & SDSS ID & RA & DEC  & Cross identifications \\
\hline
1 & Arp 240 & 587722984435351614 & 13:39:52.8 & +00:50:23.4 & NGC 5257, UGC 8641, 13395227+0050224  \\
2 & Arp 290 & 587724234257137777 & 02:03:49.7 & +14:44:19.1 & IC196, UGC 1556, 02034980+1444204  \\
3 & Arp 142 & 587726033843585146 & 09:37:44.0 & +02:45:36.5 & NGC 2936, UGC 5130, 09374413+0245394  \\
4 & Arp 318 & 587727177926508595 & 02:09:24.5 & -10:08:09.6 & NGC 835, 02092458-1008091  \\
5 & Arp 256 & 587727178988388373 & 00:18:50.1 & -10:21:41.9 & 00185015-1021414  \\
6 & UGC 11751 & 587727222471131318 & 21:28:59.4 & +11:22:55.1 & 21285942+1122574  \\
7 & Arp 104 & 587728676861051075 & 13:32:10.2 & +62:46:02.4 & NGC 5218, UGC 8529, 13321042+6246039  \\
8 & Double Ring, Heart & 587729227151704160 & 15:53:08.6 & +54:08:50.4 & 15530935+5408557  \\
9 & Arp 285 & 587731913110650988 & 09:24:02.9 & +49:12:14.1 & NGC 2854, UGC 4995, 09240315+4912156  \\
10 & Arp 214 & 587732136993882121 & 11:32:35.4 & +53:04:00.0 & NGC 3718, UGC 6524, 11323494+5304041  \\
11 & NGC 4320 & 587732772130652231 & 12:22:57.7 & +10:32:52.8 & UGC 7452, 12225772+1032540  \\
12 & UGC 7905 & 587733080814583863 & 12:43:49.4 & +54:54:16.4 & 12434940+5454181  \\
13 & Arp 255 & 587734862680752822 & 09:53:08.9 & +07:51:58.2 & UGC 5304, 09530884+0751577  \\
14 & Arp 82 & 587735043609329845 & 08:11:13.5 & +25:12:23.8 & NGC 2535, UGC 4264, 08111348+2512249  \\
15 & Arp 239 & 587735665840881790 & 13:41:39.3 & +55:40:14.6 & NGC 5278, UGC 8677, 13413961+5540146  \\
16 & Arp 199 & 587736941981466667 & 14:17:02.5 & +36:34:16.6 & NGC 5544, UGC 9142, 14170522+3634308  \\
17 & Arp 57 & 587738569246376675 & 13:16:47.5 & +14:25:39.7 & 13164737+1425399  \\
18 & Pair 18 & 587738569249390718 & 13:44:50.3 & +13:55:16.9 & 13445034+1355178  \\
19 & Arp 247 & 587739153356095531 & 08:23:34.0 & +21:20:50.3 & IC2339, UGC 4383, 08233424+2120514  \\
20 & Arp 241 & 587739407868690486 & 14:37:50.4 & +30:28:59.5 & UGC 9425, 14375117+3028472  \\
21 & Arp 313 & 587739505541578866 & 11:57:36.5 & +32:16:39.8 & NGC 3994, UGC 6936, 11573685+3216400  \\
22 & Arp 107 & 587739646743412797 & 10:52:14.8 & +30:03:28.4 & UGC 5984, 10521491+3003289  \\
23 & Arp 294 & 587739647284805725 & 11:39:42.4 & +31:54:33.4 & NGC 3786, UGC 6621, 11394247+3154337  \\
24 & Arp 172 & 587739707420967061 & 16:05:33.1 & +17:36:04.6 & IC1178, UGC 10188, 16053310+1736048  \\
25 & Arp 302 & 587739721376202860 & 14:57:00.6 & +24:37:03.3 & UGC 9618, 14570066+2437026  \\
26 & Arp 242 & 587739721900163101 & 12:46:10.2 & +30:43:52.7 & NGC 4676, UGC 7938, 12461005+3043546  \\
27 & Arp 72 & 587739810496708646 & 15:46:58.2 & +17:53:04.4 & NGC 5996, UGC 10033, 15465887+1753031  \\
28 & Arp 101 & 587739845393580192 & 16:04:31.7 & +14:49:08.9 & UGC 10169, 16043172+1449091  \\
29 & Arp 58 & 587741391565422775 & 08:31:57.6 & +19:12:40.5 & UGC 4457, 08315766+1912411  \\
30 & Arp 105 & 587741532784361481 & 11:11:12.9 & +28:42:42.4 & NGC 3561, UGC 6224, 11111301+2842423  \\
31 & Arp 97 & 587741534400217110 & 12:05:45.4 & +31:03:31.0 & UGC 7085A, 12054544+3103313  \\
32 & Arp 305 & 587741602030026825 & 11:58:45.6 & +27:27:07.4 & NGC 4017, UGC 6967, 11584562+2727084  \\
33 & Arp 106  & 587741722819493915 & 12:15:35.8 & +28:10:39.8 & NGC 4211, UGC 7277, 12153585+2810396  \\
34 & NGC 2802/3 & 587741817851674654 & 09:16:41.4 & +18:57:49.4 & UGC 4897, 09164141+1857487  \\
35 & Arp 301 & 587741829658181698 & 11:09:51.4 & +24:15:41.8 & UGC 6207, 11095147+2415419  \\
36 & Arp 89 & 587742010583941189 & 08:42:39.9 & +14:17:08.3 & NGC 2648, UGC 4541, 08423982+1417078  \\
37 & Arp 87 & 587742014353702970 & 11:40:44.0 & +22:25:45.9 & NGC 3808, UGC 6643, 11404420+2225459  \\
38 & Arp 191 & 587742571610243080 & 11:07:20.8 & +18:25:58.6 & UGC 6175, 11072082+1826018  \\
39 & Arp 237 & 587745402001817662 & 09:27:43.4 & +12:17:14.1 & UGC 5044, 09274356+1217154  \\
40 & Arp 181 & 587746029596311590 & 10:28:16.7 & +79:49:24.5 & NGC 3212, UGC 5643, 10281670+7949240  \\
41 & Arp 238 & 588011124116422756 & 13:15:31.1 & +62:07:45.1 & UGC 8335, 13153076+6207447  \\
42 & MCG +09-20-082 & 588013383816904792 & 12:04:39.5 & +52:57:25.9 & 12043959+5257265  \\
43 & Arp 297 & 588017604696408086 & 14:45:19.6 & +38:43:52.5 & NGC 5754, UGC 9505, 14451966+3843526  \\
44 & NGC 5753/5 & 588017604696408195 & 14:45:18.9 & +38:48:20.6 & UGC 9507, 14451887+3848206  \\
45 & Arp 173 & 588017702948962343 & 14:51:29.3 & +09:20:05.4 & UGC 9561, 14512928+0920058  \\
46 & Arp 84 & 588017978901528612 & 13:58:37.9 & +37:25:28.9 & NGC 5395, UGC 8900, 13583793+3725284  \\
47 & UGC 10650 & 588018055130710322 & 17:00:06.8 & +23:07:53.6 & 17000690+2307533  \\
48 & Arp 112 & 758874299603222717 & 00:01:26.7 & +31:26:00.2 & 00012677+3126016  \\
55 & Arp 274 & 587736523764334706 & 14:35:08.7 & +05:21:31.7 & NGC 5679, UGC 9383, 14350876+0521324  \\
56 & Arp 146 & 587747120521216156 & 00:06:44.7 & -06:38:13.0 & 00064479-0638136  \\
57 & Arp 143 & 588007005230530750 & 07:46:52.9 & +39:01:55.6 & NGC 2444, UGC 4016, 07465304+3901549  \\
58 & Arp 70 & 758877153600208945 & 01:23:28.3 & +30:47:04.0 & UGC 934, 01232834+3047042  \\
59 & Arp 218 & 587739720308818095 & 15:53:36.8 & +18:36:34.6 & UGC 10084, 15533695+1836349  \\
61 & Violin Clef & 1237678620102623480 & 00:04:15.4 & +03:23:01.8 &   \\
49 & Arp 148 & 588017948272099698* & 11:03:54.2 & +40:50:57.7 & 11035389+4050598 \\
50 & CGCG 436-030 & 587724232641937677* & 1:20:02.8 & +14:21:43.4 & 01200265+1421417 \\
51 & Arp 272 & 587739720846934449* & 16:05:23.4 & +17:45:25.9 &  NGC 6050, UGC 10186, 16052336+1745258 \\
52 & ESO 77-14 & * & 23:21:04.6 & -69:12:47.4 & 23210539-6912472 \\
53 & NGC 5331 & 587726015081218088* & 13:52:16.2 & +2:06:01.2 & NGC 5331, UGC 8774, 13521641+0206305 \\
54 & NGC 6786 & * & 19:10:53.9 & +73:24:37.0 & NGC 6786, UGC 11415, 19105392+7324362 \\
60 & Arp 273 & * & 2:21:28.6 & +39:22:31.0 & UGC 1810, 02212870+3922326 \\
62 & Arp 244 & * & 12:01:53.0 & -18:52:00.8 & NGC 4038 \\
\hline
\end{tabular}
 \label{tartbl1}
 }
\end{center}
\end{table*}

\section{Methodology}
\label{methodology}

\subsection{Overview}

The methodology used in this study relies on the fact that members of the general public (volunteers without a background in astronomy) can recognize morphological similarities between simulations and images of real systems using their natural visual processing abilities and guided only by minimal training. When presented with a set of simulations based on randomly chosen parameters, they can identify the ones that are plausible matches. This method of visual inspection by large numbers of volunteers was proved with the success of the original {\it Galaxy Zoo} project which led to the morphological classification of the million resolved galaxies in the SDSS (\citealt{lintott_galaxy_2010, fortson_2012}). 

To produce plausible models for a large number of interacting systems, we developed a modelling pipeline to allow the volunteer input to be used to sift through a large volume of initial conditions. For a given pair of interacting galaxies from our sample, the input to the pipeline was primarily an image of the two galaxies showing any tidal distortions. Next, an estimate of the distance to the pair and an estimate of the mass of each galaxy were required. With this minimal information some constraints on the initial conditions could be applied. This allowed us to select sets of initial conditions and run simulations. The output of the simulations were then shown to the volunteers through a custom web interface (called {\it Galaxy Zoo:~Mergers}) and the volunteers compared the simulations to the image of the target system. The quality, or fitness, of the match was evaluated by the volunteer and then the best simulation was selected. Further details of the model selection are presented in Section \ref{model_selection}.

\subsection{Target Preparation}
\label{tgt_prep}

To simplify model generation, all progenitor galaxies are assumed to be discs. It is possible that some of the initial galaxies are actually ellipticals, but only discs where modelled in this study. To prepare the target images for presentation on the interface,  an interacting pair was selected from our sample and the approximate sky coordinates determined from the Arp catalogue or a NED query. A colour thumbnail \citep{2004PASP..116..133L}  was downloaded from the SDSS server or the {\it HST} press release. The `ImgCutout' service for SDSS images uses {\it g}, {\it r}, and {\it i} bands for the three colours and the wavelengths of the HST colour thumbnails varied as most were taken from press releases and not a scientific data set. The information we need from the thumbnails is the morphology, and we can use any set of bands as long as they capture the distribution of luminous stars in the system rather than only the distribution of gas. The thumbnail was then converted to a grey scale image. A simple threshold was applied by manually raising the brightness `floor' of the image until most of the image was replaced with black pixels, leaving just the galaxy pair. This grey scale thumbnail was made available, along with the colour one, for volunteers on the {\it Galaxy Zoo:~Mergers} project interface to compare with  simulations of that interacting pair. The centre of the image was selected to be the centre of the galaxy identified as primary. The selection of primary was made based on which galaxy appeared to be larger.

Because these systems are highly distorted and often blended, the automatic measurements of the galaxy's centroids from SDSS were not always reliable. Initially, automated routines were developed to identify contiguous groups of pixels in the thresholded image. Pixel groups corresponding to the primary and secondary galaxy were fitted with a minimum bounding box. From this box a rotated and inclined ellipse was fit to each galaxy. The centre of each galaxy was estimated by a selection of either the brightest pixel in a group, or the centre of the fitted ellipse. In the case of blended galaxies or when there were nearby stars that affected the boundaries of the pixel groups, we manually fit the size, shape, and orientation of each of the galaxies.

The simulation parameters used in this study included the three dimensional position and velocity vectors needed to describe the relative orbit of the two galaxies. Two of those parameters, the x and y separation distance of the two galaxies are determined by locating the galaxy centres in the image. The next parameters to describe the simulation include two orientation angles and the size of each disc. One angle is the inclination with respect to the sky. The other angle is the position angle of the galaxy disc. These two angles, along with the size of the disc are estimated from the image by the automated process described above. The remaining two parameters to be estimated are the masses of the two galaxies. The redshift of the galaxy pairs was used to set the physical distance scale and the photometric values from SDSS were used to estimate the mass of the galaxy \citep{bell_optical_2003}. For the {\it HST} targets, NED was queried for redshift and a photometric magnitude, usually a B magnitude. The mass was estimated by converting the magnitude to a luminosity, using the redshift information to first estimate distance. The luminosity was then converted to a mass with a mass-to-light ratio of one (solar masses to solar luminosities) as a reference point for our models. This mass estimate is only an approximate initial value and the simulations used a wide range of values chosen to be both higher and lower than this value. Of the six orbit parameters, six disc parameters, and two masses, the image and database information provided reasonable estimates for ten out of fourteen.

Initial attempts were made to use the difference between the redshifts of the two galaxies to constrain the line-of-sight velocity component (the z velocity.)  However, not all pairs had observed redshifts for both galaxies. The reason for these missing redshifts is likely complex. First, the companion galaxies in these systems were often below the limit of the SDSS Redshift survey. Second, many of these galaxies were blended systems, making it less likely the automated algorithms would necessarily add these galaxies into the redshift queue. Third, the close proximity of the two galaxies in our systems made fiber collisions likely for the SDSS survey. Thus we were not able to consistently constrain the line-of-sight velocity components.

The next stage in target preparation was to determine the appropriate range over which the simulation parameters were allowed to vary. The x and y components of the relative position vector were held fixed. However, the disc orientations, masses, and all three velocity components were allowed to vary over a range of values.
\begin{itemize}
\item The masses were each allowed to vary over two orders of magnitude from 0.1 $\times$ mass to 10 $\times$ mass as determined above. The uncertainty of these measurements is based on both ambiguities in the mass-to-light ratio of our galaxies and the poor photometric accuracy in blended systems in DR7. In some cases, we saw differences in the DR7 magnitude compared to previously published results of up to three magnitudes in our sample (e.g. m=12 vs. m=15). It is important to emphasize that this variation resulted in both higher and lower luminosities for the galaxies in our systems.  
\item The x and y velocities were allowed to vary between $\pm$ the escape velocity as computed using the sum of the two maximum mass values determined in the previous step for a particle located at the current x and y separation of the two galaxies.
\item The z velocity was originally allowed to range between 0 and the line of sight velocity determined from the redshifts. However, as noted above, not all galaxies had redshift values so the z velocity varied over the same range as the x and y velocities. The assumption in the velocity measurements is that the systems are close to being bound because of dynamical friction. However, because x, y, and z components are chosen independently the total relative velocities are allowed to exceed the escape velocity of the system.
\item The z position was allowed to vary between $\pm$ 5 $\times$ the diameter of the disc of the primary galaxy.
\item The position angles were allowed to vary $\pm$ 20 degrees.
\item The inclination angle is used to describe the rotation direction. We imposed no constraints based on spiral-pattern, so with a four-fold degeneracy, the inclination angle was allowed to vary $\pm$ 20 degrees.
\item The disc radius for each galaxy was allowed to vary from 0.5 to 1.5 $\times$ the value estimated visually from the image. Again, the SDSS measurements of disc radii can be highly unreliable for blended systems and highly distorted, so we opted instead to use our ad hoc measurements from the image directly. 
\end{itemize}

Some of these limits were somewhat arbitrary. However, they provided a reasonable reduction in the phase space that was needed to search for matches while ruling out only relatively unlikely interactions. In addition, to limit the number of  simulations containing minimal tidal features, each randomly selected set of parameters was passed through a filter (described below)  that calculated a tidal distortion parameter. After the initial range of parameters were selected and passed through the filter, experts reviewed the simulation results of several hundred randomly selected input parameters. Each parameter within a generated set was selected at random from a uniform distribution scaled to the specific limits above. If, during this review phase, fewer than ten simulations resulted in at least some morphology that contained tidal features similar to the target image, the parameter ranges were adjusted manually. Usually these adjustments restricted the ranges of our parameters. Once we were able to find at least ten useful candidate simulation matches within a set of 100 to 200 sets of randomly generated parameter sets, the target was considered ready to be presented to the Citizen Scientists.

In this manner, the fourteen simulation parameters per system were assigned allowable maximum and minimum values. These ranges were stored in a simulation parameter file. 
With the specified ranges, simulation parameters could be selected by drawing a random number from a uniform distribution between 0 and 1 and scaling by the min and max values.

\subsubsection{Simulation Filter}
\label{simfilt}
As mentioned above, an additional filter was imposed on the randomly selected sets of parameters to reduce the number of simulations with no tidal features. During initial testing that included half a dozen different systems, the above ranges of parameter values allowed for a large number of randomly generated simulation inputs that resulted in simulations that showed no tidal features. In order to estimate whether a given simulation would result in tidal features we calculated a form of the tidal approximation parameter from \cite{binney_galactic_2008}. They provide an estimate for the change in velocity for a particle at a location in the primary galaxy due to tidal forces resulting from the passage of the secondary galaxy.

\begin{equation}
\Delta v \approx \frac{2 G M_2}{b^2 V_2} (x, y, 0)
\end{equation}

\noindent where $b$ is the impact parameter between the two galaxies, and $V_2$ and $M_2$ are the velocity and mass of the secondary galaxy respectively. 

We adopted a similar form for the filter parameter which we call $\beta$, represented in our simulation units as:

\begin{equation}
\beta = \frac{M_1 + M_2}{{r_{min}}^2 V_{r_{min}}}
\end{equation}

\noindent where $r_{min}$ is the closest approach distance and $V_{r_{min}}$ is the relative velocity at the time of closest approach.

The $\beta$ parameter captures two important quantities. The first is the mutual gravitational attraction. This is important because we wish to observe tidal distortions to both the primary and secondary galaxies in some systems. The second key component is the inverse velocity at the time of closest approach. This incorporates the sense of interaction time, during which one galaxy can impart an impulse on the other. Even though the units are not identical to the tidal approximation parameter we believe it contains sufficient information to predict whether there will be noticeable tidal distortions. The $\beta$ parameter increases with increasing mass, it decreases with increasing distance, and decreases with increasing velocity. More massive systems have a chance to cause greater tidal distortions. Systems that pass farther apart from one another will have less distortion. Systems that are only close for a short time due to a high relative velocity will not show as much distortion. An alternate form of this parameter was considered where the true $\Delta v$ of the outer disc particles was calculated. However, these calculations slowed down the realtime simulation too a point where volunteers were likely to lose patience. The $\beta$ parameter could be further enhanced by multiplying by a representative disc scale, but this was not written into the simulation code used in this study.

Before running a full simulation with all of the test particles, we performed a backwards integration of just the two galaxy centres of mass to determine r$_{min}$ and $V_{r_{min}}$. We then calculated values of $\beta $ for each set of simulation parameters. For our threshold, we set an arbitrary minimum value of $\beta$ = 0.5. Any set of simulation parameters with a $\beta$ greater than 0.5 was considered to have a significant chance of displaying tidal distortion. For systems with a $\beta$ value less than the specified minimum, we accepted them with an exponentially decreasing probability.

\begin{equation}
p = \exp(-0.05 \frac{\beta}{\beta_{min}} )
\end{equation}

This decreasing probability allows us to sample parameter space with sets of parameters that do not exceed our minimum $\beta$ while at the same time we avoid having to review a large number of simulations that will likely not show any tidal distortions. With these thresholds there are an average of twelve parameter sets rejected for every one that passes. Of the parameter sets where $\beta$ was below the threshold, on average 7\% are accepted after passing the probability filter.  This process was successful in reducing the number of simulations presented to the volunteers that showed no tidal distortions.

\subsection{Simulations}
\label{sims}
The gravitational potential describes the attractive force between every pair of massive objects. In a restricted three-body simulation\footnote{Sometimes also referred to as multiple three-body or restricted multi-body simulations.} the mass distribution for each galaxy is represented with a static potential with an origin at the galaxy's centre. This potential can take on several different forms, a 1/r potential for a point mass, a softened point mass potential, or a distributed mass potential such as the one used in this work. The discs of the galaxies are then populated with a set of massless test particles. The acceleration of each massless particle is the sum of the accelerations produced by each of the galaxy centres. The test particles are distributed randomly in a series of rings for each disc in a way that results in a uniform surface density out to the specified disc radius. This ensures sufficient particles in the outer ring where tidal effects are likely to be greatest. The particles are assigned an initial circular velocity. The second order differential equation for the acceleration due to gravity as a function of position is usually decomposed into two coupled, first order differential equations. The first equation sets the time derivative of the position equal to the particle velocity, and the second equation defines the time derivative of the velocity as the acceleration due to the mass of galaxies. The simulation can be advanced at each time step by using the previous velocity and computed acceleration to advance the position and velocity respectively of each particle. Many numerical techniques exist that can be applied to solve these equations such as the Euler method of leapfrog integration.

Restricted three-body simulation codes are efficient, but do they provide the ability to construct realistic models of observed galaxies?  One important reason for the impact of \cite{tt72} was the success they had in recreating the disturbed morphologies for four actual pairs of disc galaxies. Other researchers applied a similar approach to modelling elliptical galaxies \citep{borne_interacting_1984}. Simulations of interacting galaxies using the restricted three-body method produce realistic and visually appealing results with only a few thousand particles and can run in under one second on modern personal computers.

For our restricted three-body simulations we use the code called JSPAM.\footnote{JSPAM is the Java Stellar Particle Animation Module.}    Full details on this code are available elsewhere \citep{wallin_dynamical_1990, wallin2013a}. A convenient distinction of the JSPAM code (compared to other simulation tools) is that it is formulated to make the centre of mass of the primary galaxy the origin of the coordinate system. This makes it a very simple task to match up each simulation time step with a target image. It also requires that we define the relative orbit of the two galaxies in terms of the position and velocity of the secondary galaxy with respect to the primary. This is somewhat different from the usual step of setting the origin at the centre of mass computed from both galaxies together.

Another innovation in the JSPAM code is the use of a more realistic gravitational potential.  The potential is calculated by initializing an N-body simulation of a ``unit'' galaxy.  This galaxy is composed of disc, halo, and bulge components.  The relative masses are 1 for the disc, 5.8 for the halo, and 0.3333 for the bulge.  The relative scale lengths are 1 for the disc, 10 for the halo, and 0.2 for the bulge.  The mass in the ``unit'' galaxy simulation is then distributed in a similar fashion to \cite{hernquist_n-body_1993} for the disc and halo, but we modify the bulge mass distribution to be Gaussian.  Once the mass is initialized, we sample the velocity dispersion and mass in spherical shells.  This information can be used to calculate the force between the two galaxies and the acceleration felt by the test particles by interpolating between shells. The sampled values can then be scaled from the ``unit'' galaxy to the simulation-specific values based on the mass and disc radius specified for each galaxy.  The velocity profile for particles in these galaxies are much flatter and are a better match to observed profiles than normal restricted three-body simulations produce.  Realistic velocities for particles in the outer disc are crucial for accurately recreated tidally-induced morphologies.  Though we use a halo/disc/bulge model to generate this potential, the simulation retains only two massive particles, the centre of each galaxy.  The interpolated values act like a modification to the central potential exhibited by each particle. The unit potential for each galaxy is scaled by the specific mass and disc radius when calculating accelerations. Our simulations gain some of the benefits of having an extended dark matter halo, but retain the computational simplicity of a restricted three-body method. 

\subsubsection{Dynamical Friction}
Analytic predictions \citep{chandrasekhar_statistics_1943} and self-consistent N-body codes have demonstrated that the orbits of secondary galaxies will decay over time \citep{barnes_dynamics_1992}. One important process that leads to the loss of orbital energy is scattering in the form of dynamical friction. These codes can also produce other multi-body effects like violent relaxation. These effects are usually absent in restricted three-body codes. The orbital decay, even during a first passage encounter, can be significant. Dynamical friction plays a key role in galaxy evolution through other interactions such as between a bar and the dark matter halo. A parametrized version of this effect leading to orbital decay is derived in \cite{binney_galactic_2008}. 

A massive body M moving through a field of other massive particles will interact with them through the gravitational force. The field particles have individual masses much less than M. However, these field particles are part of an overall system that is very massive and large. It is customary to approximate this system as infinite and homogeneous, with the distribution of velocities taken to be Maxwellian. As the body M moves through this field of massive particles (such as stars), the field particles will be deflected resulting in an enhanced density behind the massive body sometimes referred to as a wake. The attraction of this wake on the moving body is opposite in direction compared to its velocity resulting in dynamical friction.

For a set of background masses of density $\rho$ and a Maxwellian distribution of velocities with dispersion $\sigma$, Chandrasekhar's dynamical friction formula \citep{chandrasekhar_statistics_1943}  for the acceleration becomes:
\begin{eqnarray}
\frac{d\boldsymbol{v}}{dt} & =&\frac{4 \pi G^2  M \rho  {\rm ln}\Lambda  }{v^3 } \left[ {\rm erf} \left( X \right) - \frac{  2 X }{ \sqrt{\pi}} e^ {- X ^2} \right]  \boldsymbol{v}
\end{eqnarray}

\noindent where we define $X = v/\sigma $ and $\Lambda$ is the ratio of the limits of the impact parameter.
In addition, we define the following formulae useful for calculating the velocity dispersion $\sigma$:

\begin{eqnarray}
p(r) = G \int_r^{\infty}  \frac {\rho(r)  m(r) }{r^2 } dr 
\end{eqnarray}

\begin{eqnarray}
v_r^2 = \frac{p(r)}{\rho(r)} 
\end{eqnarray}

Because this approximation of dynamical friction has a closed form, it can be reversed when setting up the initial conditions for the galaxy positions. Like the other parts of the restricted three-body code, this is an approximation. We used a fixed value of 0.001 for $ln \Lambda$ for all simulations shown to the volunteers.  This value is unitless and usually on the order of 1. To be clear $\Lambda$ is not a free parameter in the simulation. Our selection of a particular constant value for this parameter in our code includes conversion factors between simulation units and physical units and was done after comparing our results with N-body simulations. \cite{harvey} uses a similar calibration process for the approximate treatment of dynamical friction. The process involves using the restricted three-body code, with a particular value of $ln \Lambda$, and integrating the position and velocity of the secondary galaxy backwards from the current epoch to some time in the past. The calculated position and velocity are then used to initialize a new simulation which is then integrated forward using an N-body code.  If the N-body code does not place the secondary galaxy in the expected relative position at the current epoch, the value for $ln \Lambda$ is adjusted and the process repeated. We have adopted a constant value of $ln \Lambda$ of 0.001 with our approximate treatment of dynamical friction to produce a comparable amount of orbit decay in our code to match the typical behavior of the N-body simulations.

For systems with hyperbolic orbits and those with large pericentre distances, the acceleration from dynamical friction is almost always zero.  For elliptical, parabolic, and nearly parabolic orbits, the instantaneous acceleration from dynamical friction can be as great or greater than the main gravitational acceleration between the two bodies for one or more timesteps in the simulation. Although the dynamical friction acts over a short period of time only when the two galaxies are relatively close to one another, it plays a significant role in altering the orbits during some of the simulations. 

\cite{petsch} discuss several different approaches for incorporating dynamical friction into restricted three-body simulations.  They developed four methods of various complexity with their simplest matching our approach of using a constant value for $\Lambda$. \cite{petsch} also found that analytic treatments of dynamical friction did not produce accurate decay behavior for equal mass merging galaxies.  Many of our interacting pairs have mass ratios of less than 3 to 1, the limit they found for accurate reconstructions.  However, all of our interacting pairs are relatively ``young'' in that the closest approach is usually less than 1 Gyr in the past.  Furthermore, we do not follow the systems into the future where the full merger occurs.  Though it may not be possible to accurately model the full extent of the orbital decay, any included amount of decay due to dynamical friction will increase the accuracy of the simulation compared to having no decay.  We believe that tuning of the constant value of $\Lambda$ will make it possible to make close morphological matches with most if not all of our models when converting our models for use as initial conditions in full N-body simulations. See section \ref{limitations} for further discussion.  Future enhancements to the simulations could include implementation of other method of calculating $\Lambda$ as discussed in \cite{petsch}.

\section{Model Selection}
\label{model_selection}

Once the target image was selected and prepared,  the colour and grey scale thumbnail images along with the simulation parameter file was uploaded to an online system where they could be accessed by the {\it Galaxy Zoo:~Mergers} website. The {\it Galaxy Zoo:~Mergers} project had four main tasks:  {\it Explore}, {\it Enhance}, {\it Evaluate} and {\it Model Refinement}. The first task, {\it Explore}, required the volunteer to identify the most plausible matches between a target system and simulations. {\it Enhance} enabled the volunteer to tune the simulation parameters within the allowed ranges to obtain the best fit by eye. {\it Evaluate} required the volunteer to select the best of several selected ``matches" for a given target. Finally, the {\it Model Refinement} task presented the volunteer with subsequent rounds of simulations picked in {\it Explore} and {\it Enhance} to identify the best-of-the-best of each round. The {\it Model Refinement} stage itself comprised in the first instance the {\it Merger Wars}  task, then followed later by the {\it Simulation Showdown} and {\it Best of the Best} tasks as the project neared completion.

\begin{figure*}
\begin{center}
\includegraphics[scale=0.3]{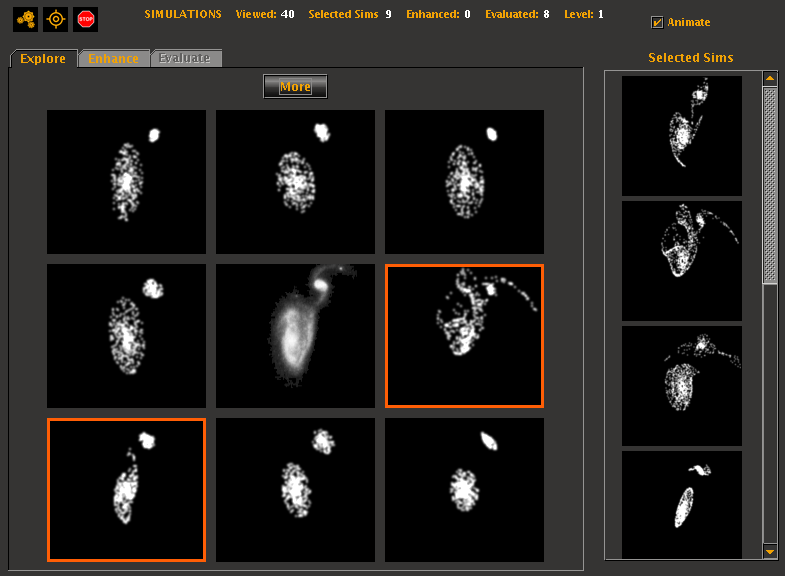}
\caption{The {\it Explore} interface as presented to the volunteers online. The  ``real'' image in the centre is the target pair of interacting galaxies. Each of the eight surrounding images are simulations generated with a restricted three-body simulation in realtime. The volunteer would select any of the simulated images that they believe match the target image characteristics. These selected images would then be posted on the right-hand column ``Selected Sims". After numerous rounds, these ``Selected Sims" would eventually be ranked in the {\it Evaluate} task.}
\label{ch4explore}
\end{center}
\end{figure*}

Upon arriving on the site, the volunteers were presented with background information about the project, including tutorials, and were offered the choice between the {\it Explore} or {\it Merger Wars} task. Participating in {\it Explore} required the volunteer to download a Java applet to enable the restricted three-body simulation code to generate simulated images while {\it Merger Wars} was a fully ``in-browser" task. The interface for the {\it Explore} task  presented the results of restricted three-body simulations run in real time based on the simulation parameter file. Eight simulation outputs surrounded the target image in the centre as seen in Figure \ref{ch4explore}. The volunteer would then click on a simulation image to indicate that they believed it was a possible match to the target in the centre. If not a match, then clicking on the simulation image could be used to indicate that it at least shared one or more important tidal features with the target. In practice, all eight of the simulation images could be selected as being possible matches. After reviewing the presented set of eight images, the volunteer clicked ``More" to see eight more simulated images and began the matching process for this set of simulations. In this fashion, a single volunteer was able to review a thousand simulations an hour.

\begin{figure*}
\begin{center}
\includegraphics[scale=0.3]{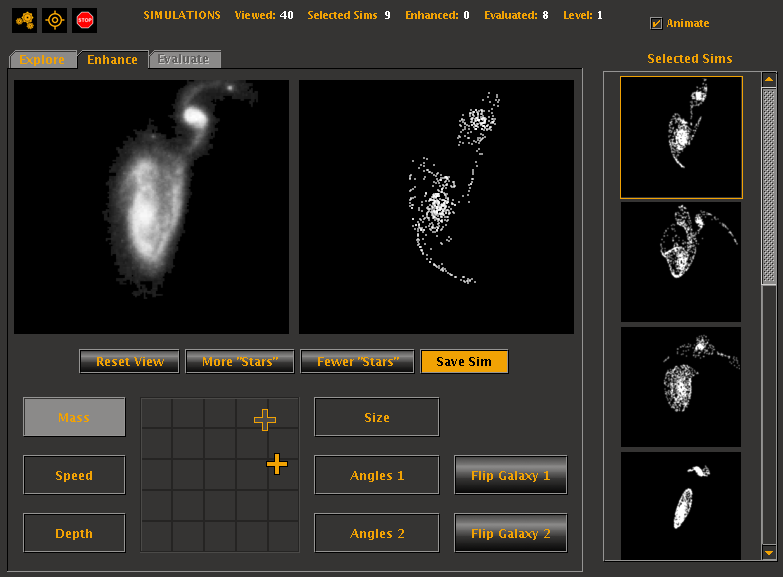}
\caption{The {\it Enhance} interface obtained after clicking on the ``Enhance" tab in the {\it Explore} interface. The volunteer could modify the input parameters of a given selected simulation by clicking on the parameter of interest and sliding the plus sign to a different location in the grid and then rerun the simulation. In this manner, the parameters could be directly manipulated to obtain the volunteer's best effort at a best fit for a given target image. }
\label{ch4enhance}
\end{center}
\end{figure*}

The {\it Enhance} interface, shown in Figure \ref{ch4enhance}, was reached through a button on the {\it Explore} interface and allowed the volunteer to adjust each of the twelve tunable simulation parameters to attempt to improve how well the simulation matched the target image. Whenever the volunteer had determined that they had found a somewhat better match, they could save the simulation to their selected set.

After the volunteer accumulated eight ``good match" simulations in the selected set,  their screen was changed to the {\it Evaluate} activity. Here they were asked to pick, in order, the three best images from the set of eight simulations that they had already selected from their session. After eight first round images were selected, the volunteer was asked to select the top three in a second level ``tournament" of sorts. Dedicated volunteers worked their way up through four or more levels of the tournament. However, most volunteers did not complete this task. Thus we developed a separate process for scoring the combined results from multiple volunteers (see Section \ref{mergerwars} below).


\subsection{Model Refinement}
\label{refinement}

The models selected by the volunteers were refined through a combination of Citizen Scientist and expert input in a series of review activities described below.

\subsubsection{Merger Wars}
\label{mergerwars}
The {\it Merger Wars} algorithm was used to sort the large lists of simulation results. For each target image, volunteers would be presented with simulation images two at a time. The volunteer simply clicked on the simulation image that was a better match to the target image. When a simulation image was presented, that counted as participating in a competition. When a simulation image was selected, that counted as a win in the competition. The overall score, or fitness, for a simulation image was the simple ratio of number of wins to number of competitions. The images with a higher winning percentage were considered better matches than those with a lower winning percentage.

Compared to standard pairwise comparison tests, the {\it Merger Wars} algorithm included two novel enhancements. The first was the inclusion of a third choice labelled ``Neither is a good match". In situations where volunteers felt that the simulation images were both rather poor, they could click the ``Neither" button to record a loss for both images. The other enhancement has to do with image selection and shallow tournaments. Rather than simply compete all the images in a single large tournament in a winner take all style, or simply apply user selections as the comparison function in a traditional sorting algorithm, the method made use of shallow tournaments. Images were competed against each other in randomly selected sets of eight. In a given tournament, the image could lose in the first round, second round or third round and accumulate zero to three wins. These shallow tournaments soften the impact of an incorrectly judged competition. For example, if a volunteer clicked on the wrong image, a good image would be scored down and a bad image scored up. However, both images would be competed again in other tournaments. There is no single elimination in the larger process. Simulation images were selected for inclusion in a tournament in such a way as to keep the total number of competitions for all images close to equal.

In \cite{holincheck2013}, we demonstrate that the {\it Merger Wars} algorithm has comparable performance to traditional sorting algorithms, $O(n \log n)$, and that in the presence of inaccurate comparisons, where volunteers click the wrong image, the algorithm is more accurate than traditional sorting algorithms. The {\it Merger Wars} method was implemented as a JavaScript interface in the {\it Galaxy Zoo:~Mergers} website. This allowed Citizen Scientists that did not have the Java plugin installed and enabled, and thus were not able to run simulations, to contribute to the project. Additionally, it gave volunteers a chance to see the types of simulations being selected by others.

\subsubsection{Final Activities}
In the last few months of the {\it Galaxy Zoo:~Mergers} project, a set of final activities were launched. For each of the 54 SDSS targets, the top 20 {\it Merger Wars} results were reviewed by the research team. We excluded the non-SDSS images from the public review because we felt the differences in image quality and resolution of the target galaxies might bias the results. The team selected four to eight simulations per target to represent the best simulations. Two activities were then launched to further rate these top simulations with the goal of selecting a single overall best simulation for each pair. The first activity was called {\it Simulation Showdown}. Here, the volunteer was presented with two sets of images. The first set included the target image and simulation image from one galaxy pair, the other set included the two images for another pair. The volunteer needed to identify which image was a better match to its respective target. This compared simulations from different pairs against each other. The second final activity was called {\it Best of the Best}. It presented the target image in the centre. The best images were distributed at random around the target image. The volunteer was asked to select the best image for each target. Each activity generated a new fitness score. The top simulations for each pair were ranked by these scores. For about half of the systems the top-ranked simulation for {\it Simulation Showdown} and {\it Best of the Best} was the same. For the other half, the research team picked a consensus best simulation from the two candidates for each target. For each of the eight non-SDSS targets from {\it HST}, the candidate simulations (between four and eight) were selected by the research team. Because these results did not participate in the final activities, the original {\it Merger Wars} score was used to rank the results for each target.

\section{Results}
\label{results}

\subsection{Contributions of Citizen Scientists}
The {\it Galaxy Zoo:~Mergers} project was launched on November 23, 2009. The last simulation submissions and {\it Merger Wars} clicks were collected on June 7, 2012. 
In the two and a half year period that the site was active,  6081 Citizen Scientists with the Zooniverse logged in\footnote{A total of 30305 registered Zooniverse volunteers visited the site, but only 6081 completed the tutorial and saved results.} and ran a combined 3.31 $\times$ 10$^{6}$  restricted three-body simulations in 4765 hours of session time. The volunteers also judged 10$^{6}$ {\it Merger Wars} competitions. In addition to the more than 3 million simulations viewed by the volunteers, the $\beta$ filter described in Section \ref{tgt_prep} was used to exclude an estimated 300 million (3$\times$10$^{8}$) sets of initial conditions that did not produce significant tidal distortions. All of these simulations were run with the Java applet using the CPU of volunteers' machines. Of the simulations that the volunteers viewed, they selected over 66000 simulations as being of potential interest and spent time trying to refine the parameters for 13000 simulations in the {\it Enhance} activity. This means on average that each pair of galaxies had 4.8 million sets of initial conditions rejected by the $\beta$ filter with over 50000 simulations reviewed by volunteers who selected, again on average, over 1000 simulations per system to be evaluated with over 16000 {\it Merger Wars} competitions.

\begin{figure*}
\begin{minipage}[b]{0.45\linewidth} 
\centering
\includegraphics[scale=0.3]{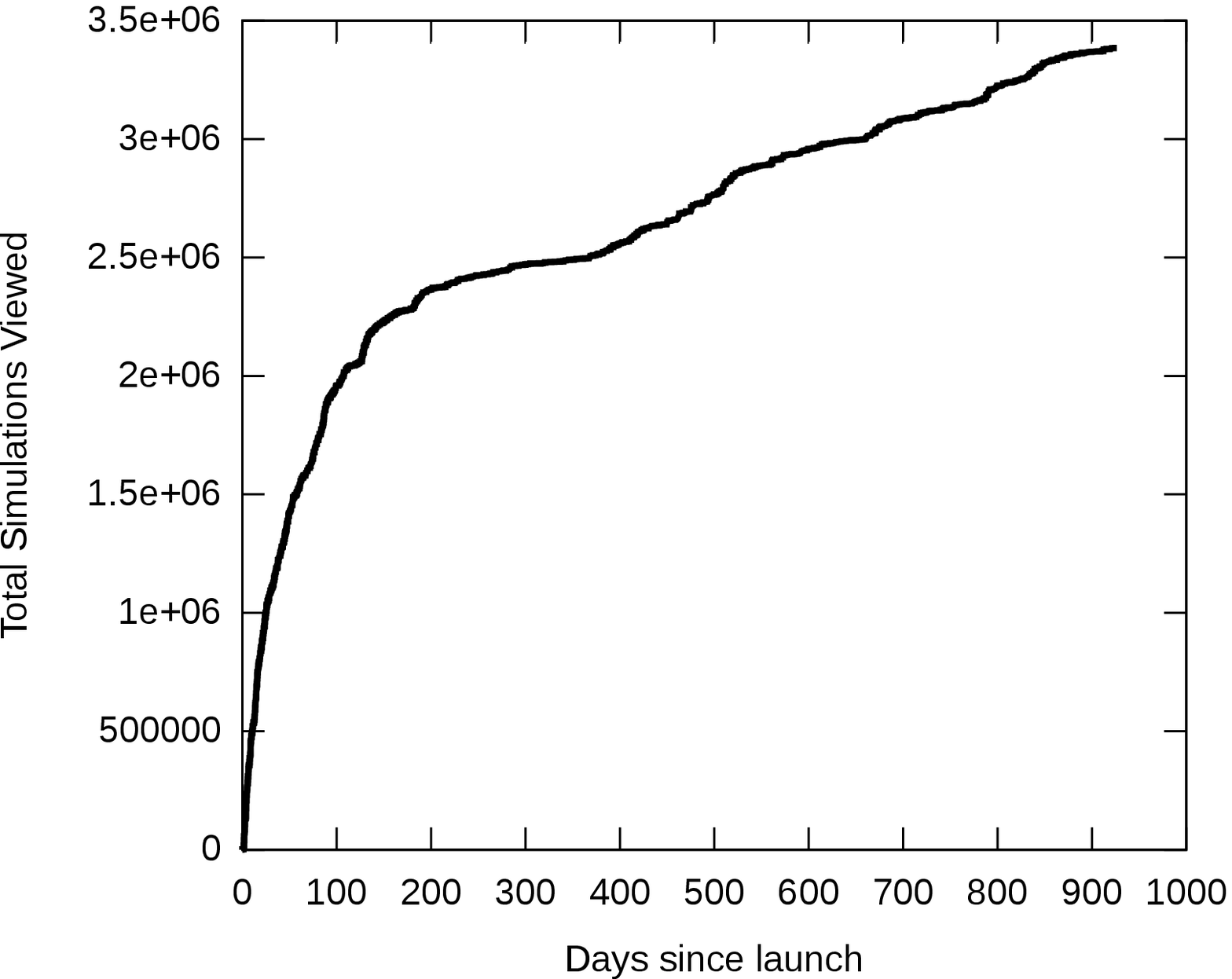}
\caption{Cumulative count of simulations viewed.}
\label{ch4sims}
\end{minipage}
\hspace{0.5cm} 
\begin{minipage}[b]{0.45\linewidth}
\centering
\includegraphics[scale=0.3]{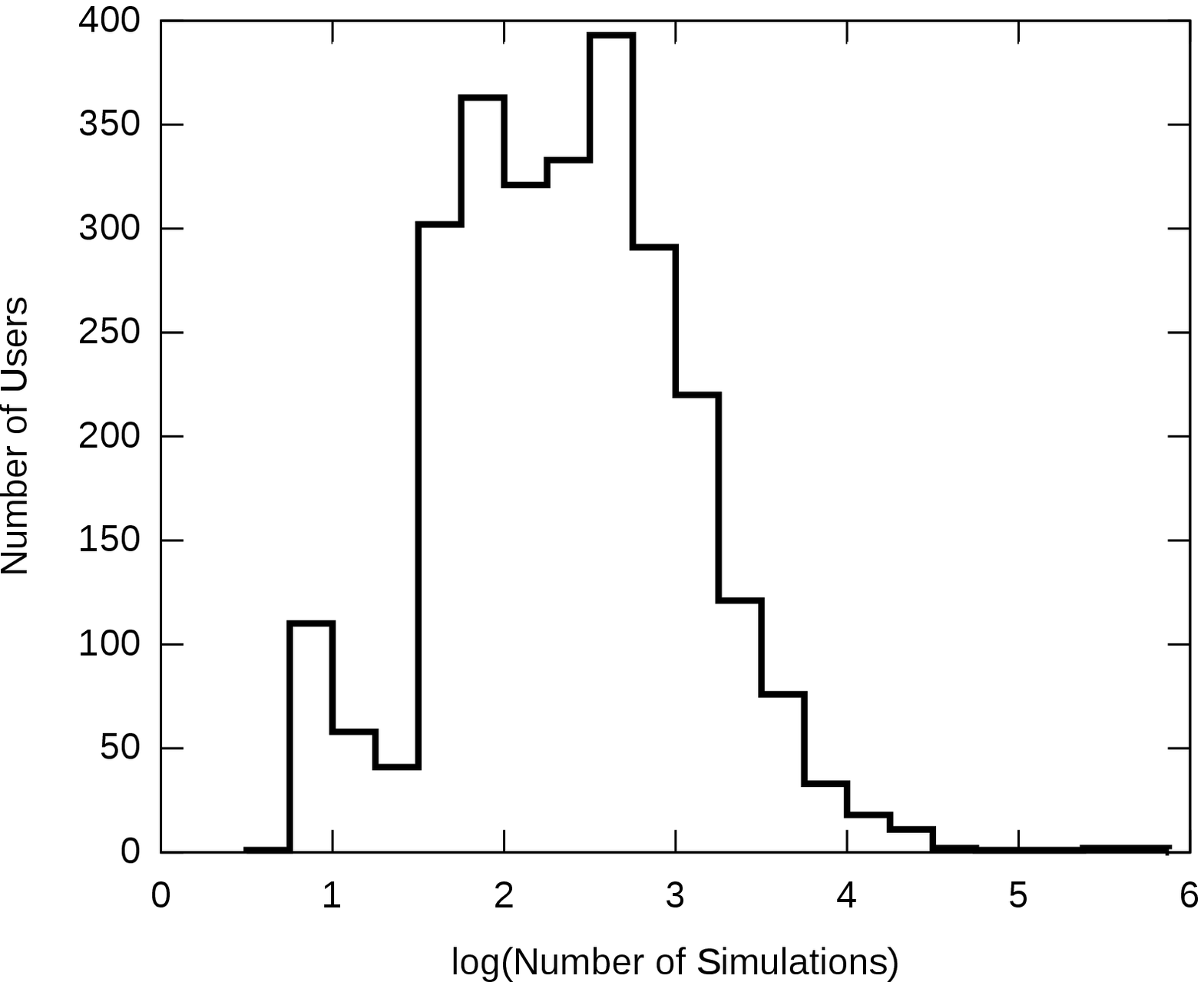}
\caption{The distribution of simulations viewed among volunteers.}
\label{ch4us}
\end{minipage}
\end{figure*}

Figure \ref{ch4sims} shows the cumulative number of simulations viewed by Citizen Scientists with respect to the time since the site launched. For the first six months the rate at which volunteers reviewed simulations was notably higher than the last two years the site was active. Figure \ref{ch4us} groups volunteers into bins by $log10$ of the number of simulations they viewed. Most volunteers viewed at least 64 simulations. There were $\sim$ 500 users that viewed at least 1000 simulations, and 34 users viewed 10000 simulations or more. The two most active volunteers viewed $\sim$ 325000 simulations over 250 hours and $\sim$ 553000 over 100 hours respectively. Based on the time zone information submitted with their results, over 90$\%$ of volunteers were from Europe or the United States.

\subsection{Results for each Pair of Galaxies}

\begin{figure*}
\begin{center}
\includegraphics[scale=0.6]{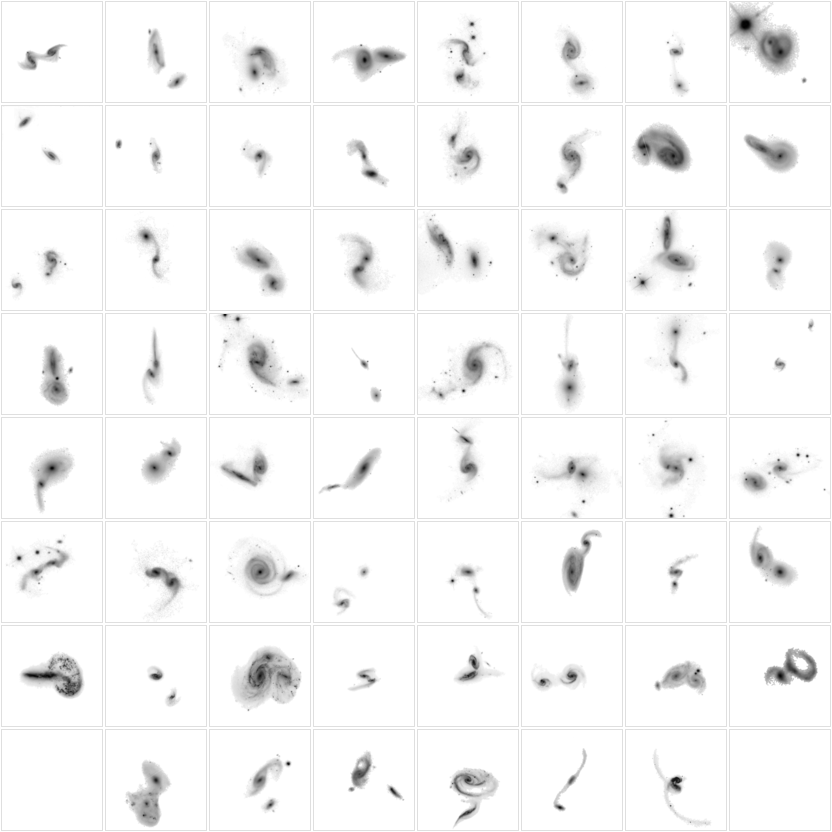}
\caption{All target images sorted by display order from Table \ref{tartbl1}.}
\label{allimg}
\end{center}
\end{figure*}

\begin{figure*}
\begin{center}
\includegraphics[scale=0.6]{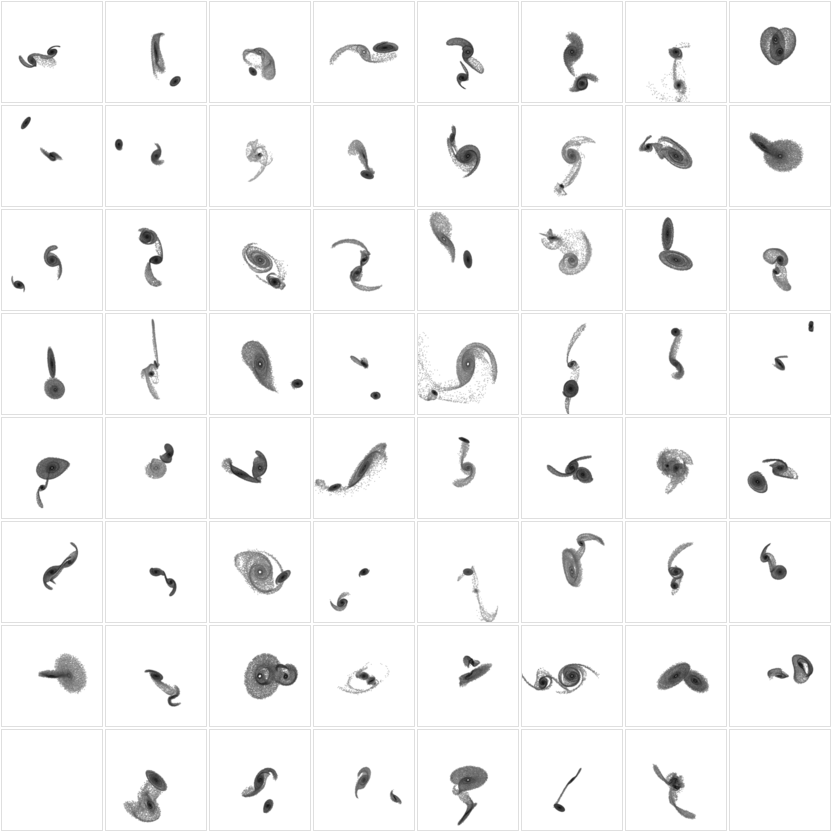}
\caption{All simulation results sorted by display order from Table \ref{tartbl1}.}
\label{allsim}
\end{center}
\end{figure*}

Figure \ref{allimg} shows all the target images of our interacting galaxy sample described in section \ref{sample}. Figure \ref{allsim} shows corresponding images of simulations representing the best models selected by the process detailed in Section \ref{model_selection}.

In the online data repository\footnote{\url{http://data.galaxyzoo.org/mergers.html}}, results for each of the 62 {\it Galaxy Zoo:~Mergers} targets are presented. For each target, we include both summary information about the simulations as well as the data to reproduce them. We include a simple histogram of the Merger Wars fitness to indicate the distribution of volunteer-judged ``quality" for the simulations. Next we attempt to demonstrate the convergence of the best-fit orbit through several means. The first is to present the target image along with the simulation results of the best three targets. Next, we present plots of the trajectories from the simulations for several different fitness populations. Similar trajectories indicate convergence. As the fitness level is increased, the diversity of trajectories should decrease if the model has converged. The next set of plots include information about how much of the total parameter space remains for each fitness population. We believe that we have developed a succinct method for presenting information about several populations of a dozen parameters for tens of thousands of simulations for each system. We will present one example of this summary in this paper to explain the layout and plots associated with each system.

 We would like to be able to present an accurate, quantitative measure of fitness of these models. However, no such metric has yet been identified. Simply doing image subtraction between models and the target galaxy leads to large errors because of the differences in the radial profiles of the models and the stretches of the images. Doing contour fits of isophotes also fails because a slight misalignment of a tidal feature by just a few degrees may actually be a ``better model" for an interaction than a model with no tidal features. In fact, the reason we are not doing this project automatically is exactly because we don't have this kind of objective measurement. Thanks to this project, we have the data to start creating such an objective fitness function. Further discussion about this is found in section \ref{fitness}.

\subsection{Arp 240 - An example system from {\it Galaxy Zoo:~Mergers}}

Our sample target was selected from the list of targets in the SDSS. Table \ref{mzsumtgt}  summarizes the {\it Galaxy Zoo:~Mergers} activity for the target. It lists the total number of simulations viewed by all volunteers, how many they rejected, how many they selected, and the number that were enhanced. The next three columns describe the {\it Merger Wars} outcome for the simulation images for this target. There were over 22000 Merger Wars competitions, but only $\sim$ 7000 winners. That means that for more than 15000 {\it Merger Wars} competitions, the volunteers clicked the neither button. All simulation results were included in approximately the same number of  {\it Merger Wars} competitions.

\begin{center}
\begin{table*}
\begin{center}
\caption{{\it Galaxy Zoo:~Mergers} summary for Arp 240.}
\begin{tabular}{ | c | c | c | c | c | c | c | }
\hline
Viewed & Rejected & Selected & Enhanced & MW Comps & MW Wins & Neither \\
\hline
74697 & 71868 & 2829 & 603 & 22745 & 7463 & 15282 \\ \hline
\end{tabular}
\label{mzsumtgt}
\end{center}
\end{table*}
\end{center}

\begin{figure*}
\begin{center}
\includegraphics*[width=0.35\textwidth]{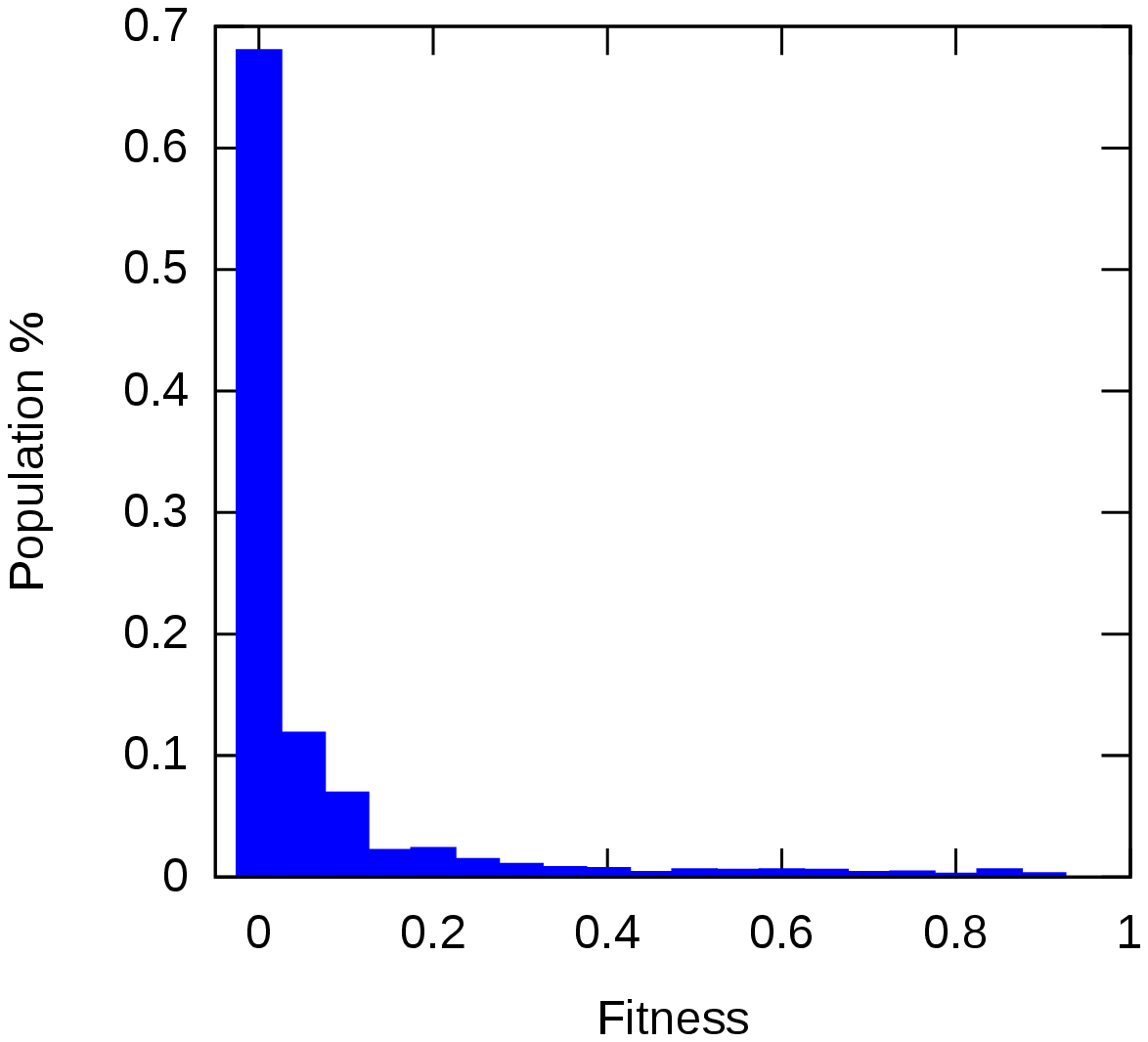}\\
\caption{Relative frequency of fitness for all 2829 selected states of Arp 240}
\label{fig:587722984435351614fhist}
\end{center}
\end{figure*}

The low number of {\it Merger Wars} winners resulted in a large number of simulations receiving a low {\it Merger Wars} fitness score. We see in Figure \ref{fig:587722984435351614fhist} that almost 70$\%$ of simulations were assigned a fitness of 0. Looking towards the higher fitness values, we see a relatively low fraction of states with fitness scores above 0.4, and only a few above 0.8. The distribution of fitness values is different for each target. 

\begin{figure*}
\begin{center}
\setlength{\tabcolsep}{0mm}
\subfloat{%
\begin{tabular}{c}
\includegraphics*[width=0.2\textwidth]{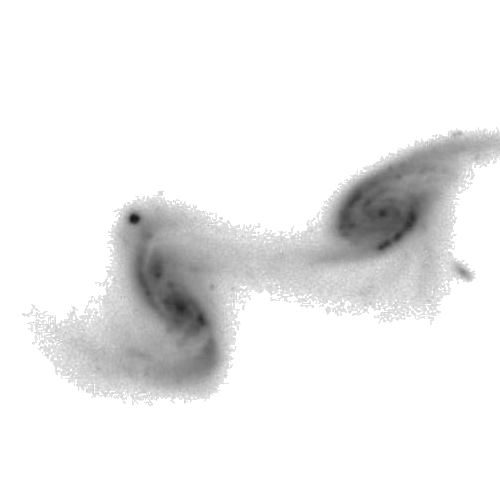}\\
\includegraphics*[width=0.2\textwidth]{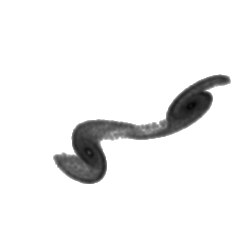}
\end{tabular}
}%
\subfloat{%
\begin{tabular}{c}
\includegraphics*[width=0.2\textwidth]{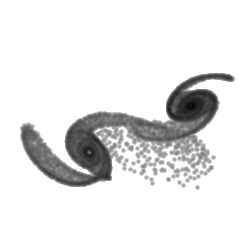}\\
\includegraphics*[width=0.2\textwidth]{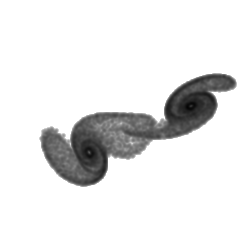}
\end{tabular}
}%
\caption{Target image and top three simulations for Arp 240}
\label{fig:587722984435351614match}
\end{center}
\end{figure*}

Figure \ref{fig:587722984435351614match} is a panel of four images, with the target image used on {\it Galaxy Zoo:~Mergers} for the pair of interacting galaxies presented in the upper left corner. The single best simulation for this target is located at the upper right panel. The next two highest fitness simulations occupy the bottom row, left to right. In this manner we can view how well the best simulations match the tidal features and overall morphology of the target image. For this particular pair of galaxies, the volunteers have done an excellent job. Each galaxy has symmetric tails that are recreated in the simulation with the proper size and orientation. 

\begin{figure*}
\begin{center}
\setlength{\tabcolsep}{0mm}
\subfloat{%
\begin{tabular}{c}
\includegraphics*[width=0.2\textwidth]{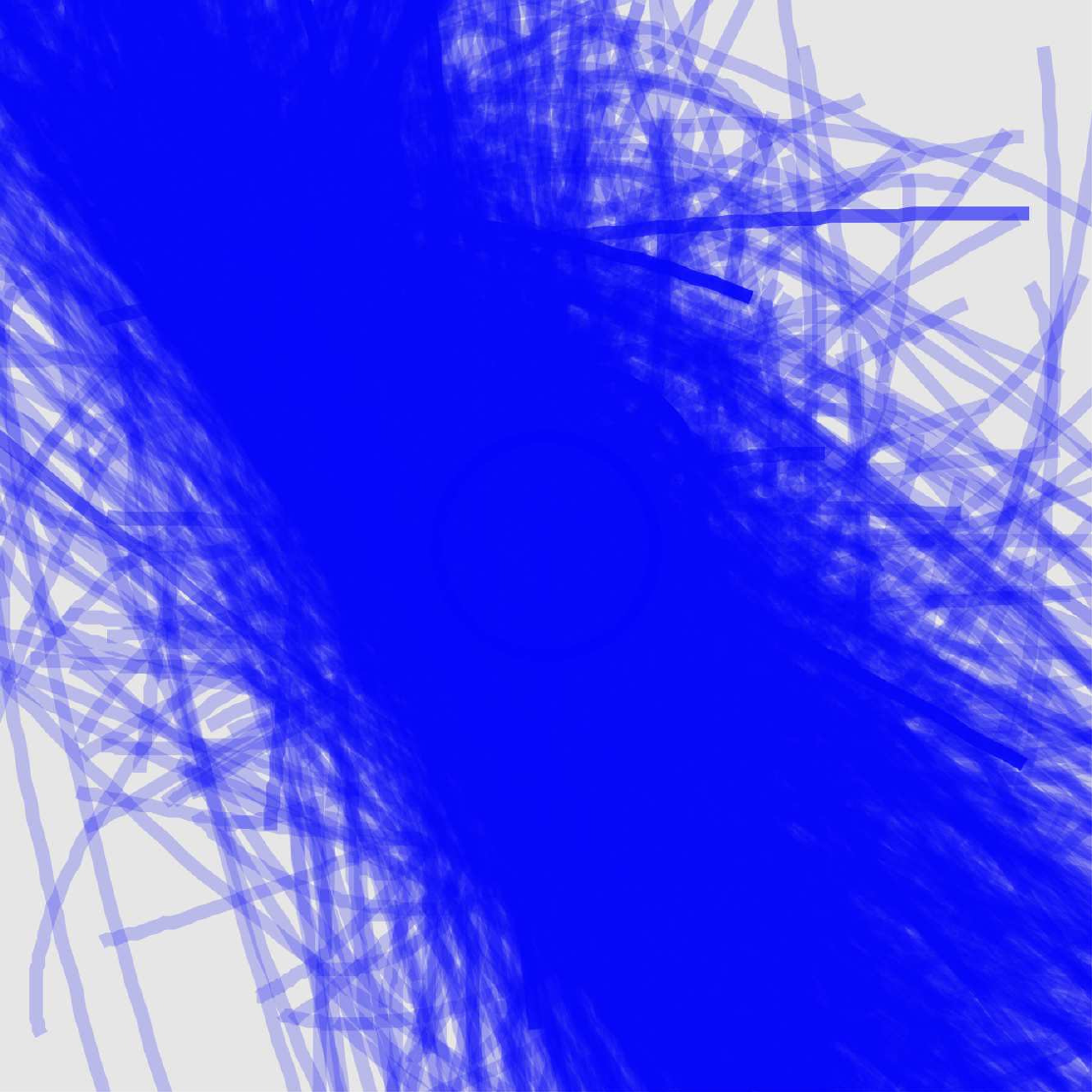}\\
\includegraphics*[width=0.2\textwidth]{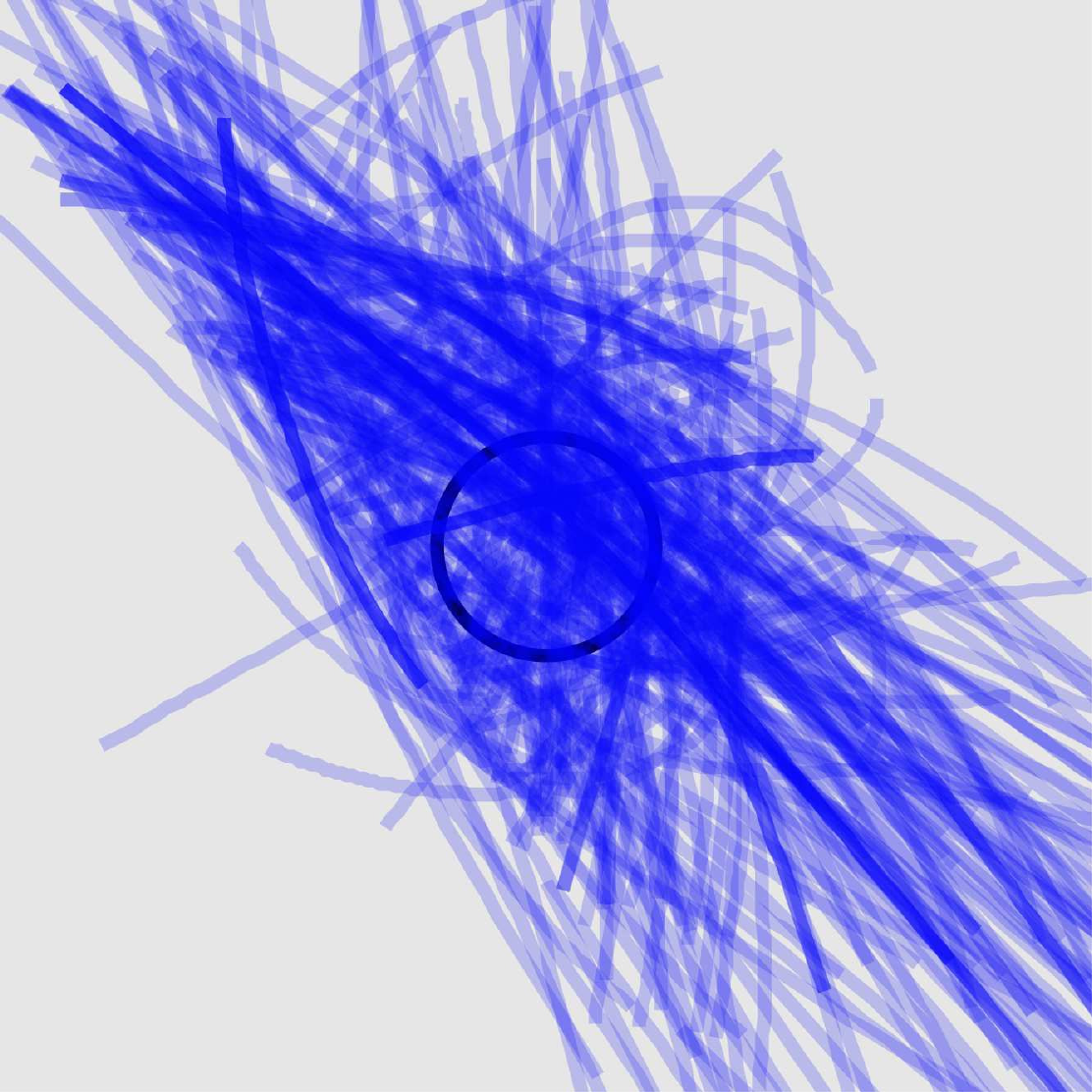}
\end{tabular}
}%
\subfloat{%
\begin{tabular}{c}
\includegraphics*[width=0.2\textwidth]{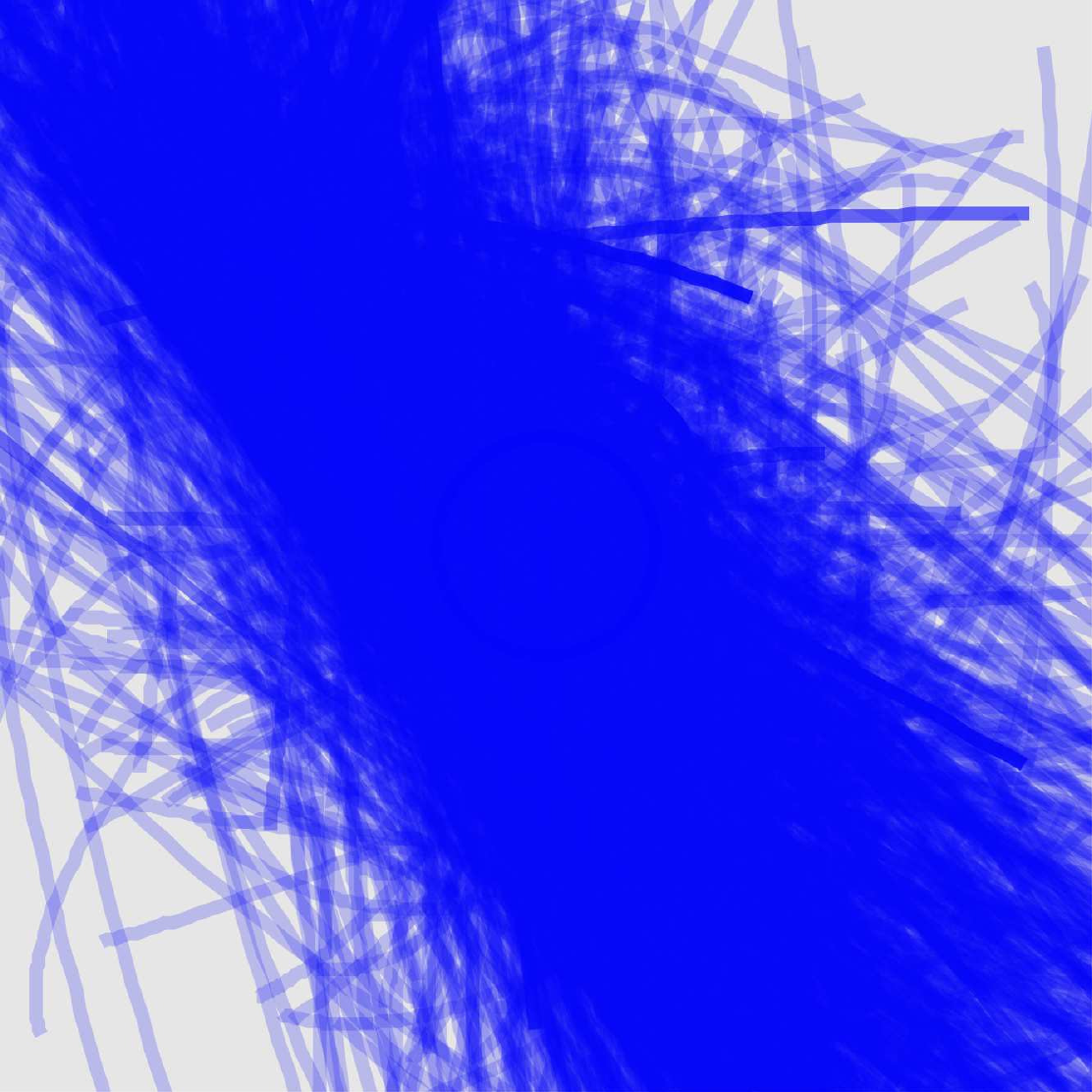}\\
\includegraphics*[width=0.2\textwidth]{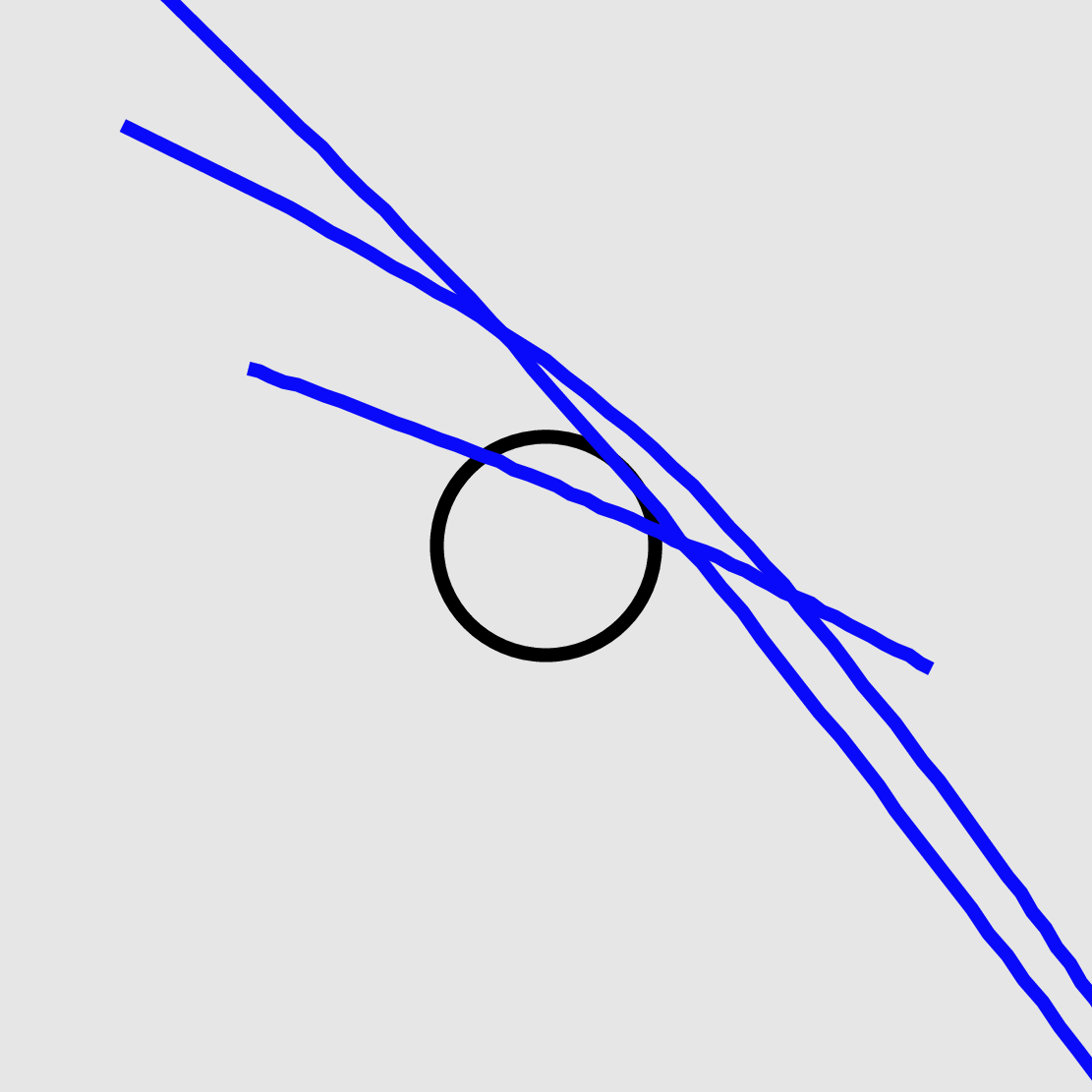}
\end{tabular}
}%
\caption{Trajectories for all selected states, the top 50$\%$, the top 10$\%$, and the top three states for Arp 240.}
\label{fig:587722984435351614traj}
\end{center}
\end{figure*}

 Figure \ref{fig:587722984435351614traj} comprises four panels that each show a set of simulated trajectories for the secondary galaxy relative to the primary galaxy. The trajectories are calculated as part of the simulation. They are rotated from the plane of the sky to be in the plane of the primary disc. This is a different rotation for each simulation because the orientation angles for the primary disc, $\theta_1$ and $\phi_1$, are allowed to vary. The black circle represents the size of the primary disc computed from the average of all $r_1$ values for the set of trajectories plotted in that panel. The circle is the same size in each plot, so the overall scale for each panel is adjusted accordingly. An individual blue line traces the path of the secondary galaxy for a single simulation. The top left panel shows the paths of all simulations that the volunteers selected. The top right panel shows the trajectories for the top 50$\%$ of the population, by fitness. The lower left panel shows the paths for the top 10$\%$. Finally, on the lower right, we see the trajectories for the top three simulations. These are the same simulations plotted in Figure \ref{fig:587722984435351614match}. The top three trajectories pass the primary disc in roughly the same location. They have similar shapes, but one trajectory appears somewhat shorter in this projection than the other two. This means it has a different inclination relative to the plane of the primary disc than the others. Trajectories that are very different from one another in the bottom right panel indicate a non-unique orbit and a system that likely has different dynamical solutions.

The final pair of plots for this system describes the level of convergence for each parameter. The parameters are broken up into two different sets. The first set of parameters are the twelve simulation parameters that were varied as part of the {\it Galaxy Zoo:~Mergers} process. This occurred either by random selection by the software during the {\it Explore} activity or by the volunteers' selections during the {\it Enhance} activity. The twelve parameters are:

\begin{itemize}
\item $r_z$ - the z-component of the orbit position vector in plane-of-the-sky frame;
\item $v_x$, $v_y$, and $v_z$ - the components of the orbit velocity vector in plane-of-the sky frame;
\item $m_1$, $m_2$ - the mass of the primary and secondary galaxies;
\item $r_1$, $r_2$ - the disc radius of the primary and secondary galaxies;
\item $\phi_1$, $\phi_2$ - the position angle of the primary and secondary galaxies;
\item $\theta_1$, $\theta_2$ - the disc inclination angle of the primary and secondary galaxies.
\end{itemize}

\begin{figure*}
\begin{center}
\setlength{\tabcolsep}{0mm}
\subfloat{%
\begin{tabular}{c}
\includegraphics*[width=0.4\textwidth]{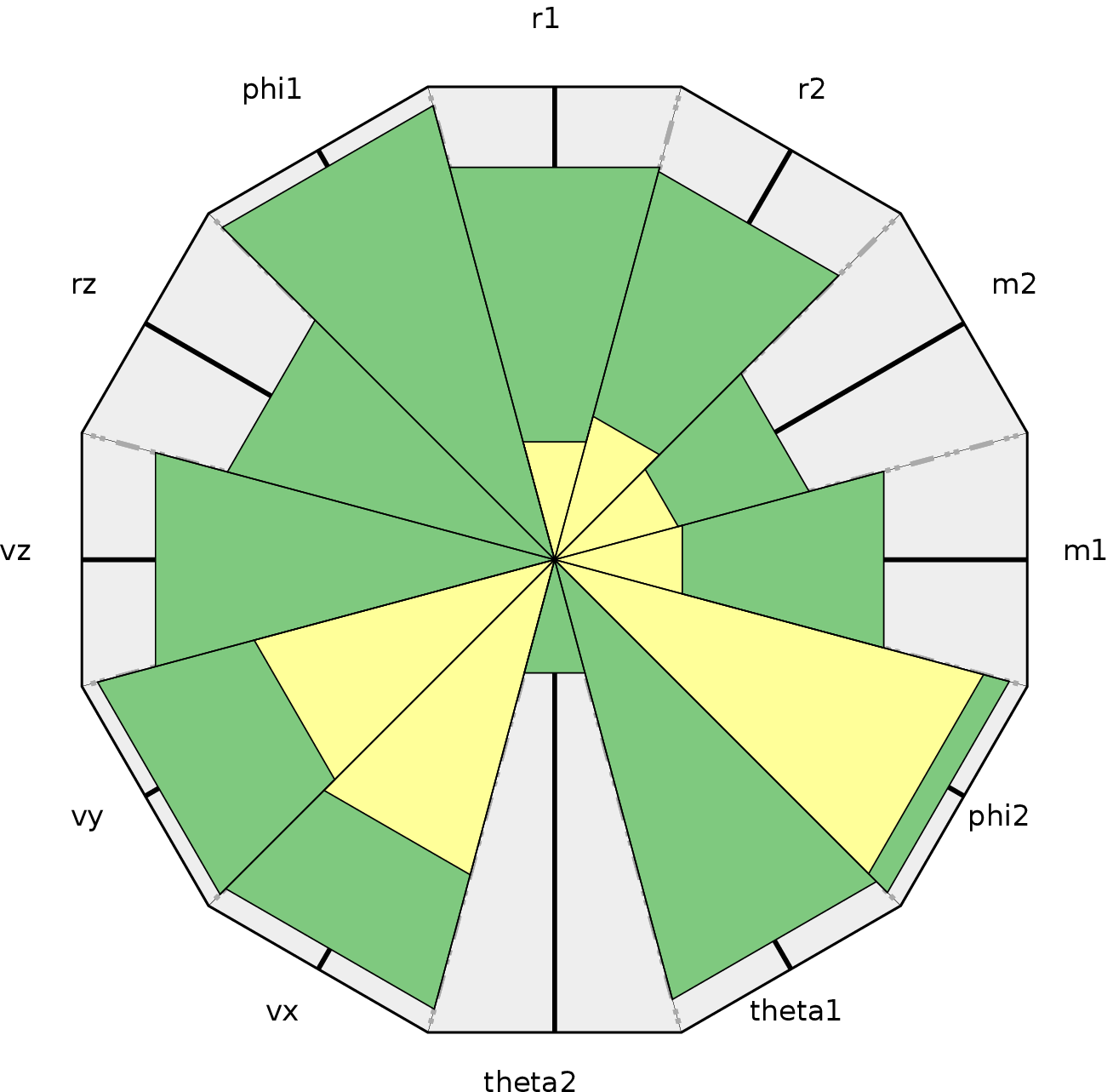}\\
\end{tabular}
}%
\subfloat{%
\begin{tabular}{c}
\includegraphics*[width=0.4\textwidth]{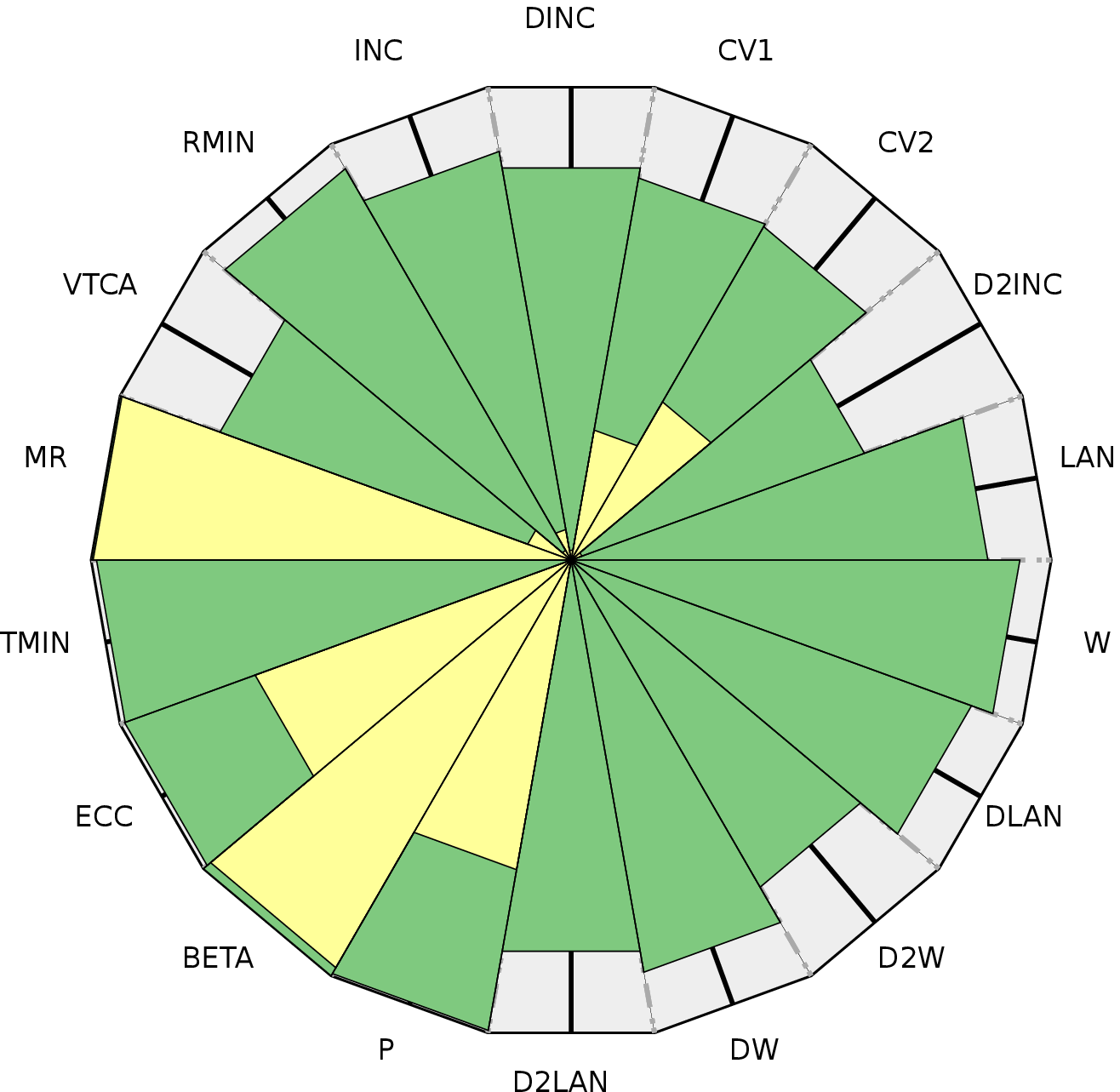}\\
\end{tabular}
}%
\caption{Glyph plots for convergence of simulation and orbit parameters for Arp 240. A larger radius represents better convergence for a parameter. The yellow shows the fraction of parameter space eliminated for all selected simulations and the green shows the parameter space eliminated for the top simulations.}
\label{fig:587722984435351614glyph}
\end{center}
\end{figure*}

For each parameter, two populations are considered: all selected states and the top states selected by the experts. The set of expert states were chosen from the {\it Merger Wars} winners and presented to the Citizen Scientists for further review in the {\it Galaxy Zoo:~Mergers} final activities ({\it Simulation Showdown} and {\it Best of the Best}) and included between four and eight high-fitness simulations for each target. For each population, the remaining fraction is computed by dividing the full range of parameters in that population by the full range of parameters for all simulations viewed for that target. For example, consider a target where $r_z$ was allowed to vary between -10 and 10 simulation units. For the selected states, the value ranged between -5 and 5. The fraction of that parameter remaining would be 10 divided by 20 or 0.5. If the set of expert states had $r_z$ values between 0 and 2, then the fraction of parameter space remaining for that parameter would be 0.1. We plot the convergence information for each parameter along its own radial line in the glyph plot (see Figure \ref{fig:587722984435351614glyph}). Here we have chosen to represent the radial distance in a given direction not by how much of the parameter space remains, but by how much was eliminated. Well converged values have a large radius. This means a well converged population will have a large area. We plot two populations for each glyph. The bottom, green layer is the population of the expert states. The top, yellow layer is the population of all selected simulations. The area of the glyph for all selected simulations is smaller than the area for the top simulations. This is consistent with the top fitness population having better convergence than the larger population.

The right panel in Figure \ref{fig:587722984435351614glyph} contains a similar plot for the orbit parameters. These values included classical orbit elements such as eccentricity as well as orientation angles relative to the plane of the sky and the plane of each disc. The orbit parameters shown in the plot are:

\begin{itemize}
\item $t_{min}$ - the time since closest approach of the two galaxies;
\item $r_{min}$ - the closest approach distance;
\item $p$ - the orbital semi-parameter for conic section orbits;\footnote{For elliptical orbits $p$ is related to the semi-major axis, $a$, by $p = a(1-e^2).$}
\item $ecc$- the eccentricity of the orbit;
\item $inc$, $lan$, $\omega$ - the inclination angle for the orbit, longitude of ascending node, and argument of pericentre in plane-of-the-sky frame;
\item $dinc$, $dlan$, $d\omega$ - the inclination angle for the orbit, longitude of ascending node, and argument of pericentre in frame of the primary disc;
\item $dinc2$, $dlan2$, $d\omega2$ - the inclination angle for the orbit, longitude of ascending node, and argument of pericentre in frame of the secondary disc;
\item $mr$ - mass ratio;
\item $vtca$ - the velocity at time of closest approach;
\item $cv_1$, $cv_2$ - the orbital velocity of a particle at the edge of the disc of the primary and secondary galaxy;
\item $\beta$ - the interaction parameter.
\end{itemize}

\subsection{Convergence of Simulation Parameters}

For each of the twelve simulation parameters we can compute the remaining fraction of parameter space. This is done by computing the range of parameter values for the top four to eight states selected by the experts and then dividing by the range of parameter values for all states shown to the Citizen Scientists. This ratio indicates how much of parameter space remains, and by inversion, how much of the parameter space was ruled out for producing a good match to the target image. The distribution of remaining fraction values are shown in Table \ref{rfdist}. Overall, the components of the relative velocity vector were best constrained, along with the z-component of the relative position vector. This means the process places tight constraints on the path through space that the secondary galaxy travels with respect to the primary. The disc radii are the next best constrained along with the disc inclination angle. The two masses typically have remaining fractions near 0.3. Eliminating 70$\%$ of a parameter that spans two orders of magnitude is a significant reduction. The least well constrained values are the two position angles, $\theta$. These values are considered over linear ranges.

\begin{table}
\begin{center}
\caption{The mean, median, min, and max remaining fraction (RF) for the twelve simulation parameters.}
\begin{tabular}{ | c | c | c | l | c | }
\hline
Name & Mean RF & Median RF & Best RF &  Worst RF \\
\hline
$r_z$ & 0.12 & 0.09 & 0.0 & 0.54 \\
$v_x$ & 0.04 & 0.02 & 0.003 & 0.36 \\
$v_y$ & 0.05 & 0.02 & 0.005 & 0.54 \\
$v_z$ & 0.06 & 0.03 & 0.0 & 0.44 \\
$m_1$ & 0.33 & 0.30 & 0.0007 & 0.94 \\
$m_2$ & 0.32 & 0.28 & 0.0 & 0.87 \\
$r_1$ & 0.29 & 0.22 & 0.01 & 0.97 \\
$r_2$ & 0.31 & 0.22 & 0.007 & 0.97 \\
$\phi_1$ & 0.29 & 0.17 & 0.004 & 0.97 \\
$\phi_2$ & 0.37 & 0.27 & 0.0 & 0.97 \\
$\theta_1$ & 0.53 & 0.65 & 0.01 & 0.95 \\
$\theta_2$ & 0.56 & 0.62 & 0.0 & 0.97 \\
\hline
\end{tabular}
\label{rfdist}
\end{center}
\end{table}

\subsection{Range of Model Parameters}
\label{modelparam}

In Table \ref{restbl2}, we present the best fit parameters for the 62 models of interacting galaxies. Full details for these models are available on-line\footnote{\url{http://data.galaxyzoo.org/mergers.html}}. In the on-line data repository, we have also linked to the JSPAM software to allow people the ability to run and reproduce the results from this project. In addition to the best fit models, we have included a table with all of the models ranked by the {\it Merger Wars} interface including their human derived fitness function. The mass ratios, minimum distance, time since closest approach, orbital eccentricity, and $\beta$ are included. Table \ref{restbl2} includes the value from the best-fit simulation as well as the minimum and maximum values from all simulations presented to the Citizen Scientists. The mass ratios cover a large range of values due to the large range in individual mass values that were sampled. The minimum separation distance varies approximately two orders of magnitude. The time since closest approach ranged from 0, the current epoch, to over 8 Gyr. However, the oldest best-fit $t_{min}$ is less than 0.57 Gyr. Orbit eccentricities range from circular, 0, through parabolic, 1, and very hyperbolic, greater than 1000. The $\beta$ parameter had a wide range from a minimum less than 0.001 to a maximum value of over 50000 for the most extreme case. The uncertainty for simulation values was estimated by sampling the values from the set of ``top" fitness states 10000 times in accordance to their fitness values from the two final activities. The variance in the sampled population is calculated in order to estimate the one $\sigma$ errors for each simulation, and derived orbit, parameter. Table \ref{tblsimparam} includes sample rows describing the key simulation parameters available in the data repository. 

\begin{landscape}
\begin{table}
\caption[Simulation results and orbit parameters]{\small Simulation results for each target include the mass ratio, minimum distance, time since closest approach, orbital eccentricity, and $\beta$ parameters. The best value, standard deviation of the best simulations, and the full min and max of all simulated values are included. The $\ast$ indicates a negative kurtosis value for the fitness distribution.}
\begin{center}
\scalebox{0.70}{
\begin{tabular}{| c | r | r | r | r | r | r | r | r | r | r | r | r | r | r | r |}
\hline
Target & \MyHead{1cm}{Best \\ MR} & \MyHead{1cm}{Min \\ MR} & \MyHead{1cm}{Max \\ MR}  & \MyHead{1.5cm}{$r_{min}$ \\ (kpc)}   & \MyHead{1.5cm}{Min \\ $r_{min}$ \\ (kpc)}   & \MyHead{1.5cm}{Max \\ $r_{min}$ \\ (kpc)}  & \MyHead{1.5cm}{$t_{min}$ \\ (Myr)} & \MyHead{1.5cm}{Min \\ $t_{min}$ \\ (Myr)}  & \MyHead{1.5cm}{Max \\ $t_{min}$ \\ (Myr)} & \MyHead{1cm}{$ecc$} & \MyHead{1cm}{Min \\ $ecc$} & \MyHead{1cm}{Max \\ $ecc$} & \MyHead{1cm}{$\beta$} & \MyHead{1cm}{Min \\ $\beta$} & \MyHead{1cm}{Max \\ $\beta$} \\
\hline
Arp 240 & 1.167 $\pm$ 0.391 & 0.307 & 889.801 & 81.443 $\pm$ 10.206 & 0.402 & 409.274 & 249.449 $\pm$ 31.503 & 0.000 & 8602.542 & 3.700 $\pm$ 2.349 & 0.001 & 677.694 & 0.291 $\pm$ 0.056 & 0.001 & 3920.725 \\
Arp 290* & 1.427 $\pm$ 0.415 & 0.020 & 272.820 & 18.813 $\pm$ 5.501 & 0.128 & 199.207 & 479.844 $\pm$ 117.278 & 0.000 & 8612.936 & 0.790 $\pm$ 0.352 & 0.001 & 3080.303 & 0.499 $\pm$ 1.282 & 0.001 & 4361.565 \\
Arp 142 & 0.712 $\pm$ 0.200 & 0.147 & 15.843 & 8.947 $\pm$ 1.142 & 0.105 & 142.032 & 77.953 $\pm$ 6.293 & 0.000 & 8657.975 & 0.442 $\pm$ 0.141 & 0.001 & 168.073 & 2.937 $\pm$ 0.505 & 0.009 & 7040.448 \\
Arp 318 & 0.358 $\pm$ 0.074 & 0.018 & 217.774 & 15.036 $\pm$ 2.608 & 0.190 & 383.156 & 169.764 $\pm$ 7.645 & 0.000 & 8656.243 & 1.579 $\pm$ 0.340 & 0.001 & 838.669 & 0.732 $\pm$ 0.214 & 0.002 & 2763.580 \\
Arp 256* & 1.036 $\pm$ 0.403 & 0.372 & 5.458 & 17.250 $\pm$ 2.751 & 0.102 & 231.411 & 122.992 $\pm$ 24.529 & 0.000 & 8657.975 & 1.628 $\pm$ 0.335 & 0.002 & 400.671 & 0.801 $\pm$ 0.351 & 0.002 & 12355.246 \\
UGC 11751 & 1.234 $\pm$ 0.105 & 0.185 & 8.147 & 16.299 $\pm$ 4.509 & 0.207 & 379.329 & 502.363 $\pm$ 53.766 & 0.000 & 8628.526 & 0.991 $\pm$ 0.376 & 0.012 & 2272.374 & 0.369 $\pm$ 0.195 & 0.000 & 606.179 \\
Arp 104 & 1.845 $\pm$ 0.564 & 0.076 & 16.571 & 1.069 $\pm$ 0.831 & 0.057 & 148.908 & 67.559 $\pm$ 26.077 & 0.000 & 8635.455 & 0.640 $\pm$ 0.112 & 0.001 & 464.457 & 101.600 $\pm$ 37.321 & 0.002 & 31287.255 \\
Double Ring & 0.798 $\pm$ 0.931 & 0.074 & 29.031 & 1.527 $\pm$ 0.196 & 0.075 & 141.077 & 65.827 $\pm$ 9.220 & 0.000 & 8628.526 & 0.657 $\pm$ 0.332 & 0.007 & 483.573 & 32.806 $\pm$ 6.321 & 0.003 & 14902.455 \\
Arp 285* & 0.050 $\pm$ 0.284 & 0.050 & 29.772 & 18.315 $\pm$ 5.899 & 0.285 & 83.746 & 242.520 $\pm$ 127.640 & 0.000 & 8614.668 & 2.533 $\pm$ 0.740 & 0.002 & 841.036 & 0.228 $\pm$ 0.250 & 0.003 & 496.124 \\
Arp 214 & 1.151 $\pm$ 0.147 & 0.183 & 3.646 & 9.324 $\pm$ 2.821 & 0.170 & 152.956 & 495.434 $\pm$ 16.927 & 0.000 & 8635.455 & 1.007 $\pm$ 0.346 & 0.004 & 3181.084 & 0.675 $\pm$ 0.994 & 0.001 & 833.729 \\
NGC 4320 & 0.709 $\pm$ 0.255 & 0.060 & 50.696 & 4.117 $\pm$ 6.748 & 0.077 & 155.099 & 133.386 $\pm$ 12.646 & 0.000 & 8651.046 & 0.365 $\pm$ 0.466 & 0.002 & 1180.444 & 12.037 $\pm$ 5.096 & 0.002 & 10825.800 \\
UGC 7905 & 0.177 $\pm$ 0.109 & 0.004 & 79.719 & 8.363 $\pm$ 2.480 & 0.112 & 140.481 & 102.205 $\pm$ 10.118 & 0.000 & 8652.778 & 0.218 $\pm$ 0.068 & 0.001 & 176.206 & 2.548 $\pm$ 8.124 & 0.006 & 6797.398 \\
Arp 255* & 5.266 $\pm$ 0.559 & 1.578 & 15.792 & 20.493 $\pm$ 4.082 & 0.188 & 150.749 & 247.717 $\pm$ 65.567 & 0.000 & 8657.975 & 0.375 $\pm$ 0.142 & 0.002 & 309.546 & 0.885 $\pm$ 0.668 & 0.005 & 3510.090 \\
Arp 82 & 1.986 $\pm$ 0.454 & 0.021 & 416.918 & 6.374 $\pm$ 1.638 & 0.047 & 151.156 & 131.654 $\pm$ 20.967 & 0.000 & 8657.975 & 0.155 $\pm$ 0.134 & 0.001 & 931.630 & 5.021 $\pm$ 1.694 & 0.004 & 47950.387 \\
Arp 239 & 4.957 $\pm$ 1.582 & 0.314 & 11.597 & 4.592 $\pm$ 0.708 & 0.138 & 129.854 & 131.654 $\pm$ 47.964 & 0.000 & 8652.778 & 0.119 $\pm$ 0.071 & 0.000 & 118.606 & 5.899 $\pm$ 5.295 & 0.009 & 3637.425 \\
Arp 199* & 1.569 $\pm$ 0.391 & 0.177 & 6.620 & 12.080 $\pm$ 2.682 & 0.140 & 141.443 & 3.465 $\pm$ 8.720 & 0.000 & 8595.613 & 7.365 $\pm$ 2.416 & 0.005 & 244.303 & 0.257 $\pm$ 0.621 & 0.002 & 2238.622 \\
Arp 57 & 1.441 $\pm$ 0.252 & 0.382 & 5.886 & 14.738 $\pm$ 2.852 & 0.648 & 231.592 & 289.292 $\pm$ 42.933 & 0.000 & 8654.510 & 4.378 $\pm$ 1.513 & 0.008 & 182.603 & 0.683 $\pm$ 0.143 & 0.002 & 337.144 \\
Pair 18* & 0.395 $\pm$ 0.273 & 0.204 & 6.945 & 24.142 $\pm$ 4.360 & 0.223 & 325.138 & 263.308 $\pm$ 73.111 & 0.000 & 8657.975 & 0.748 $\pm$ 0.318 & 0.001 & 394.143 & 0.665 $\pm$ 0.425 & 0.002 & 2593.574 \\
Arp 247* & 2.069 $\pm$ 0.700 & 0.154 & 7.558 & 6.605 $\pm$ 5.446 & 0.147 & 95.170 & 320.473 $\pm$ 128.951 & 0.000 & 8657.975 & 0.413 $\pm$ 0.230 & 0.001 & 780.897 & 3.025 $\pm$ 5.967 & 0.002 & 3912.572 \\
Arp 241 & 0.677 $\pm$ 0.244 & 0.130 & 28.202 & 7.389 $\pm$ 1.926 & 0.064 & 150.308 & 329.135 $\pm$ 121.723 & 0.000 & 8637.188 & 0.219 $\pm$ 0.048 & 0.001 & 328.179 & 2.049 $\pm$ 2.282 & 0.002 & 17007.528 \\
Arp 313 & 1.887 $\pm$ 1.087 & 0.150 & 6.748 & 17.512 $\pm$ 3.274 & 0.226 & 127.950 & 109.134 $\pm$ 37.686 & 0.000 & 8602.542 & 6.253 $\pm$ 3.008 & 0.002 & 1065.720 & 0.261 $\pm$ 0.256 & 0.001 & 2176.819 \\
Arp 107* & 1.353 $\pm$ 0.474 & 0.167 & 21.220 & 6.741 $\pm$ 5.656 & 0.203 & 201.900 & 420.946 $\pm$ 86.802 & 0.000 & 8657.975 & 0.400 $\pm$ 0.246 & 0.001 & 732.706 & 3.651 $\pm$ 1.185 & 0.001 & 2552.681 \\
Arp 294* & 0.648 $\pm$ 0.220 & 0.146 & 4.625 & 62.351 $\pm$ 18.874 & 0.231 & 142.814 & 119.528 $\pm$ 56.493 & 0.000 & 8656.243 & 1.491 $\pm$ 2.354 & 0.001 & 325.053 & 0.078 $\pm$ 0.136 & 0.003 & 1459.349 \\
Arp 172* & 1.400 $\pm$ 0.343 & 0.103 & 10.923 & 2.146 $\pm$ 0.519 & 0.111 & 148.919 & 155.906 $\pm$ 19.514 & 0.000 & 8656.243 & 0.523 $\pm$ 0.150 & 0.001 & 151.476 & 37.757 $\pm$ 8.052 & 0.013 & 8819.703 \\
Arp 302* & 2.692 $\pm$ 1.014 & 0.143 & 30.496 & 68.384 $\pm$ 27.017 & 0.098 & 141.524 & 100.473 $\pm$ 844.061 & 0.000 & 8649.314 & 28.310 $\pm$ 12.875 & 0.002 & 668.570 & 0.017 $\pm$ 7.741 & 0.002 & 8194.417 \\
Arp 242 & 0.597 $\pm$ 0.522 & 0.174 & 6.803 & 15.990 $\pm$ 5.001 & 0.081 & 150.089 & 429.607 $\pm$ 192.852 & 0.000 & 8657.975 & 0.702 $\pm$ 0.274 & 0.001 & 198.824 & 0.655 $\pm$ 2.415 & 0.005 & 12756.872 \\
Arp 72 & 1.379 $\pm$ 0.956 & 0.599 & 10.777 & 14.658 $\pm$ 1.651 & 0.213 & 157.070 & 162.835 $\pm$ 83.128 & 0.000 & 8652.778 & 4.699 $\pm$ 1.909 & 0.002 & 5138.944 & 0.207 $\pm$ 0.191 & 0.000 & 736.442 \\
Arp 101 & 0.864 $\pm$ 0.031 & 0.097 & 9.110 & 4.873 $\pm$ 1.353 & 0.137 & 71.771 & 214.804 $\pm$ 20.872 & 0.000 & 8642.384 & 0.636 $\pm$ 0.158 & 0.014 & 680.411 & 4.460 $\pm$ 2.730 & 0.003 & 3324.229 \\
Arp 58* & 5.944 $\pm$ 1.206 & 0.088 & 9.070 & 21.807 $\pm$ 1.751 & 0.245 & 148.497 & 256.379 $\pm$ 10.929 & 0.000 & 8657.975 & 0.580 $\pm$ 0.169 & 0.001 & 304.237 & 0.713 $\pm$ 0.042 & 0.004 & 1843.127 \\
Arp 105 & 0.211 $\pm$ 0.035 & 0.019 & 393.639 & 21.427 $\pm$ 3.171 & 0.069 & 120.486 & 278.898 $\pm$ 30.897 & 0.000 & 8652.778 & 1.784 $\pm$ 0.160 & 0.002 & 804.968 & 0.649 $\pm$ 0.223 & 0.002 & 8977.519 \\
Arp 97 & 0.781 $\pm$ 0.113 & 0.415 & 9.752 & 17.261 $\pm$ 3.428 & 0.085 & 146.805 & 325.670 $\pm$ 35.396 & 0.000 & 8656.243 & 0.824 $\pm$ 0.331 & 0.001 & 1434.159 & 0.281 $\pm$ 0.202 & 0.002 & 3633.394 \\
Arp 305* & 1.809 $\pm$ 0.097 & 0.194 & 27.380 & 12.595 $\pm$ 2.398 & 0.164 & 91.844 & 472.915 $\pm$ 45.624 & 0.000 & 6423.324 & 1.549 $\pm$ 0.556 & 0.006 & 1323.764 & 0.590 $\pm$ 0.219 & 0.001 & 1759.599 \\
Arp 106 & 6.686 $\pm$ 0.432 & 1.194 & 14.136 & 7.088 $\pm$ 1.134 & 0.083 & 147.802 & 124.725 $\pm$ 45.391 & 0.000 & 8657.975 & 0.626 $\pm$ 0.214 & 0.000 & 179.424 & 2.988 $\pm$ 1.579 & 0.004 & 11209.197 \\
NGC 2802 & 2.436 $\pm$ 0.861 & 0.035 & 6.708 & 8.734 $\pm$ 0.840 & 0.150 & 140.689 & 91.811 $\pm$ 22.209 & 0.000 & 8657.975 & 2.821 $\pm$ 0.761 & 0.014 & 383.936 & 1.026 $\pm$ 0.572 & 0.002 & 3378.716 \\
Arp 301 & 0.759 $\pm$ 0.524 & 0.016 & 284.456 & 16.455 $\pm$ 2.599 & 0.092 & 197.674 & 65.827 $\pm$ 59.043 & 0.000 & 8657.975 & 1.301 $\pm$ 0.558 & 0.000 & 882.459 & 0.930 $\pm$ 0.328 & 0.002 & 11681.497 \\
Arp 89* & 1.655 $\pm$ 5.332 & 0.185 & 21.114 & 25.408 $\pm$ 4.280 & 0.153 & 147.430 & 452.127 $\pm$ 128.352 & 0.000 & 8657.975 & 0.987 $\pm$ 0.441 & 0.001 & 1136.358 & 0.292 $\pm$ 0.242 & 0.001 & 2150.949 \\
Arp 87 & 0.868 $\pm$ 0.261 & 0.136 & 19.277 & 11.147 $\pm$ 2.234 & 0.122 & 138.027 & 329.135 $\pm$ 103.689 & 0.000 & 8538.447 & 1.806 $\pm$ 0.146 & 0.003 & 2021.211 & 0.522 $\pm$ 0.105 & 0.001 & 2485.946 \\
Arp 191 & 0.602 $\pm$ 0.077 & 0.053 & 8.506 & 9.236 $\pm$ 2.566 & 0.160 & 230.897 & 147.244 $\pm$ 32.344 & 0.000 & 8657.975 & 1.295 $\pm$ 0.517 & 0.000 & 126.375 & 1.916 $\pm$ 2.567 & 0.003 & 4413.198 \\
Arp 237 & 2.413 $\pm$ 0.317 & 0.214 & 6.025 & 3.621 $\pm$ 1.768 & 0.172 & 137.550 & 103.937 $\pm$ 33.587 & 0.000 & 8656.243 & 0.386 $\pm$ 0.119 & 0.001 & 241.766 & 13.322 $\pm$ 8.121 & 0.004 & 3083.534 \\
Arp 181 & 0.563 $\pm$ 0.182 & 0.109 & 6.013 & 2.169 $\pm$ 0.136 & 0.065 & 148.946 & 20.787 $\pm$ 4.777 & 0.000 & 8657.975 & 0.336 $\pm$ 0.044 & 0.001 & 109.481 & 24.801 $\pm$ 3.903 & 0.002 & 21124.506 \\
Arp 238 & 1.098 $\pm$ 0.364 & 0.197 & 5.019 & 13.540 $\pm$ 2.952 & 0.174 & 275.404 & 58.898 $\pm$ 13.132 & 0.000 & 8657.975 & 0.106 $\pm$ 0.336 & 0.001 & 192.612 & 1.900 $\pm$ 1.683 & 0.003 & 5201.787 \\
Pair 42 & 2.099 $\pm$ 0.375 & 0.197 & 4.140 & 4.953 $\pm$ 1.116 & 0.110 & 263.861 & 50.236 $\pm$ 16.381 & 0.000 & 8657.975 & 0.471 $\pm$ 0.089 & 0.001 & 596.985 & 5.908 $\pm$ 8.829 & 0.002 & 11168.961 \\
Arp 297 & 2.322 $\pm$ 0.170 & 0.068 & 77.628 & 18.567 $\pm$ 1.280 & 0.069 & 146.923 & 562.994 $\pm$ 54.050 & 0.000 & 8656.243 & 0.587 $\pm$ 0.026 & 0.001 & 937.365 & 0.576 $\pm$ 0.094 & 0.002 & 14854.786 \\
NGC 5753/5 & 1.761 $\pm$ 0.963 & 0.082 & 18.461 & 8.824 $\pm$ 3.845 & 0.103 & 132.723 & 337.796 $\pm$ 93.307 & 0.000 & 8657.975 & 0.794 $\pm$ 0.292 & 0.000 & 2069.538 & 1.374 $\pm$ 1.092 & 0.001 & 5754.884 \\
Arp 173 & 30.202 $\pm$ 55.784 & 0.138 & 1075.938 & 4.775 $\pm$ 5.144 & 0.223 & 98.984 & 174.961 $\pm$ 323.403 & 0.000 & 8637.188 & 0.341 $\pm$ 0.186 & 0.000 & 171.560 & 16.792 $\pm$ 6.935 & 0.007 & 4792.330 \\
Arp 84* & 4.089 $\pm$ 0.959 & 0.037 & 271.690 & 4.606 $\pm$ 3.279 & 0.078 & 133.241 & 230.394 $\pm$ 31.216 & 0.000 & 8657.975 & 0.054 $\pm$ 0.309 & 0.004 & 2237.097 & 4.418 $\pm$ 1.349 & 0.001 & 10030.263 \\
UGC 10650* & 0.945 $\pm$ 0.410 & 0.910 & 6.015 & 13.918 $\pm$ 3.706 & 0.067 & 143.050 & 452.127 $\pm$ 59.788 & 0.000 & 8654.510 & 0.604 $\pm$ 0.163 & 0.000 & 345.020 & 0.788 $\pm$ 2.122 & 0.003 & 19540.015 \\
Arp 112 & 0.950 $\pm$ 0.229 & 0.007 & 149.897 & 5.867 $\pm$ 1.610 & 0.063 & 149.720 & 58.898 $\pm$ 59.145 & 0.000 & 8657.975 & 0.050 $\pm$ 0.141 & 0.000 & 171.140 & 5.684 $\pm$ 12.056 & 0.002 & 21353.400 \\
Arp 274* & 3.239 $\pm$ 1.114 & 0.067 & 14.916 & 59.210 $\pm$ 19.783 & 0.312 & 142.069 & 0.000 $\pm$ 16.127 & 0.000 & 8638.920 & 16.797 $\pm$ 19.740 & 0.007 & 658.300 & 0.025 $\pm$ 0.076 & 0.003 & 773.381 \\
Arp 146 & 0.833 $\pm$ 0.120 & 0.049 & 1.376 & 2.776 $\pm$ 1.767 & 0.180 & 146.276 & 102.205 $\pm$ 17.051 & 0.000 & 8657.975 & 0.233 $\pm$ 0.732 & 0.003 & 82.315 & 14.192 $\pm$ 7.284 & 0.004 & 2833.376 \\
Arp 143 & 2.979 $\pm$ 0.910 & 0.063 & 13.720 & 12.117 $\pm$ 6.622 & 0.153 & 138.341 & 230.394 $\pm$ 79.302 & 0.000 & 8657.975 & 0.694 $\pm$ 0.526 & 0.000 & 1996.395 & 0.792 $\pm$ 0.220 & 0.001 & 2912.730 \\
Arp 70 & 1.289 $\pm$ 1.232 & 0.016 & 319.669 & 19.671 $\pm$ 2.587 & 0.045 & 212.174 & 271.969 $\pm$ 37.856 & 0.000 & 8652.778 & 1.181 $\pm$ 0.290 & 0.001 & 1842.282 & 0.468 $\pm$ 0.162 & 0.001 & 47999.308 \\
Arp 218 & 4.121 $\pm$ 1.406 & 0.149 & 43.927 & 6.002 $\pm$ 13.070 & 0.187 & 328.818 & 285.828 $\pm$ 163.671 & 0.000 & 8657.975 & 0.457 $\pm$ 0.333 & 0.001 & 1183.856 & 2.704 $\pm$ 6.180 & 0.001 & 1910.213 \\
Violin Clef & 0.643 $\pm$ 0.154 & 0.128 & 1.807 & 25.472 $\pm$ 6.627 & 0.262 & 148.960 & 159.370 $\pm$ 148.045 & 0.000 & 8652.778 & 2.523 $\pm$ 1.038 & 0.005 & 1369.829 & 0.240 $\pm$ 0.587 & 0.001 & 1213.433 \\
Arp 148* & 0.388 $\pm$ 4.235 & 0.274 & 19.651 & 8.497 $\pm$ 5.781 & 0.123 & 146.496 & 38.110 $\pm$ 11.462 & 0.000 & 8495.140 & 12.908 $\pm$ 2.523 & 0.001 & 597.817 & 0.239 $\pm$ 0.311 & 0.001 & 3202.354 \\
CGCG 436-030 & 3.722 $\pm$ 1.704 & 0.323 & 11.068 & 19.466 $\pm$ 7.886 & 0.134 & 95.978 & 263.308 $\pm$ 67.920 & 0.000 & 8647.581 & 0.977 $\pm$ 0.703 & 0.011 & 835.493 & 0.426 $\pm$ 0.930 & 0.002 & 2848.546 \\
Arp 272* & 1.838 $\pm$ 1.976 & 0.140 & 18.966 & 8.654 $\pm$ 9.887 & 0.180 & 151.840 & 363.780 $\pm$ 113.956 & 0.000 & 8543.644 & 0.810 $\pm$ 1.474 & 0.003 & 913.581 & 1.002 $\pm$ 0.301 & 0.001 & 1273.594 \\
ESO 77-14 & 1.311 $\pm$ 0.729 & 0.112 & 21.132 & 3.018 $\pm$ 0.957 & 0.026 & 149.208 & 93.544 $\pm$ 74.895 & 0.000 & 8656.243 & 0.782 $\pm$ 0.135 & 0.002 & 80.297 & 15.930 $\pm$ 12.601 & 0.002 & 51774.985 \\
NGC 5331* & 1.608 $\pm$ 0.237 & 0.259 & 7.622 & 7.681 $\pm$ 23.754 & 0.220 & 173.170 & 119.528 $\pm$ 96.979 & 0.000 & 8590.416 & 1.998 $\pm$ 6.565 & 0.010 & 1432.837 & 1.202 $\pm$ 0.434 & 0.001 & 971.091 \\
NGC 6786 & 2.438 $\pm$ 0.452 & 0.039 & 25.521 & 13.392 $\pm$ 2.421 & 0.064 & 346.368 & 511.025 $\pm$ 106.321 & 0.000 & 8657.975 & 1.108 $\pm$ 0.231 & 0.001 & 980.623 & 0.763 $\pm$ 0.229 & 0.003 & 18724.424 \\
Arp 273 & 2.004 $\pm$ 2.250 & 0.040 & 23.218 & 37.833 $\pm$ 12.978 & 0.221 & 408.622 & 284.095 $\pm$ 214.862 & 0.000 & 8652.778 & 4.036 $\pm$ 1.309 & 0.007 & 1215.151 & 0.114 $\pm$ 0.246 & 0.001 & 1590.700 \\
Arp 244 & 0.758 $\pm$ 0.397 & 0.076 & 16.571 & 1.572 $\pm$ 0.692 & 0.063 & 148.908 & 55.433 $\pm$ 14.895 & 0.000 & 2929.299 & 0.493 $\pm$ 0.139 & 0.001 & 464.457 & 36.021 $\pm$ 33.981 & 0.002 & 18465.811 \\

\hline
\end{tabular}
}
 \label{restbl2}
\end{center}
\end{table}
\end{landscape}

\begin{landscape}
\begin{table}
\caption[Simulation Parameters and Merger Wars results]{The input parameters for each simulation were stored as a comma-separated list.  We combined a unique identifier and {\it Merger Wars} information into a single string.  All simulations for a given target are combined into a single file.  The simulation parameters at the top of the file are sorted in decreasing order of {\it Merger Wars} fitness.  The simulations parameters for simulations that were rejected by the volunteers are at the end of the file.  The 14 simulation parameters include position and velocity vectors, masses, disc sizes, and disc orientations.  The {\it Merger Wars} information includes fitness, number of competitions won, and number of competitions scored. Additional values, not shown here, are used for less important simulation parameters, such as scaling lengths, that were constant across all simulations, and for storing additional metadata about the strength of the interaction.}
\begin{center}
\scalebox{0.7}{
\begin{tabular}{ | c | c | c | c | c | c | c | c | c | c | c | c | c | c | c | c | c | c |}

\hline
UID & MW Fitness & MW Wins & MW Scored & $r_x$ & $r_y$ & $r_z$ & $v_x$ & $v_y$ & $v_z$ & $m_1$ & $m_2$ & $r_1$ & $r_2$ & $\phi_1$ & $\phi_2$ & $\theta_1$ & $\theta_2$ \\
\hline
4e9u3835u2gq3vrvcra{\_}0009 & 0.9285714285714286 & 26 & 28 & -0.07728 & -0.80372 & -0.13938 & -0.31746 & -0.75224 & -0.08412 & 0.33607 & 0.44353 & 0.91692 & 0.89357 & 79.55898 & 35.13127 & 141.02645 & 306.89238 \\
1zxs3c6z3jfsenmv4e9h{\_}0009 & 0.8888888888888888 & 24 & 27 & -0.07728 & -0.80372 & 0.12298 & -0.2236 & -0.67937 & 0.0343 & 0.29311 & 0.35925 & 0.70052 & 0.99425 & 80.2 & 38.42956 & 230.4 & 306.34797 \\
5lr32ofly6gf1cp6hfs{\_}0023 & 0.8620689655172413 & 25 & 29 & -0.07728 & -0.80372 & 0.01638 & -0.048 & -0.79299 & -0.12945 & 0.36928 & 0.38522 & 0.68775 & 0.78284 & 94.53363 & 50.72619 & 224.21979 & 58.01574 \\
7ap8fo1f9w5552y6r4v{\_}0001 & 0.8571428571428571 & 24 & 28 & -0.07728 & -0.80372 & -0.04396 & -0.48198 & -0.62875 & -0.0582 & 0.38897 & 0.3297 & 0.88374 & 0.81503 & 93.61012 & 39.00614 & 231.51898 & 54.58154 \\
2ezye5ilu97slwjfxtla{\_}r0096 & & & & -0.07728 & -0.80372 & -0.00173 & 0.35375 & -0.18376 & 0.14257 & 0.15478 & 0.31143 & 0.68311 & 0.78965 & 93.1902 & 46.449 & 131.85486 & 228.78565 \\
8c3zp78jhbuogiq8ynf{\_}r0022 & & & & -0.07728 & -0.80372 & 0.10037 & 0.74383 & -0.47462 & 0.18233 & 0.08506 & 0.24445 & 0.79865 & 0.96232 & 79.76905 & 47.21791 & 309.77606 & 48.38752 \\
\hline
\end{tabular}
}
 \label{tblsimparam}
\end{center}
\end{table}

\end{landscape}

\section{Analysis}
\label{analysis}

\subsection{Convergence of Model Parameters}

\begin{figure*}
\begin{center}
\includegraphics[scale=0.6]{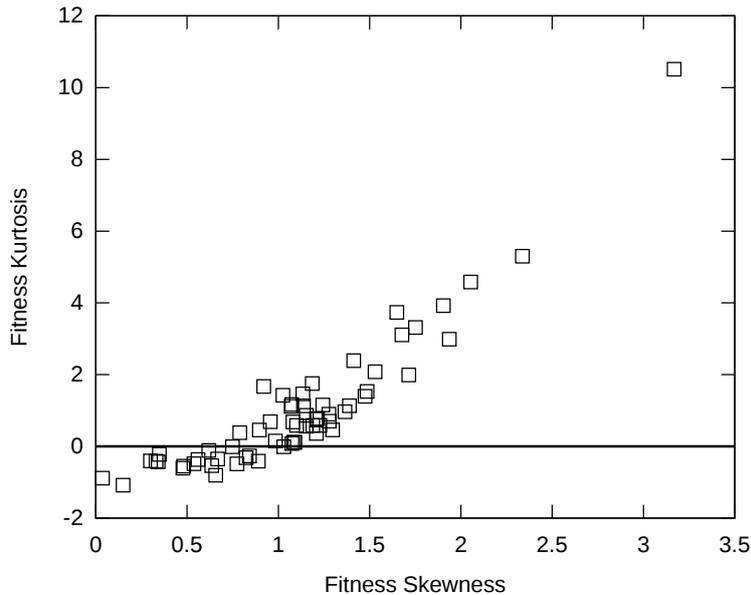}
\caption{Kurtosis vs. Skewness of the fitness distribution.}
\label{ch5skkr}
\end{center}
\end{figure*}

For estimating the convergence of model parameters we suggest several different factors be evaluated. The first is a visual inspection of the simulation model to characterize how well it matches the target image. We believe we have found a plausible visible match for all of our targets. The next convergence factor to consider is how much parameter space has been eliminated from consideration. The remaining fraction values we presented in Table \ref{rfdist} illustrates that some parameters have converged where others have not. By eliminating large portions of the full range of each parameter, we have constrained the likely values\footnote{In addition to considering reduction in the ranges and variances of each parameter individually, it may be useful to consider covariance information between the parameters. It is possible that for some parameters that individually have large remaining fractions the covariance with another parameter could indicate a small area of the two-dimensional parameter space. We plan to explore covariance behaviour in future work analysing the properties of our proposed fitness function.}. Finally, we present a possible relationship between model convergence and the distribution of {\it Merger Wars} fitness values for simulations of each target. The pairs where the fitness distributions were L-shaped (strong peak at low fitness with long, thin tail extending to high fitness) tended to have a good simulation match to the target image. They also tended to have a high level of convergence. This L-shape is characteristic of power law distributions with long tails. In order to quantify the shape of the fitness distributions we computed the first four statistical moments for the fitness population of each galaxy pair. From those moments we calculated the skewness and kurtosis. Figure \ref{ch5skkr}  shows the kurtosis versus skewness for each fitness distribution.

\begin{figure*}
\begin{center}
\setlength{\tabcolsep}{0mm}
\subfloat{%
\begin{tabular}{c}
\includegraphics*[width=0.13\textwidth]{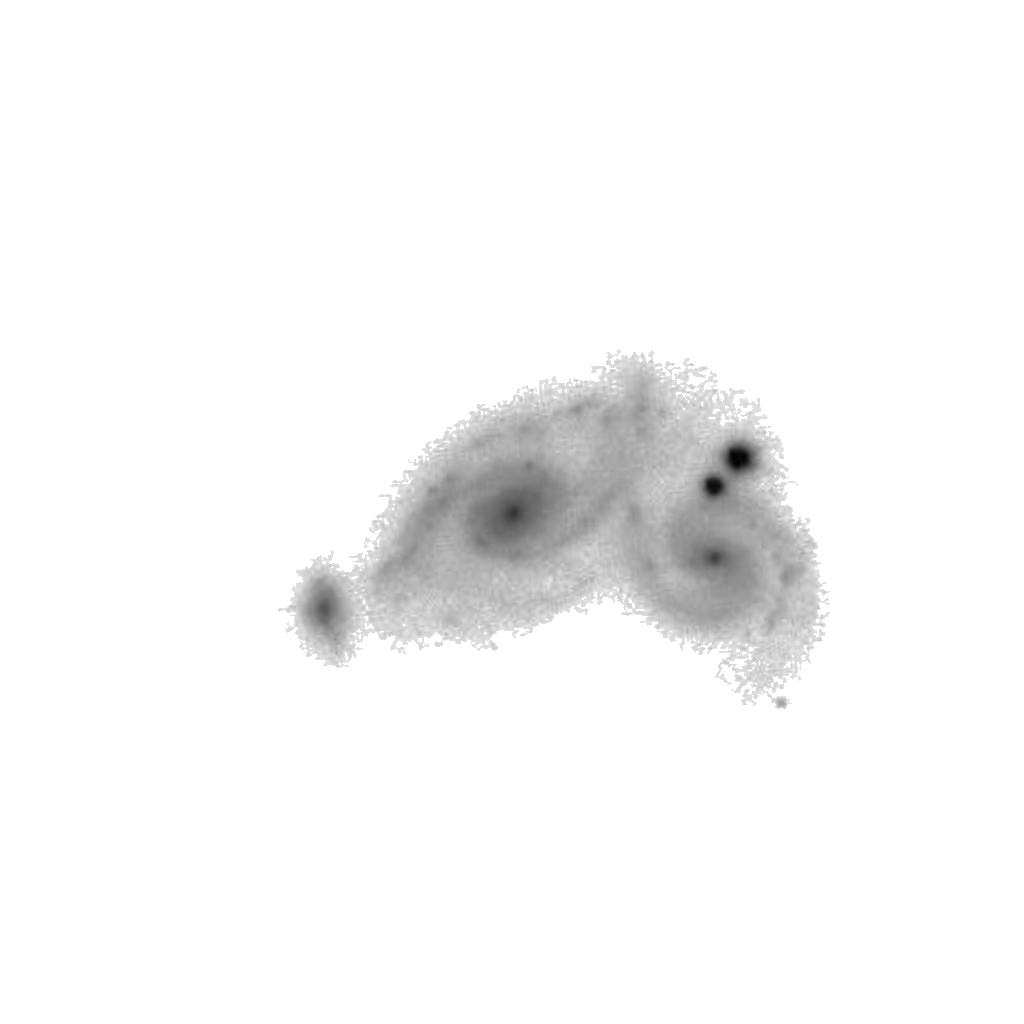}\\
\includegraphics*[width=0.13\textwidth]{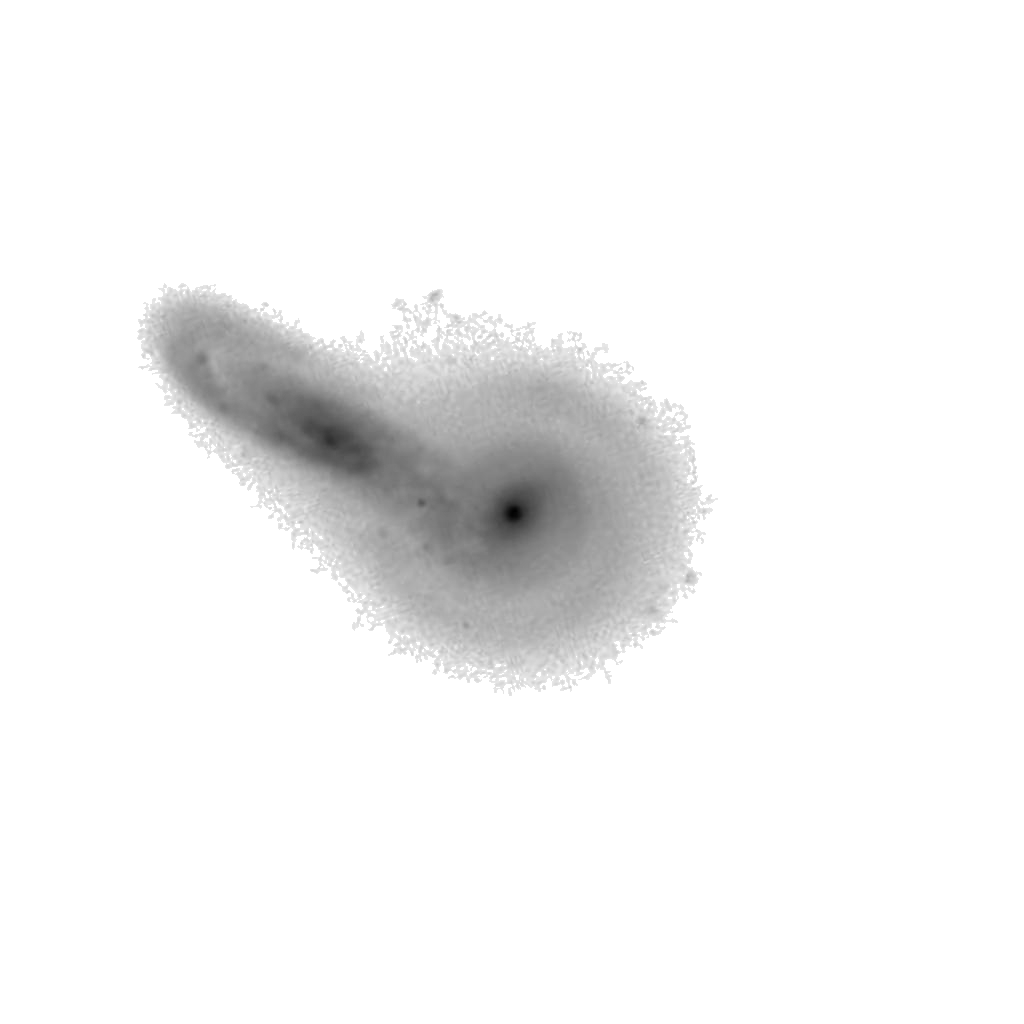}\\
\includegraphics*[width=0.13\textwidth]{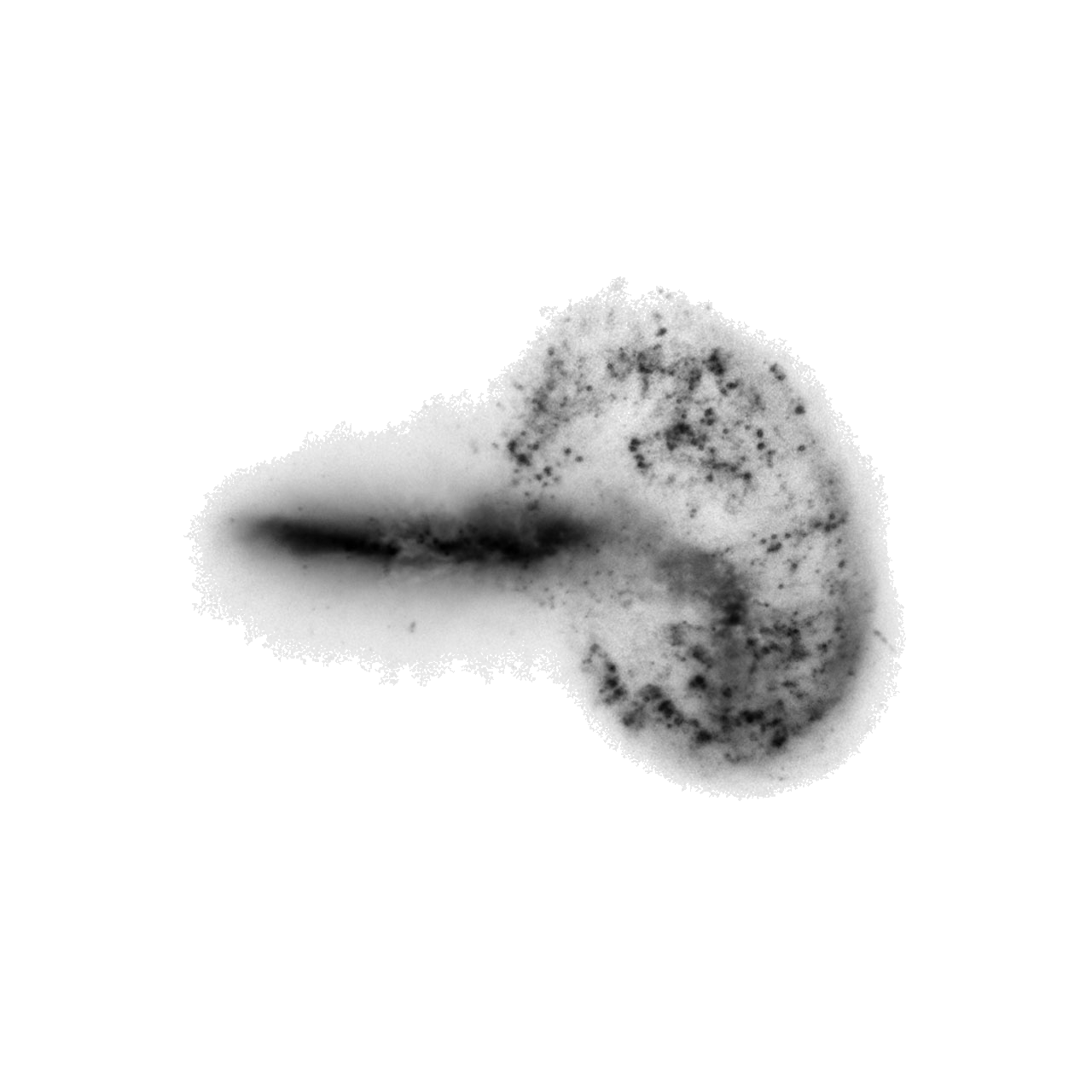}
\end{tabular}
}%
\subfloat{%
\begin{tabular}{c}
\includegraphics*[width=0.13\textwidth]{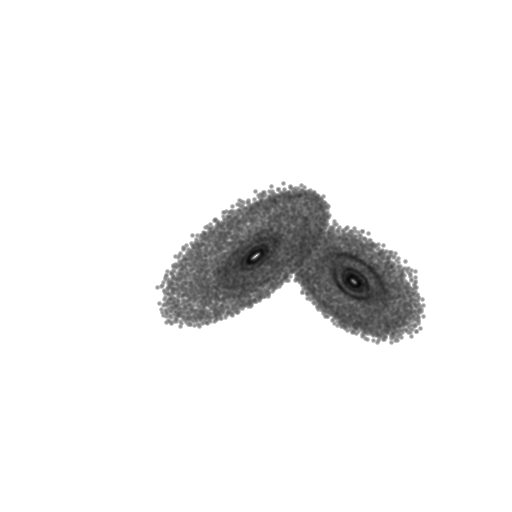}\\
\includegraphics*[width=0.13\textwidth]{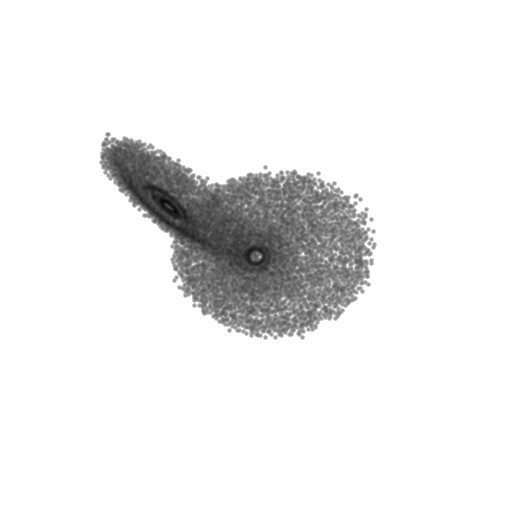}\\
\includegraphics*[width=0.13\textwidth]{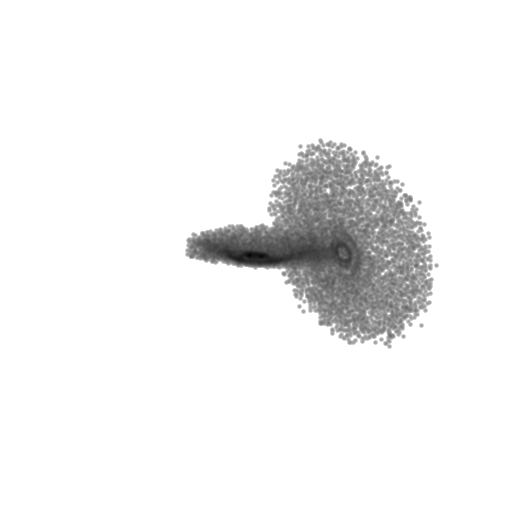}
\end{tabular}
}%
\subfloat{%
\begin{tabular}{c}
\includegraphics*[width=0.13\textwidth]{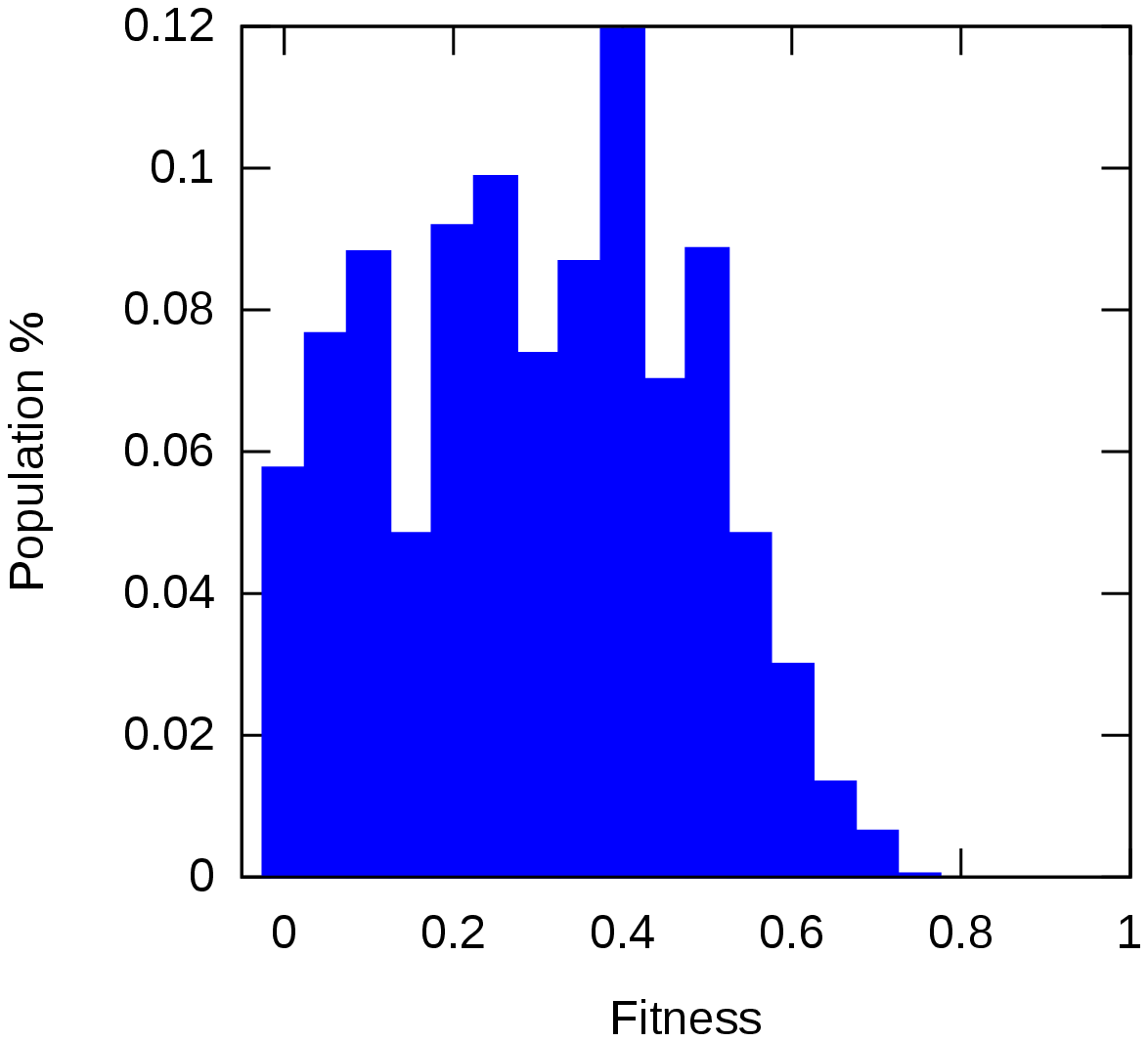}\\
\includegraphics*[width=0.13\textwidth]{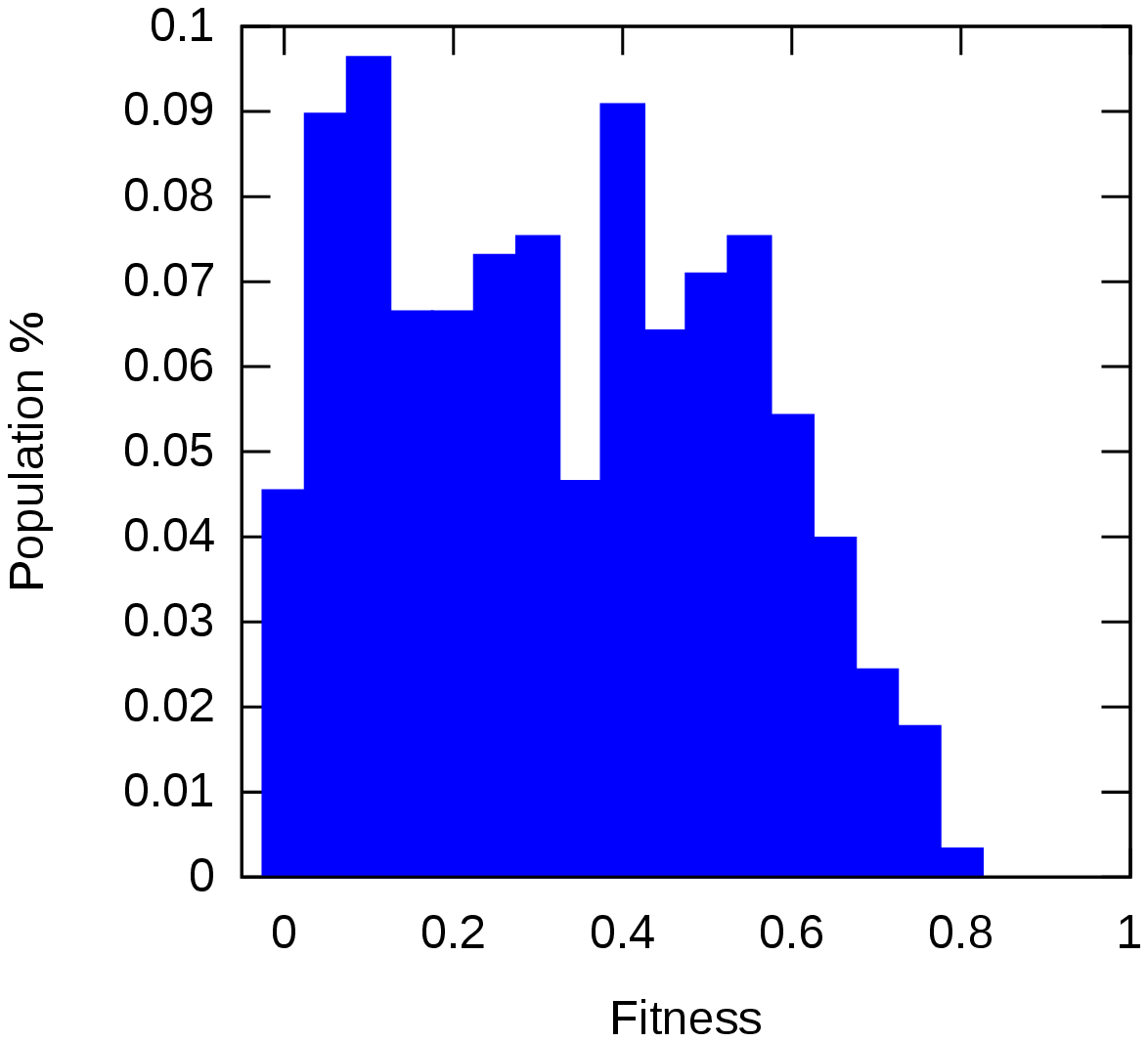}\\
\includegraphics*[width=0.13\textwidth]{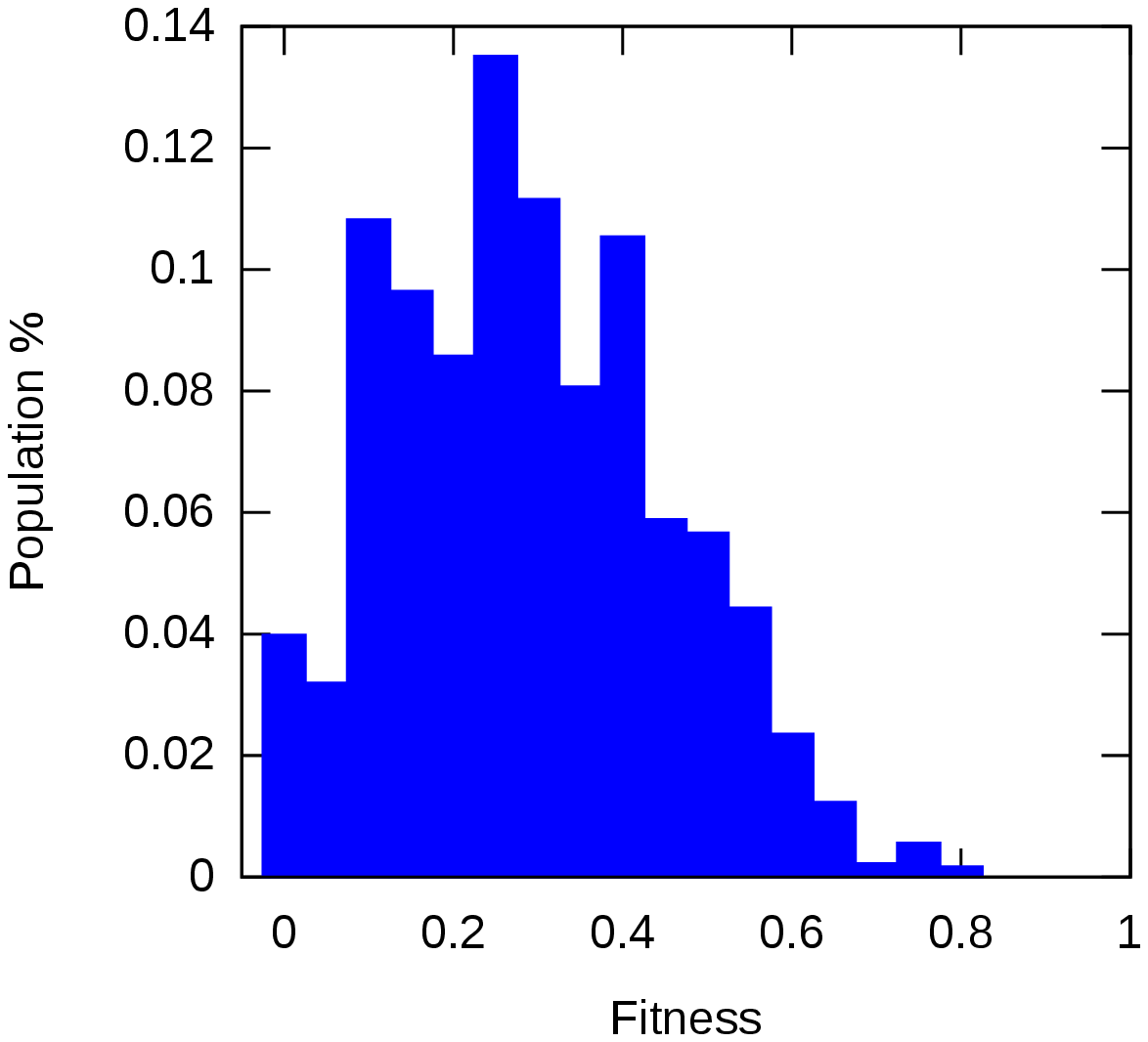}
\end{tabular}
}%
\subfloat{%
\begin{tabular}{c}
\includegraphics*[width=0.13\textwidth]{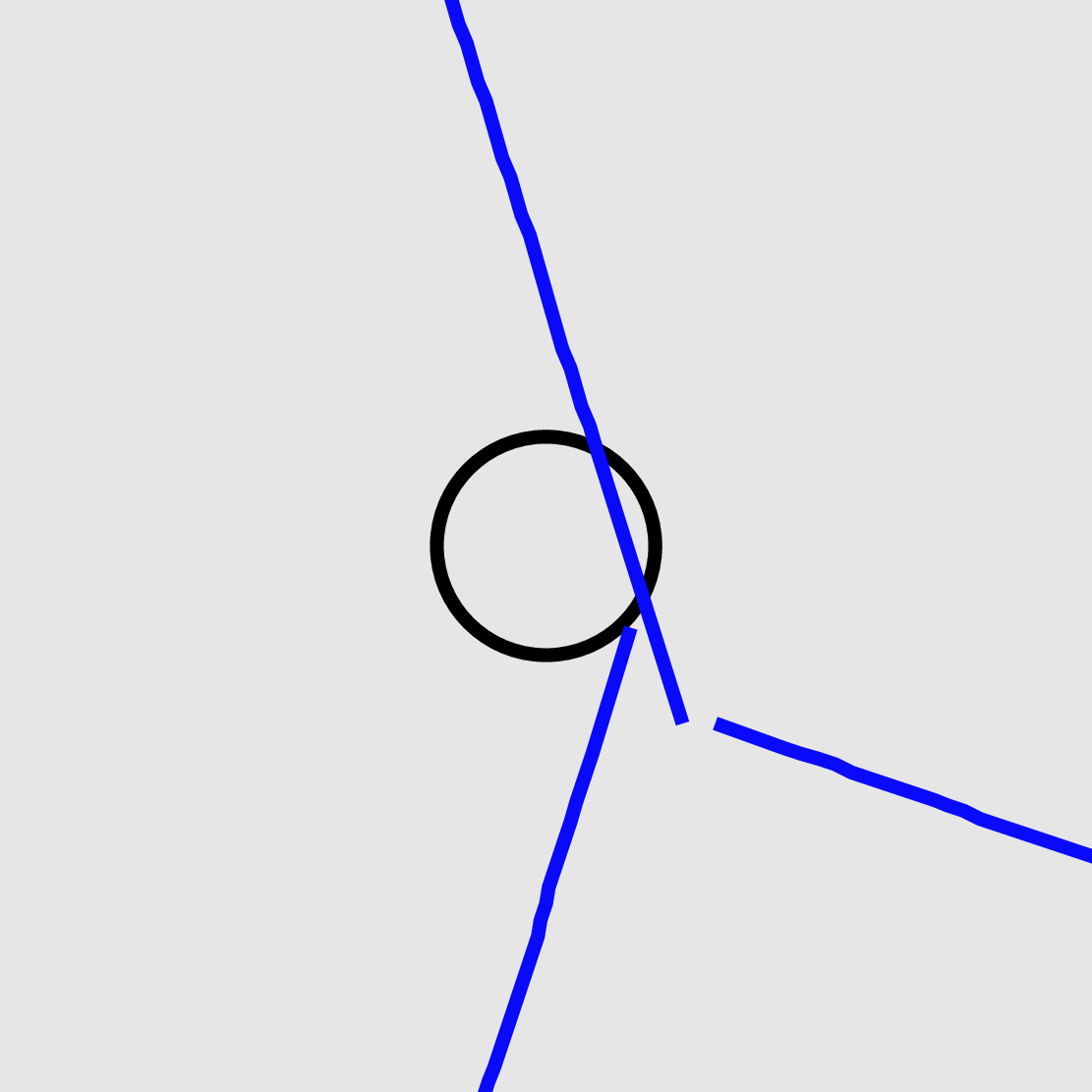}\\
\includegraphics*[width=0.13\textwidth]{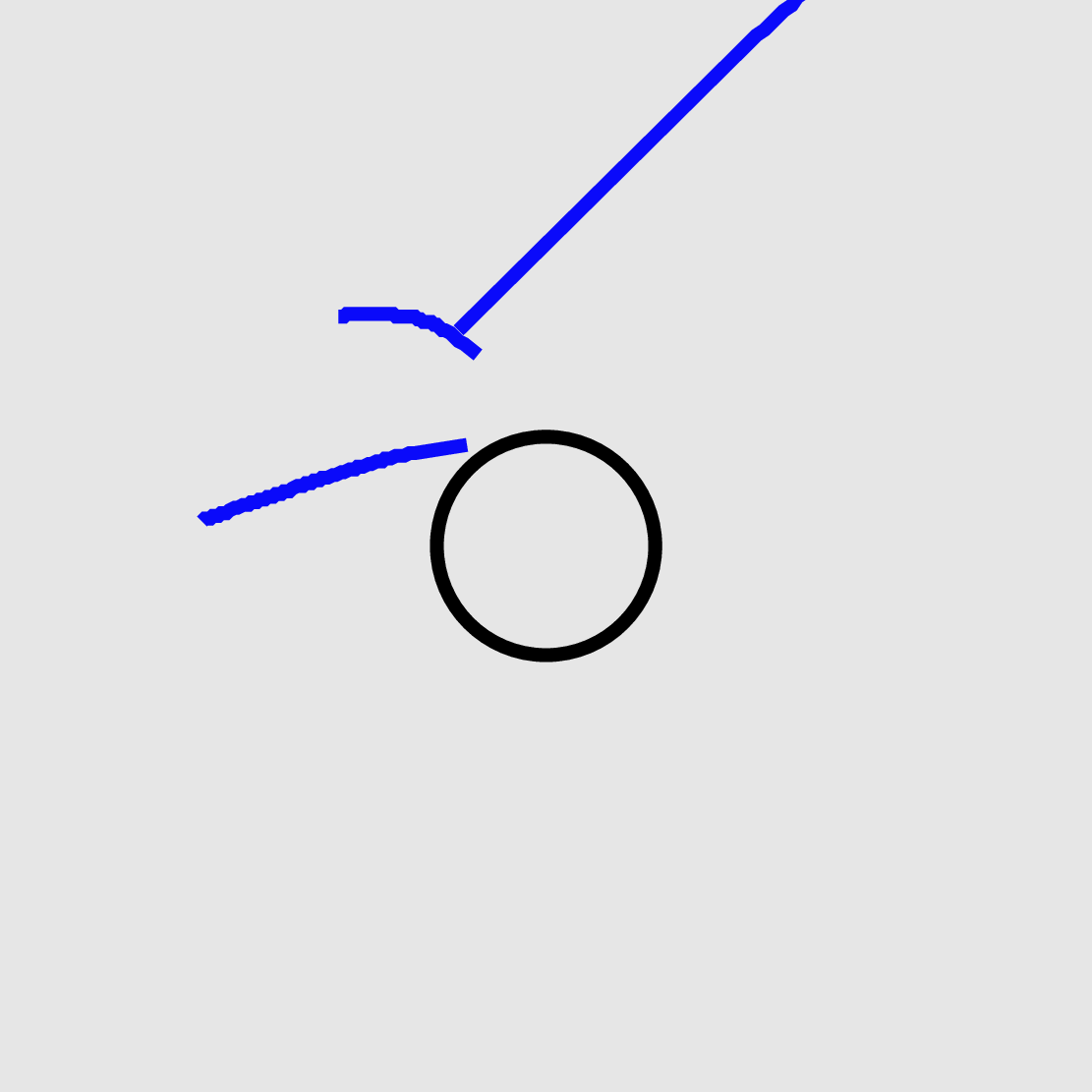}\\
\includegraphics*[width=0.13\textwidth]{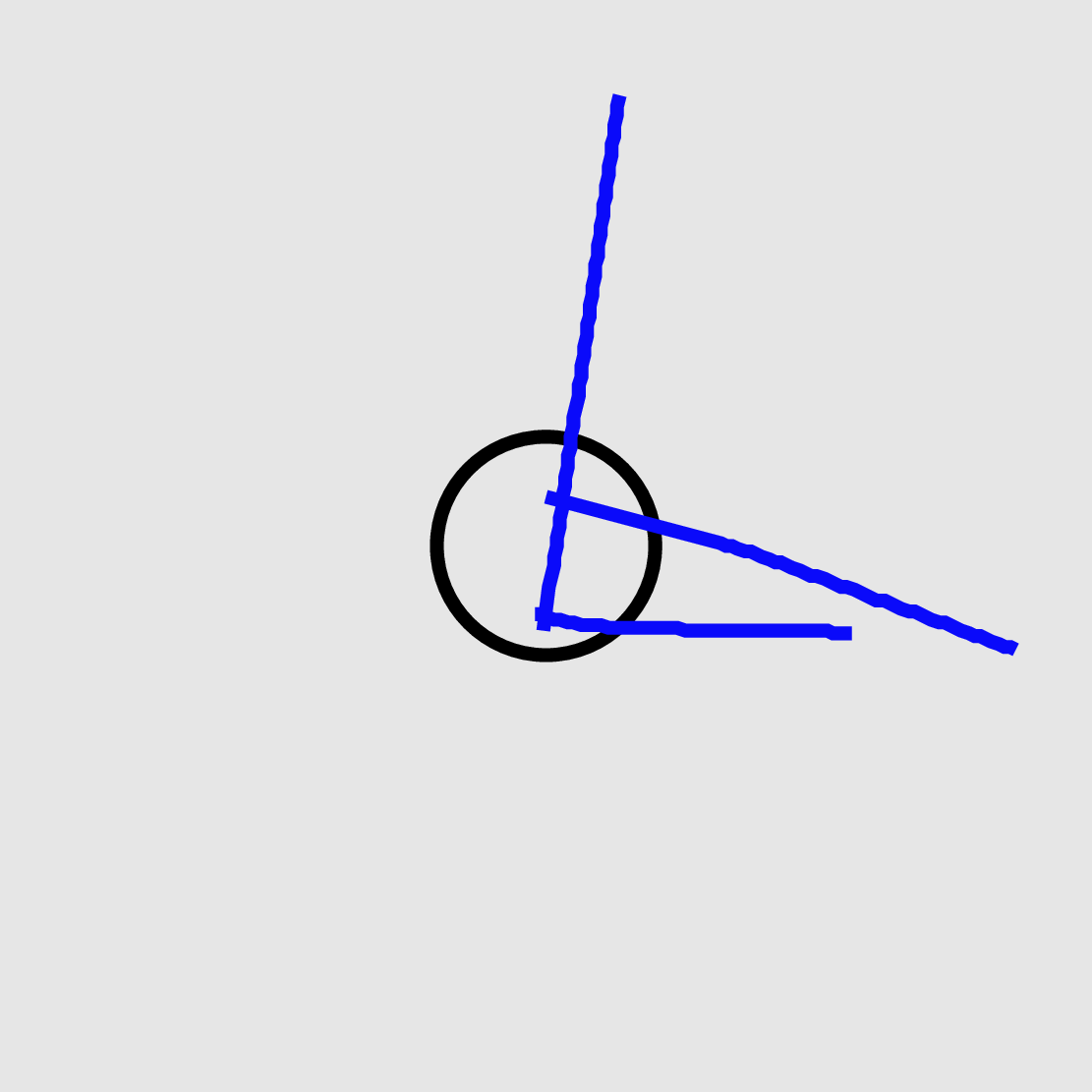}
\end{tabular}
}%
\subfloat{%
\begin{tabular}{c}
\includegraphics*[width=0.13\textwidth]{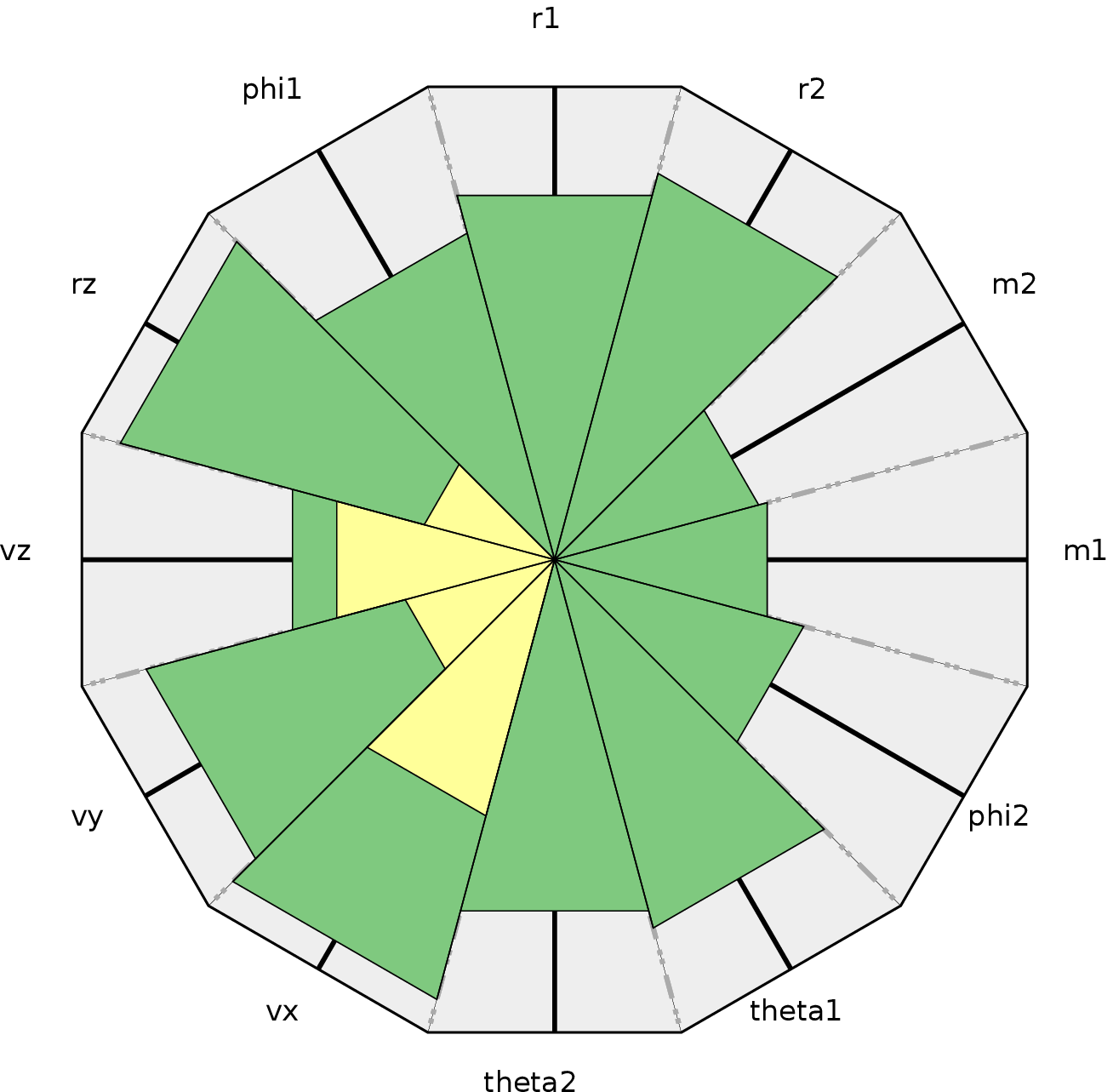}\\
\includegraphics*[width=0.13\textwidth]{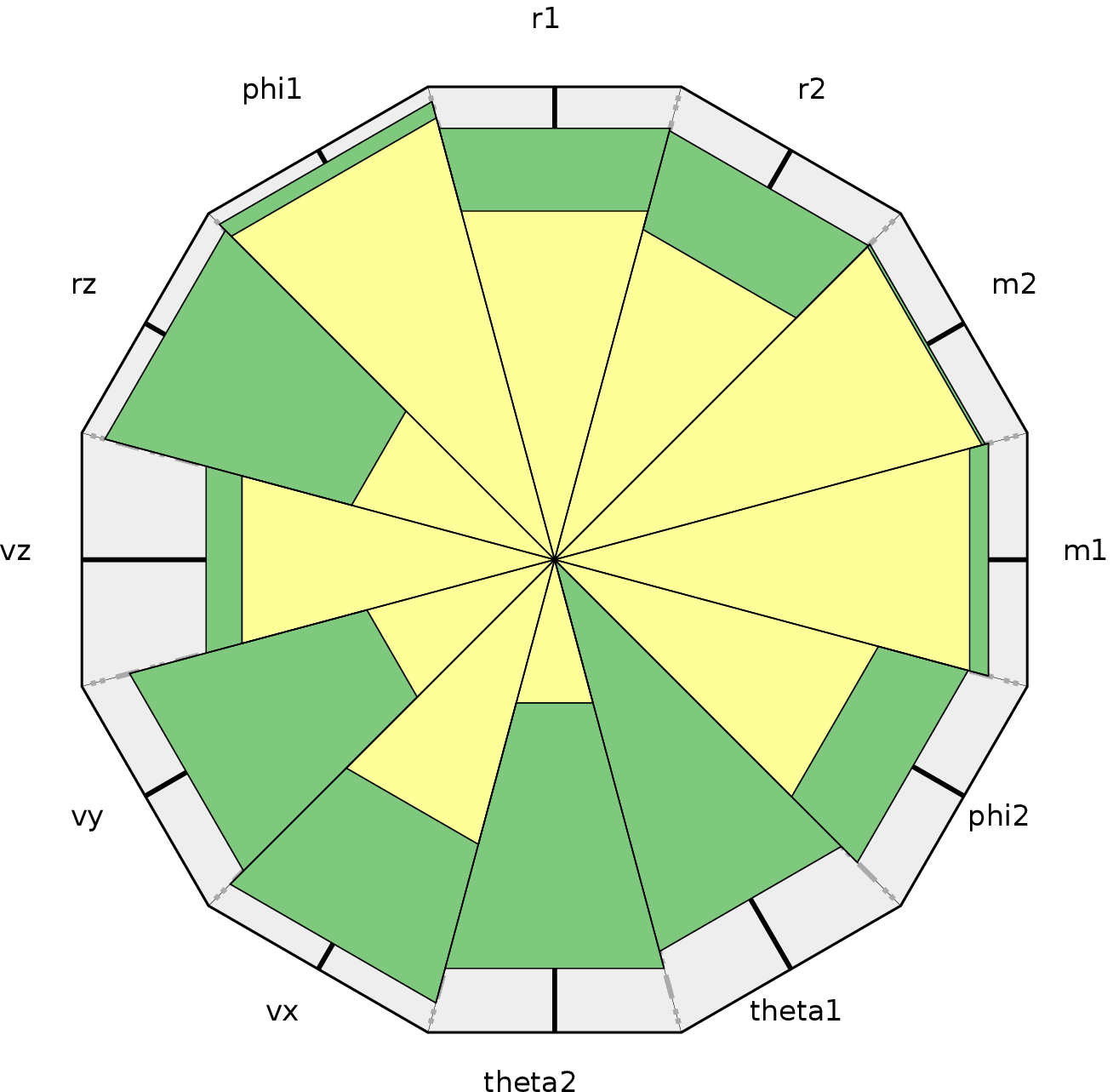}\\
\includegraphics*[width=0.13\textwidth]{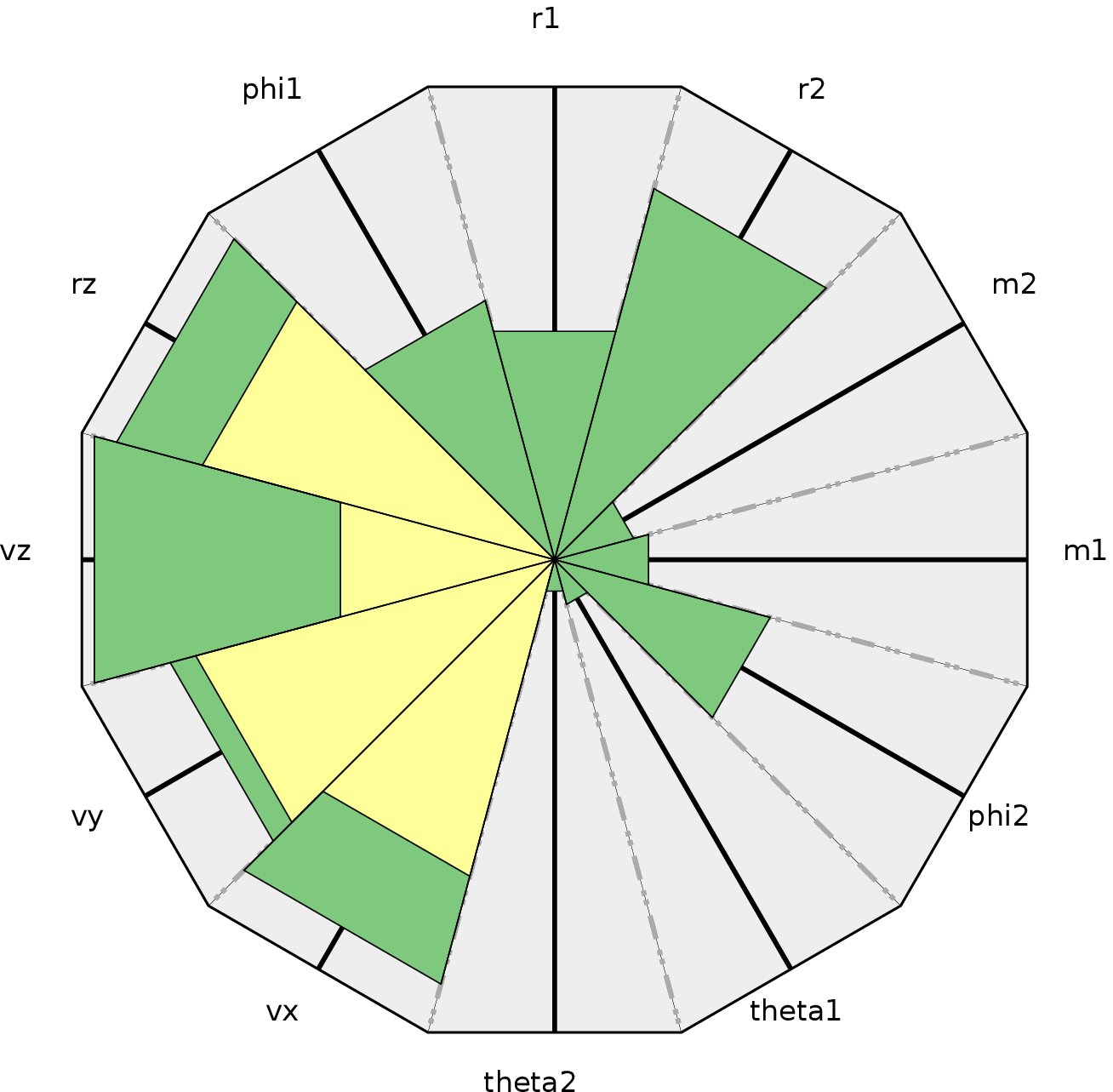}
\end{tabular}
}%
\subfloat{%
\begin{tabular}{c}
\includegraphics*[width=0.13\textwidth]{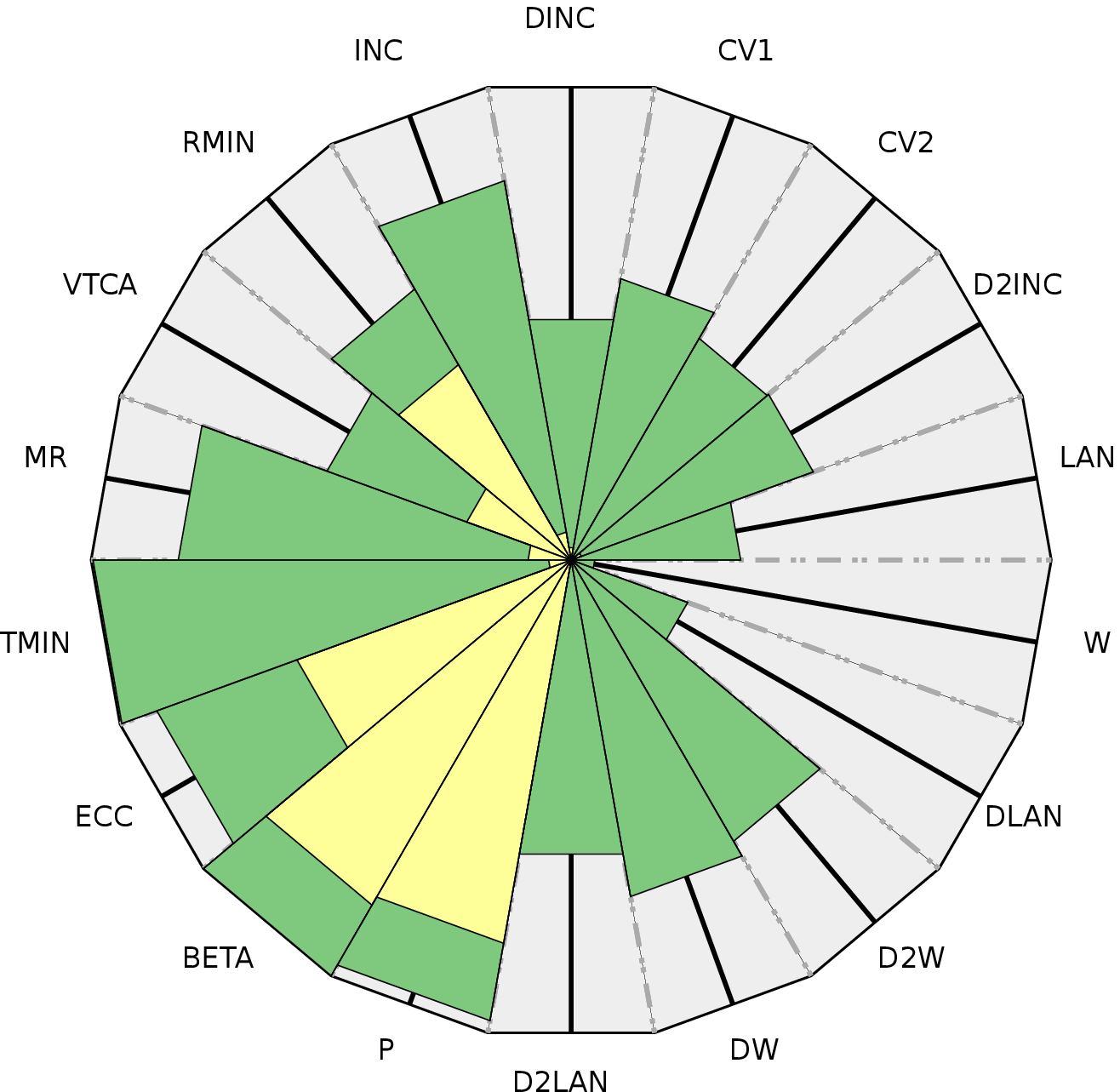}\\
\includegraphics*[width=0.13\textwidth]{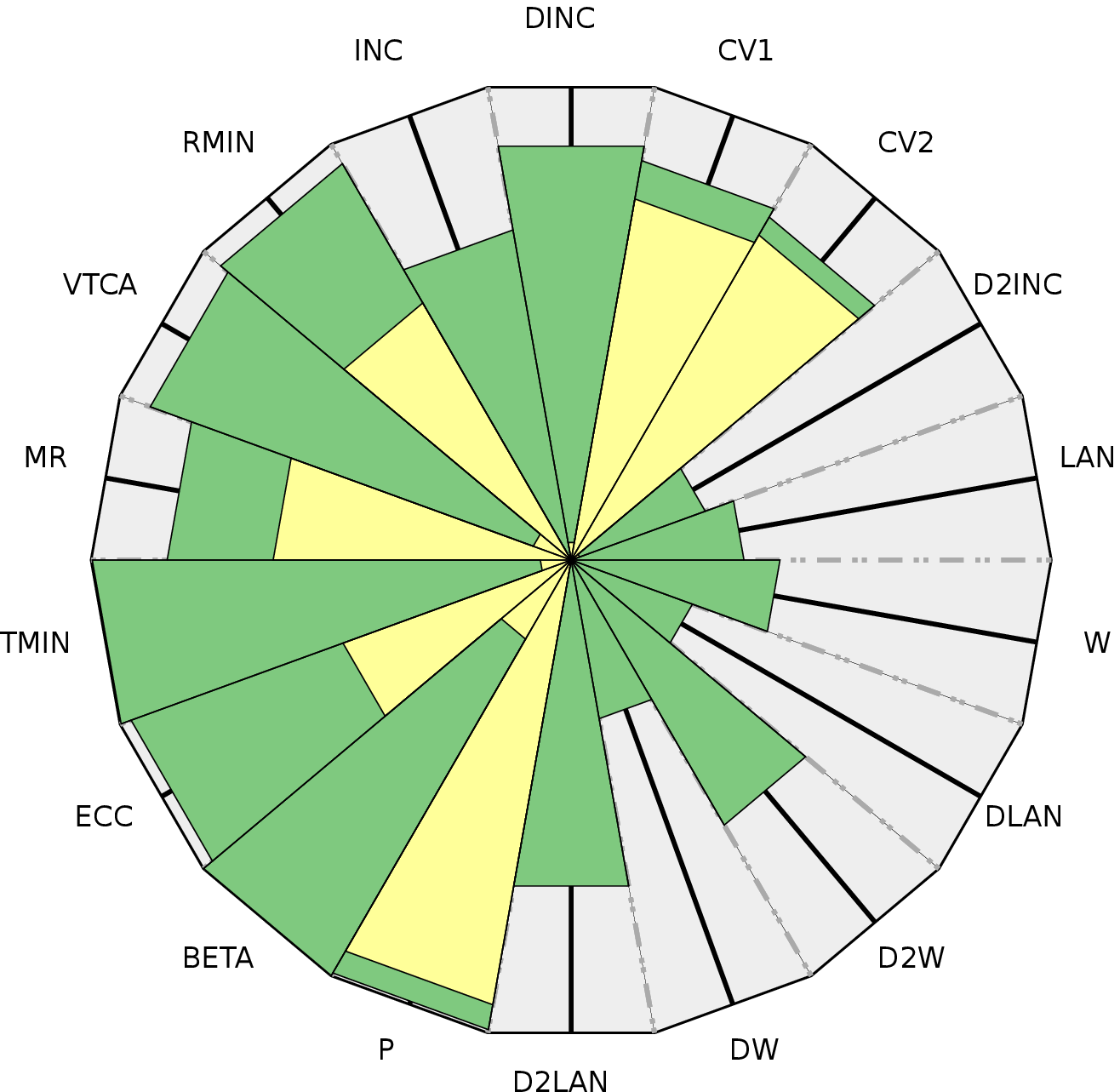}\\
\includegraphics*[width=0.13\textwidth]{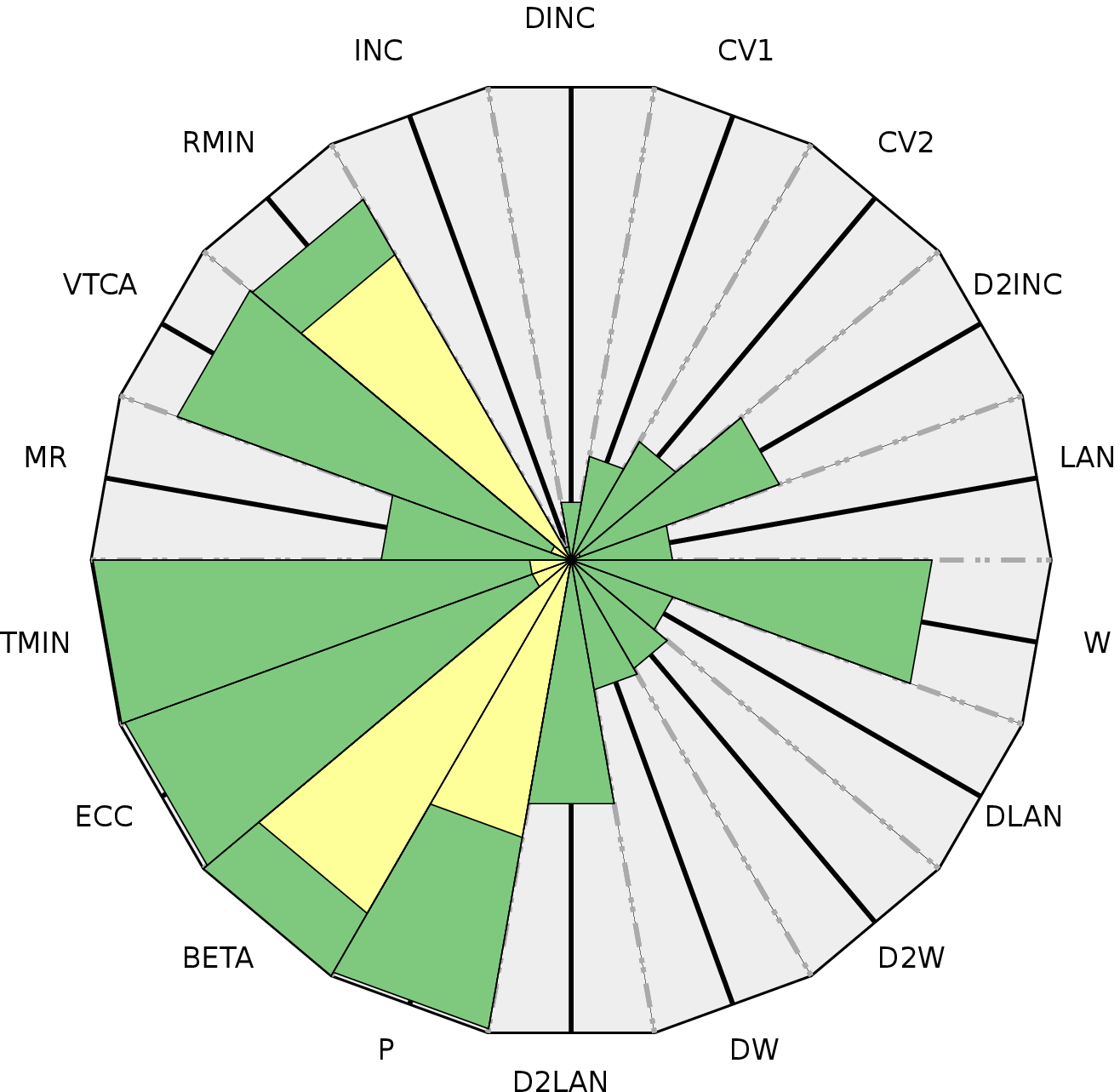}
\end{tabular}
}%
\caption{The galaxy pairs with the three lowest fitness distribution skewnesses: Arp 274, Arp 199, and Arp 148.}
\label{figlowskew}
\end{center}
\end{figure*}

The kurtosis value is negative for the fitness distribution of 18 pairs. These are labeled with an $\ast$ in Table \ref{restbl2}. This indicates that their fitness distributions are flatter than a normal distribution. These distributions also have low skewness resulting in less differentiation between low- and high- fitness simulations. For comparison we show the target image, best fit simulation, fitness distribution, top three trajectories and glyph plots for several galaxy pairs. Figure \ref{figlowskew} shows the three pairs with the lowest skewness. All three have almost no distinguishable tidal features. The fitness distributions are relatively flat. The trajectories are divergent and the glyph plots do not indicate a high level of convergence for simulation and orbit parameters.

\begin{figure*}
\begin{center}
\setlength{\tabcolsep}{0mm}
\subfloat{%
\begin{tabular}{c}
\includegraphics*[width=0.13\textwidth]{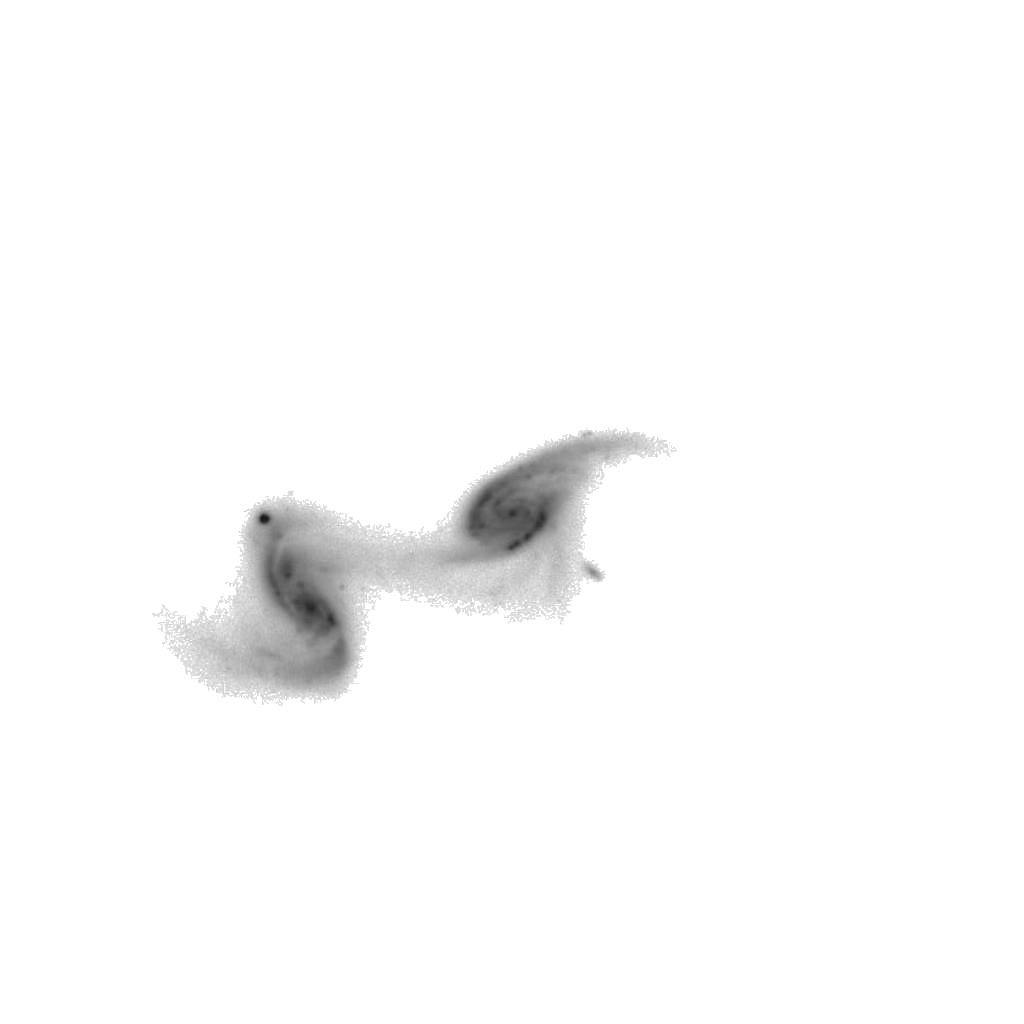}\\
\includegraphics*[width=0.13\textwidth]{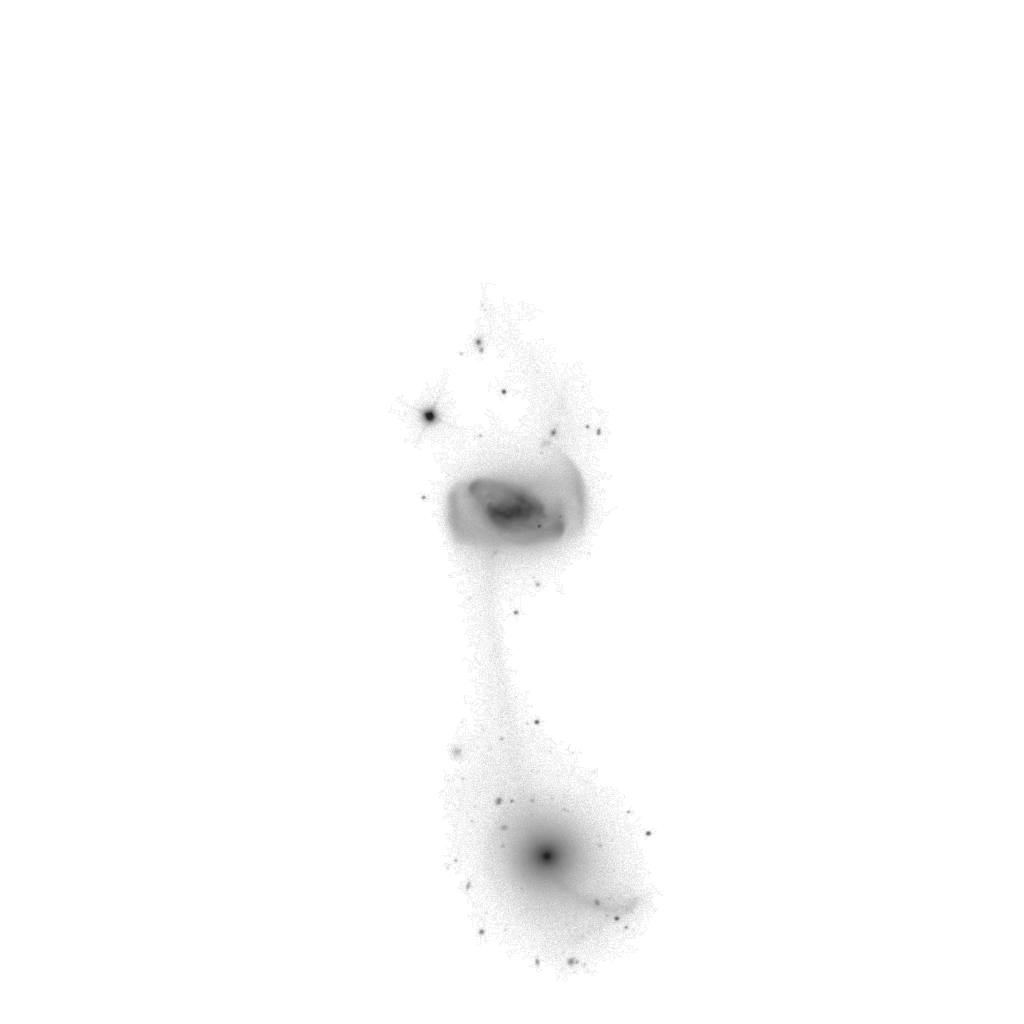}\\
\includegraphics*[width=0.13\textwidth]{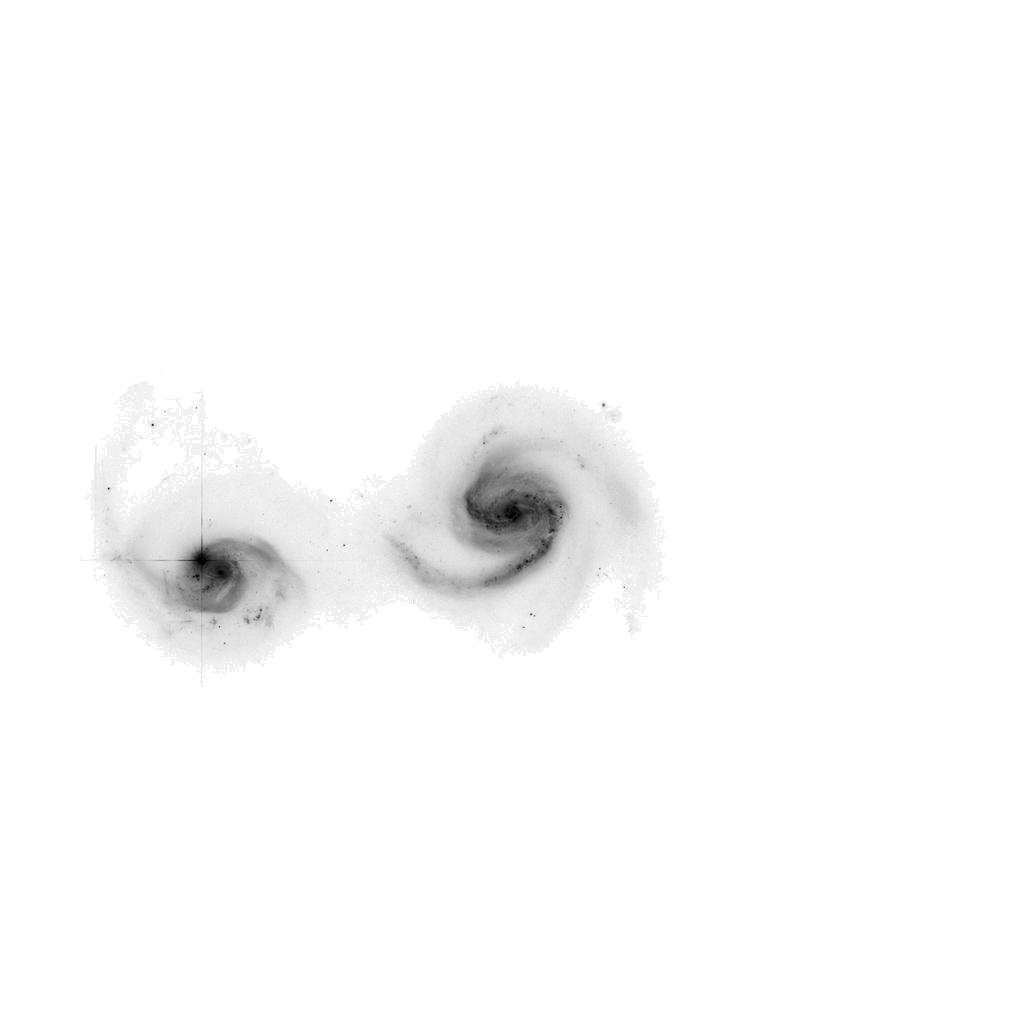}
\end{tabular}
}%
\subfloat{%
\begin{tabular}{c}
\includegraphics*[width=0.13\textwidth]{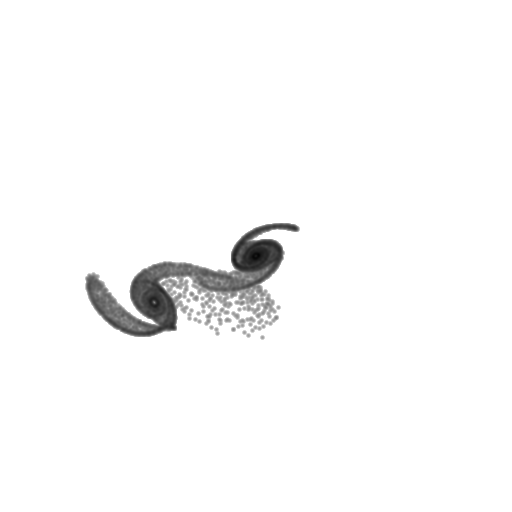}\\
\includegraphics*[width=0.13\textwidth]{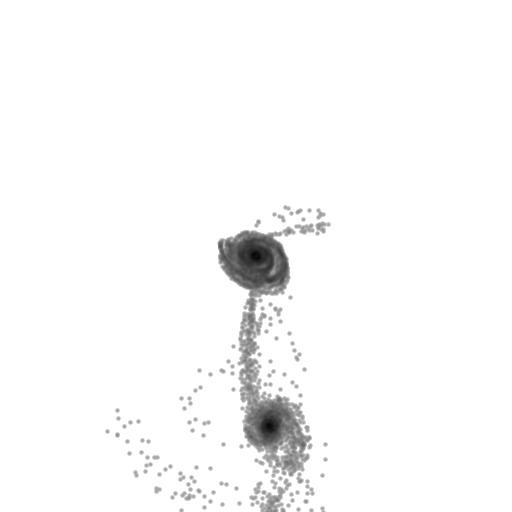}\\
\includegraphics*[width=0.13\textwidth]{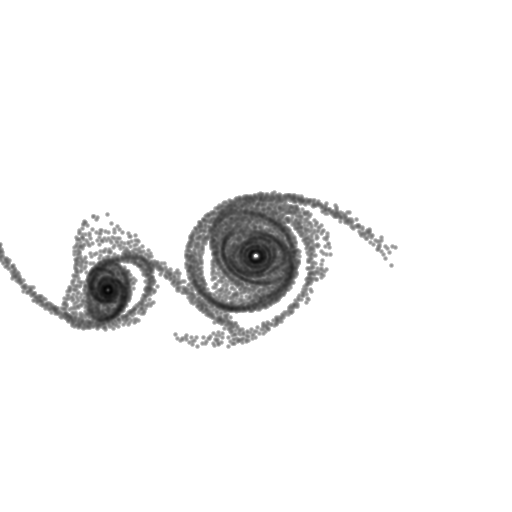}
\end{tabular}
}%
\subfloat{%
\begin{tabular}{c}
\includegraphics*[width=0.13\textwidth]{./figures/results/587722984435351614/587722984435351614_f_hist.eps}\\
\includegraphics*[width=0.13\textwidth]{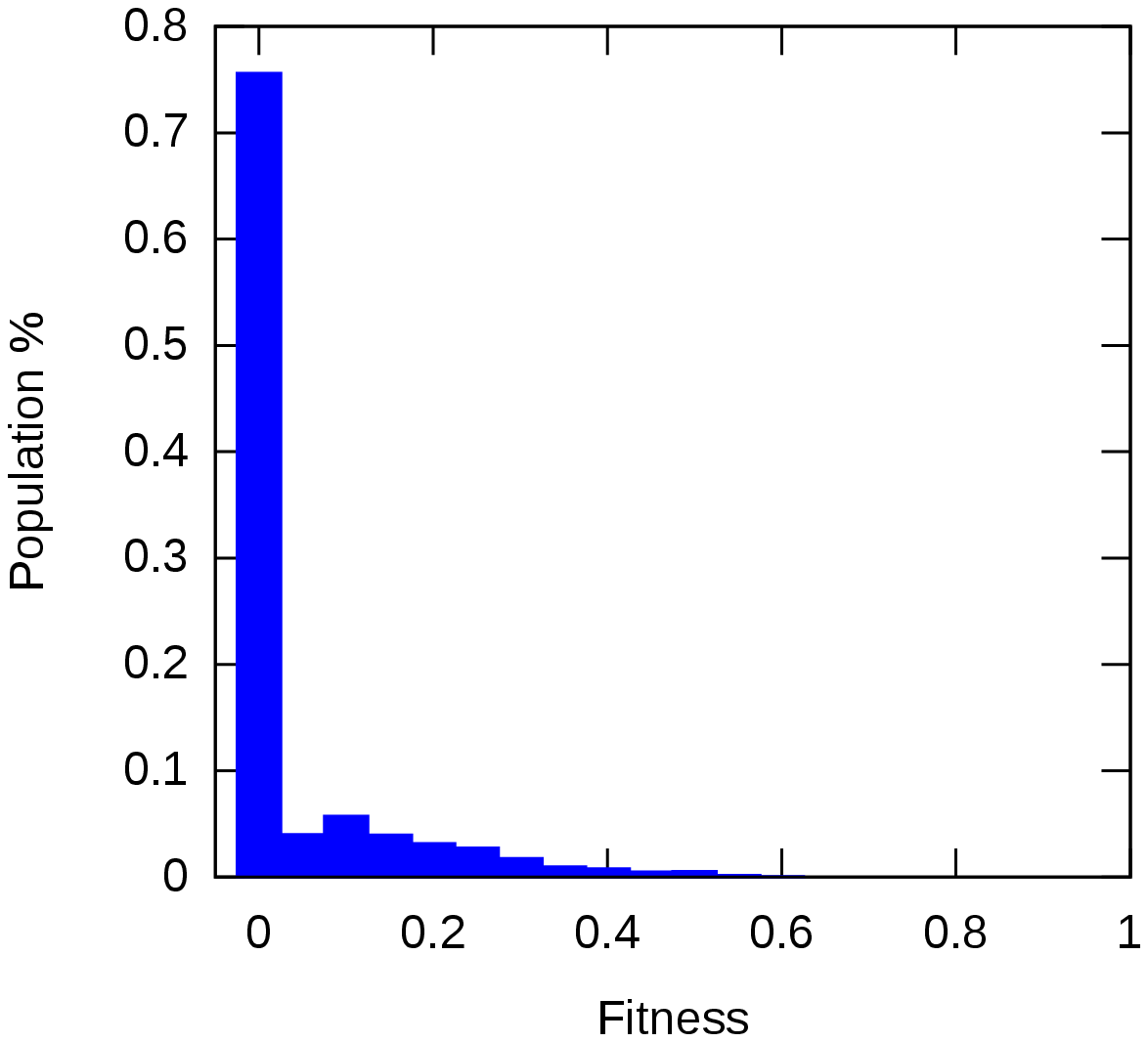}\\
\includegraphics*[width=0.13\textwidth]{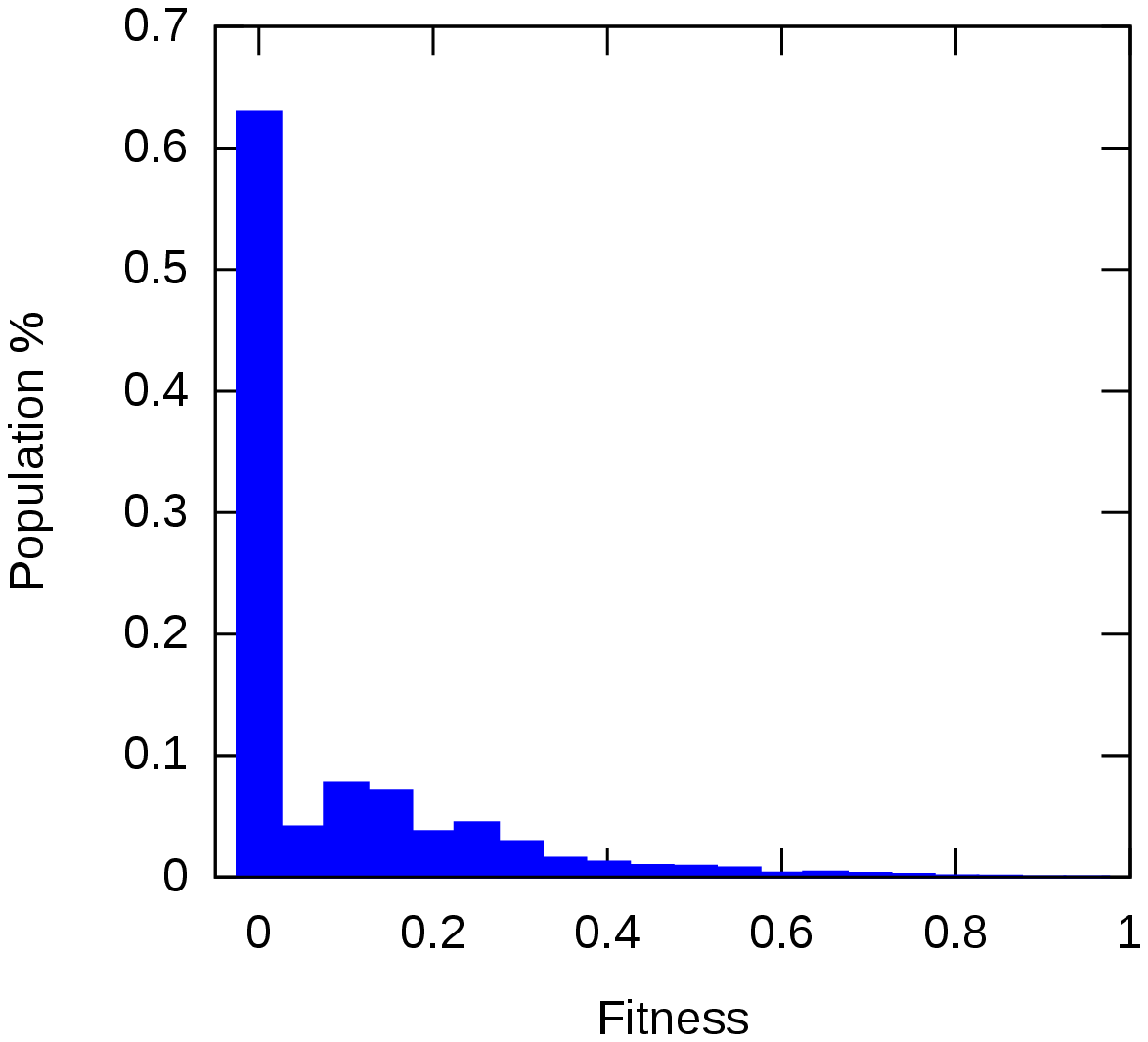}
\end{tabular}
}%
\subfloat{%
\begin{tabular}{c}
\includegraphics*[width=0.13\textwidth]{./figures/results/587722984435351614/587722984435351614_top3_traj.eps}\\
\includegraphics*[width=0.13\textwidth]{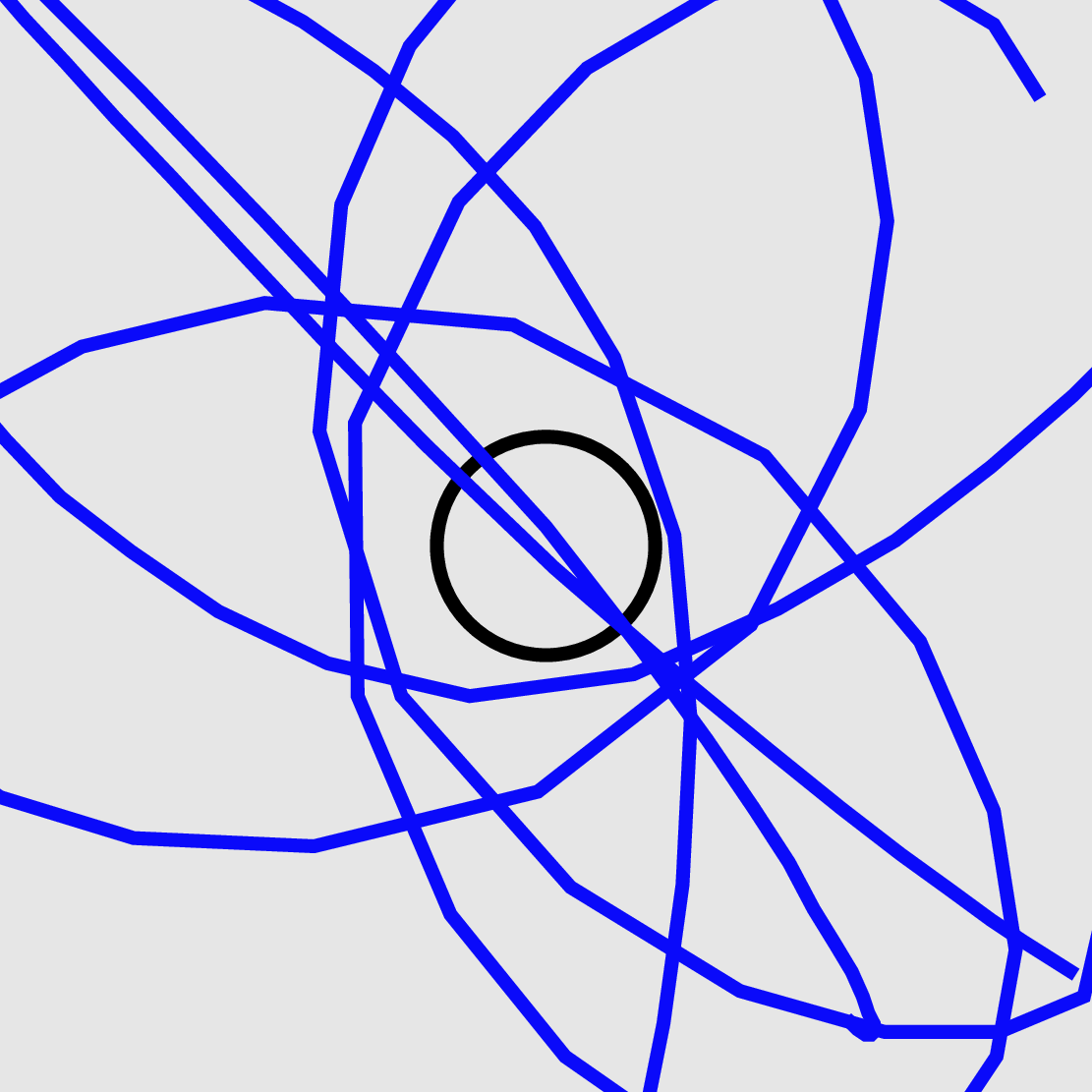}\\
\includegraphics*[width=0.13\textwidth]{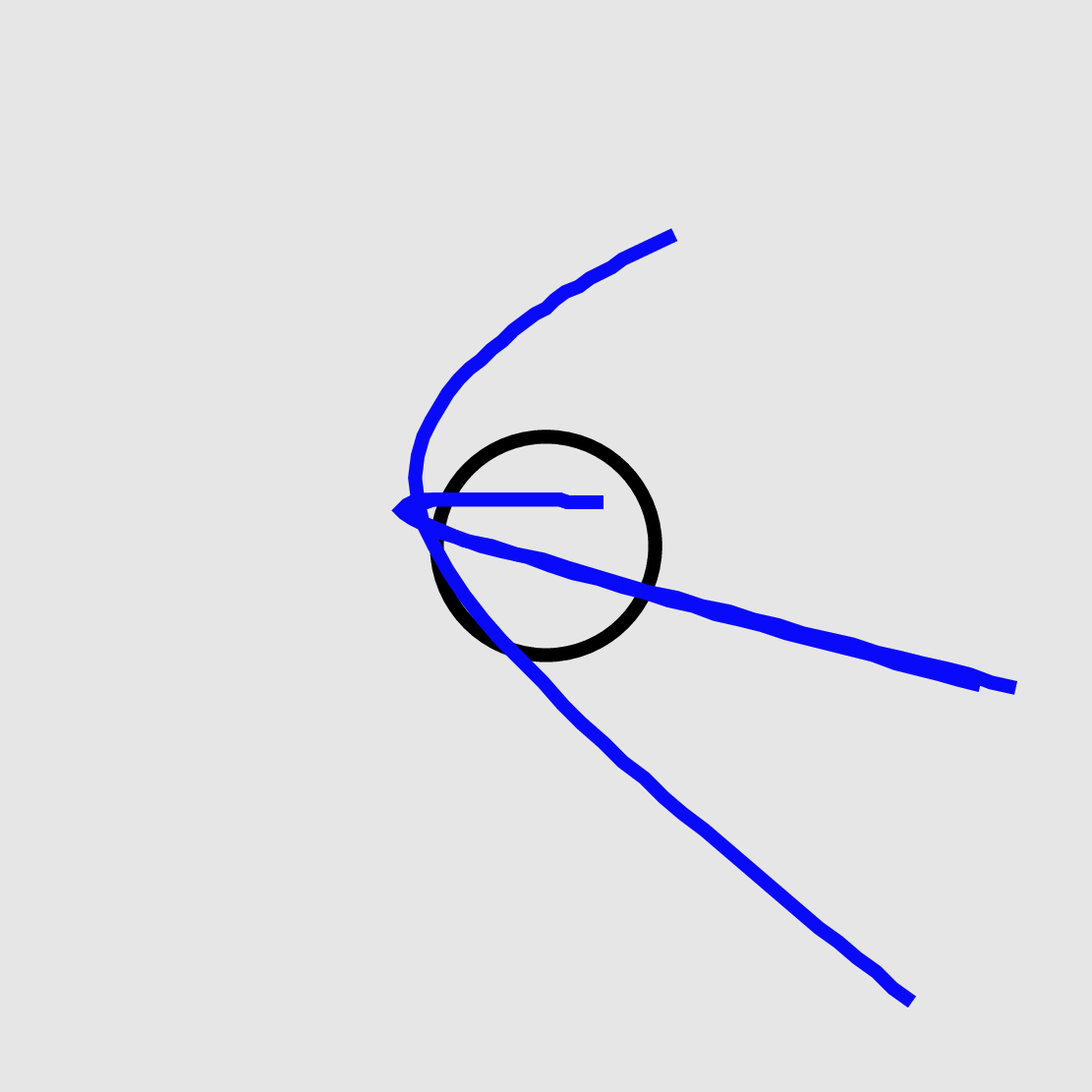}
\end{tabular}
}%
\subfloat{%
\begin{tabular}{c}
\includegraphics*[width=0.13\textwidth]{./figures/results/587722984435351614/587722984435351614_sim.eps}\\
\includegraphics*[width=0.13\textwidth]{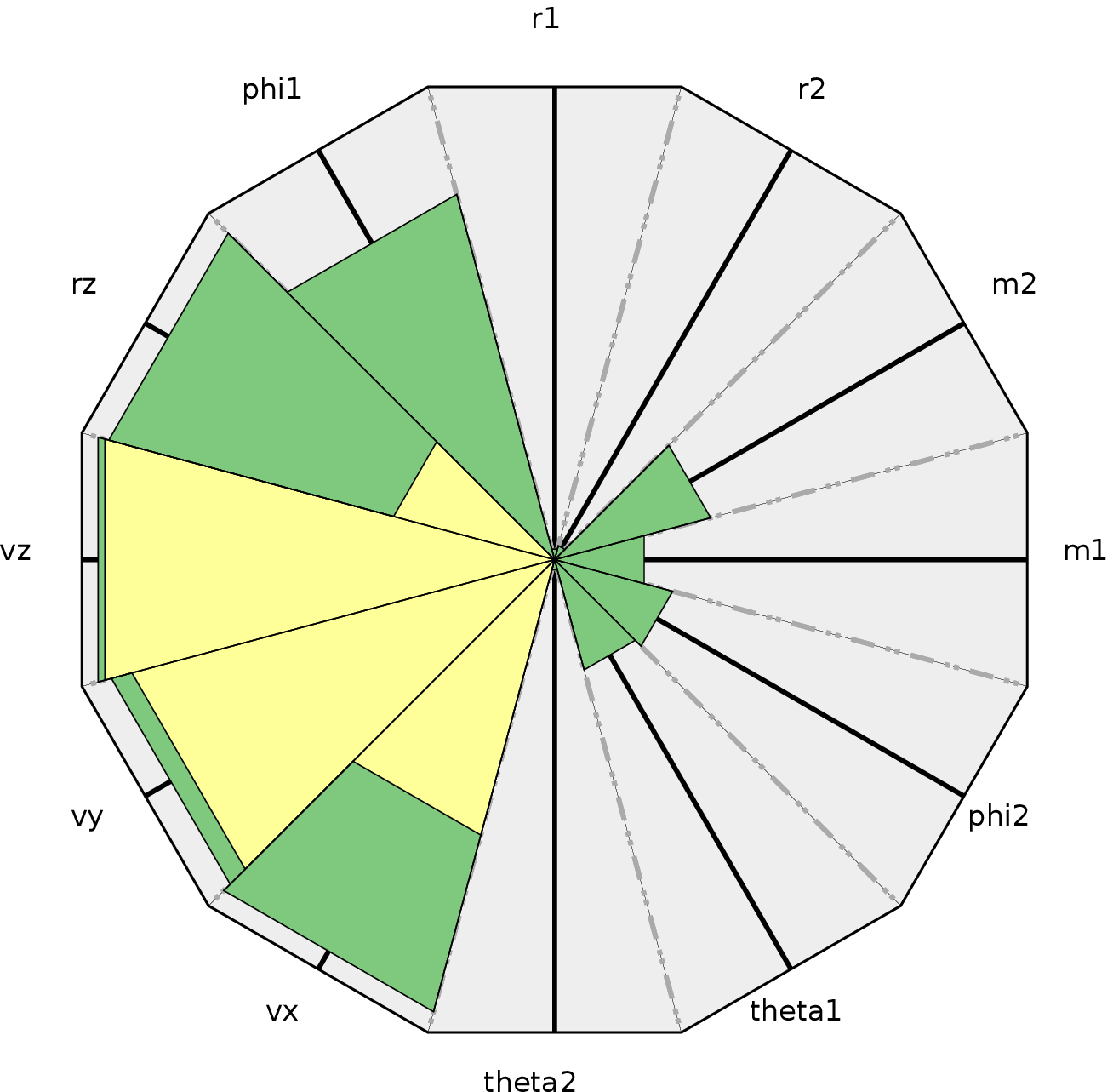}\\
\includegraphics*[width=0.13\textwidth]{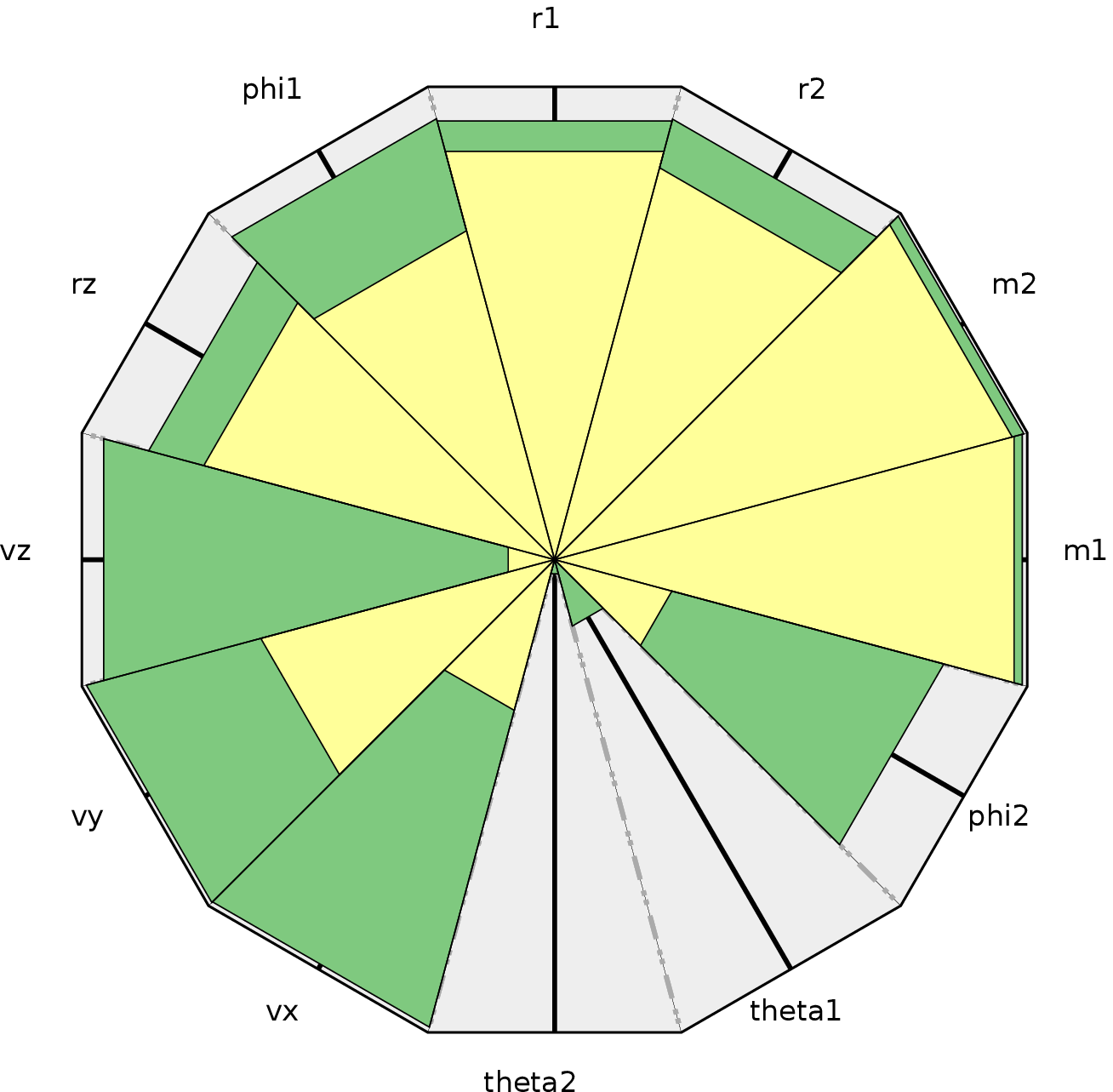}
\end{tabular}
}%
\subfloat{%
\begin{tabular}{c}
\includegraphics*[width=0.13\textwidth]{./figures/results/587722984435351614/587722984435351614_orb.eps}\\
\includegraphics*[width=0.13\textwidth]{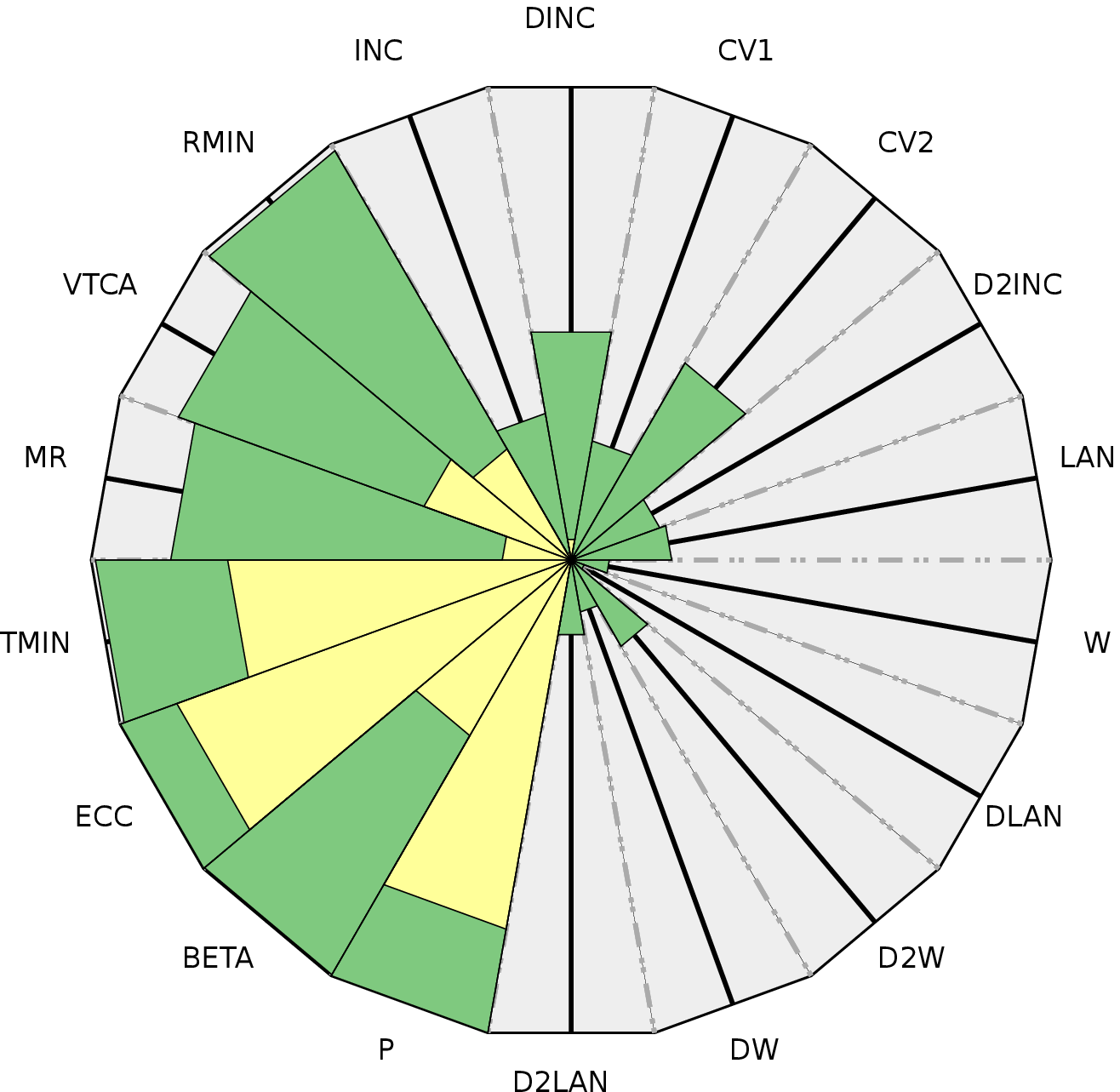}\\
\includegraphics*[width=0.13\textwidth]{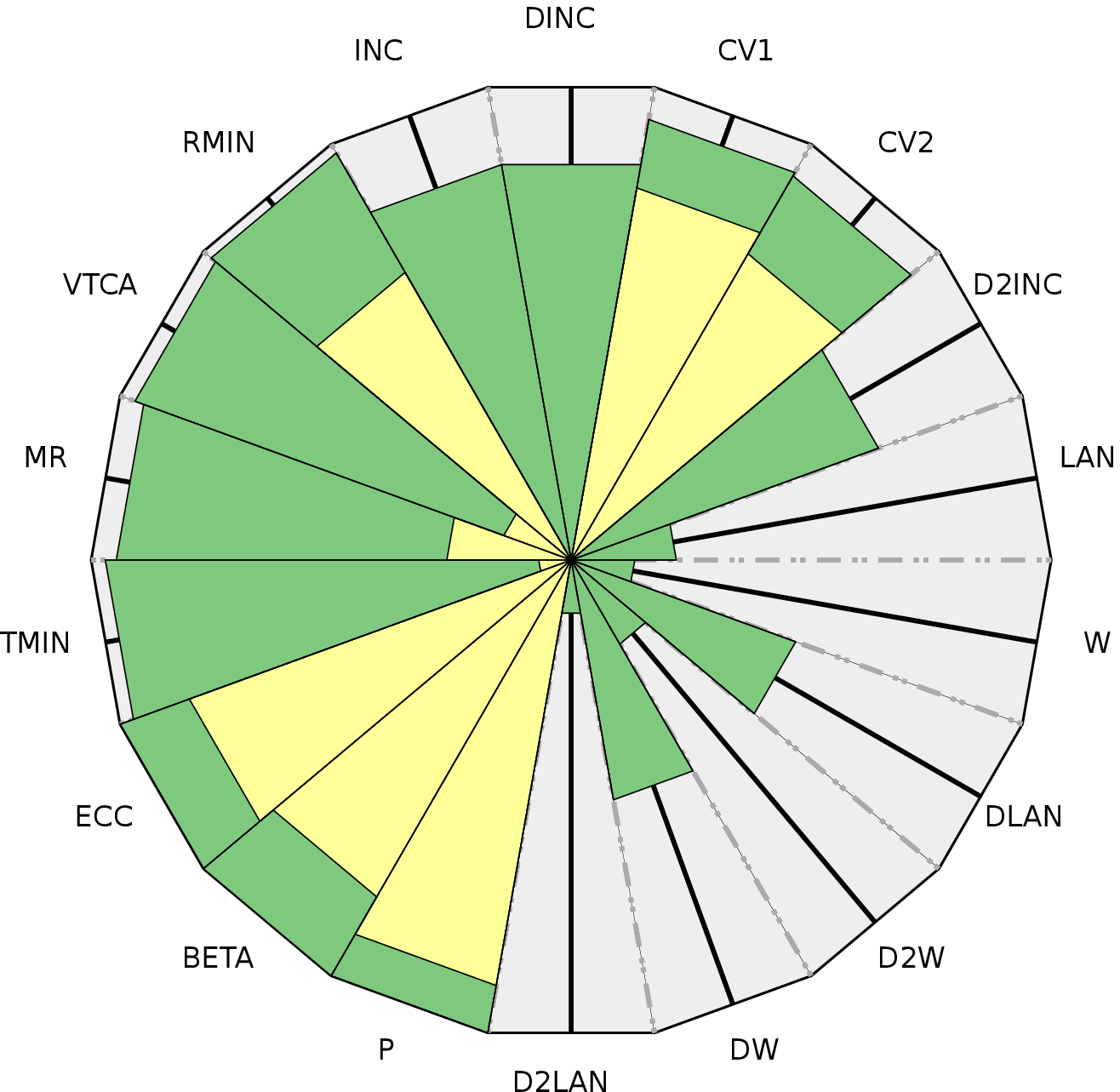}
\end{tabular}
}%
\caption{The galaxy pairs with the three highest fitness distribution skewnesses: Arp 240, Arp 104, and NGC 6786.}
\label{fighighskew}
\end{center}
\end{figure*}

Figure \ref{fighighskew} shows the three pairs with the highest skewness. All three have very obvious tidal features. The fitness distributions are very skewed with most simulations given a fitness score of zero. The trajectories are consistent. However, the glyph plots are not remarkably different from those of the low skewness galaxy pairs. This seems to indicate that the parallel coordinate and glyph plots are not a good indicator of which sets of simulations have converged to a high fitness match of the simulation. One possible explanation is that not every target was prepared identically. In some instances simulation parameter ranges were edited by hand. For example, the double-ring galaxy in our sample had the ranges for $v_x$ and $v_y$ restricted to keep their relative magnitudes low compared to $r_z$ and $v_z$. This was done to ensure more simulation parameter sets would be selected with an overall velocity perpendicular to the plane of the sky in order to encourage formation of rings\footnote{Rings are likely to form when one galaxy passes perpendicularly through the plane, and within the disc, of the other galaxy. Here both galaxy's discs had relatively low inclination angles with respect to the plane of the sky, so a velocity vector perpendicular to the sky is likely to be perpendicular to each disc as well.}. Other simulation ranges were edited on an ad hoc basis. Also, because some of the parameters are based on intrinsic properties of each galaxy pair (velocity ranges were constrained by estimated mass), not every simulation parameter was sampled across the same range of values for each galaxy. An alternative would be to plot all parallel coordinates and glyph plots on a common set of axes based on constant units such as kpc and deg.

\begin{figure*}
\begin{center}
\includegraphics[scale=0.5]{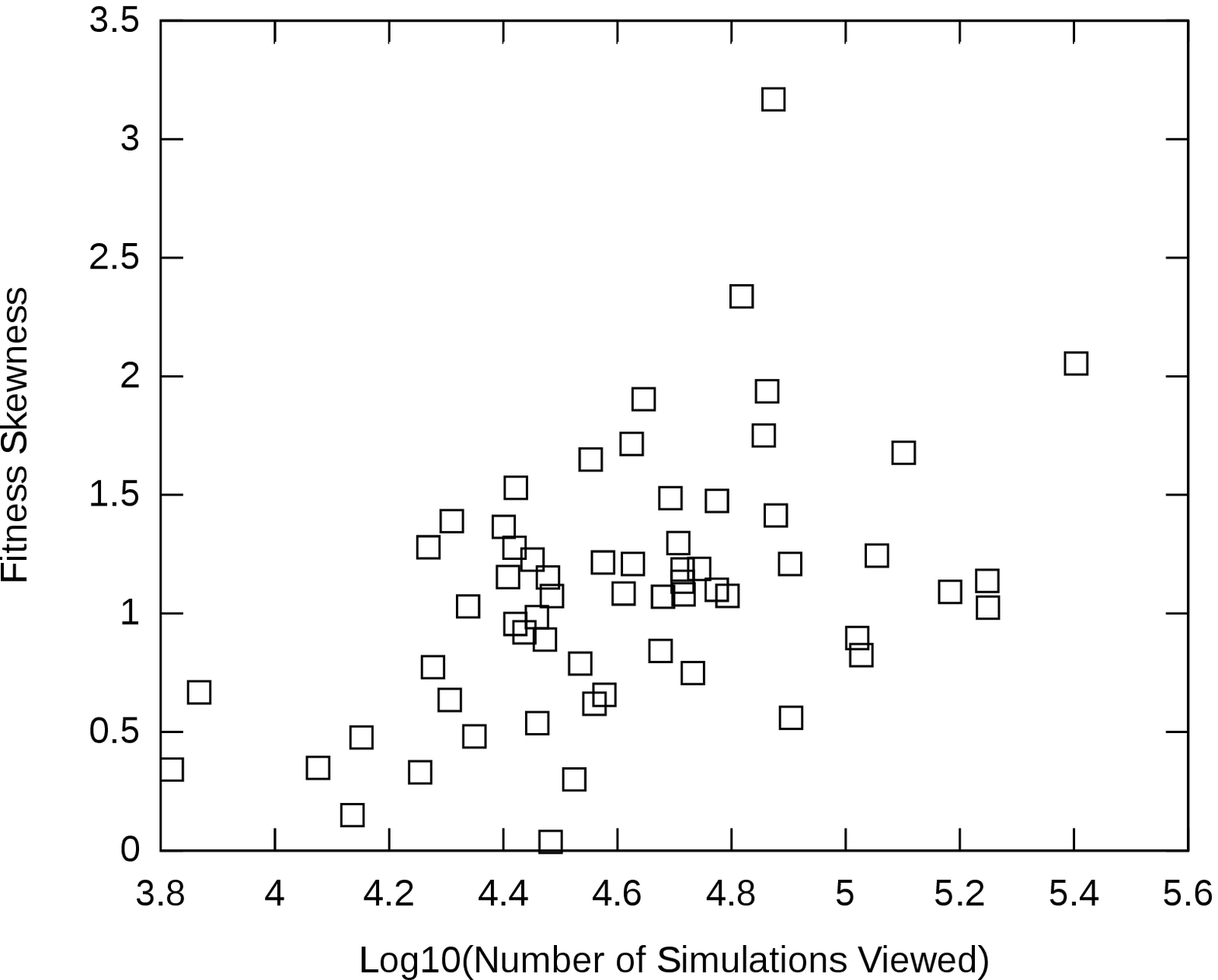}
\caption{Skewness vs. Number of Simulations Viewed.}
\label{ch5skew}
\end{center}
\end{figure*}

Another trend hinted at during the qualitative review of all of the galaxy pairs was that target images with larger, more distinct tidal features tended to be more popular. We wanted to know if the activity level for a target influenced the skewness. In Figure \ref{ch5skew} we show how the skewness varies with the number of simulations reviewed. Systems with the most number of simulations do not have the highest skewness, though a claim could be made that the targets with the fewest number of simulations viewed tended to have smaller skewness values.

If the activity level does not heavily influence the fitness skewness, then it remains possible that the distribution of fitness values is related more to the inherent properties of the interacting galaxies and not the amount of attention paid to each system by the volunteers.

\subsection{Distribution of Simulation Parameters}

\begin{figure*}
\begin{center}
\setlength{\tabcolsep}{0mm}
\subfloat{%
\begin{tabular}{c}
\includegraphics*[width=0.3\textwidth]{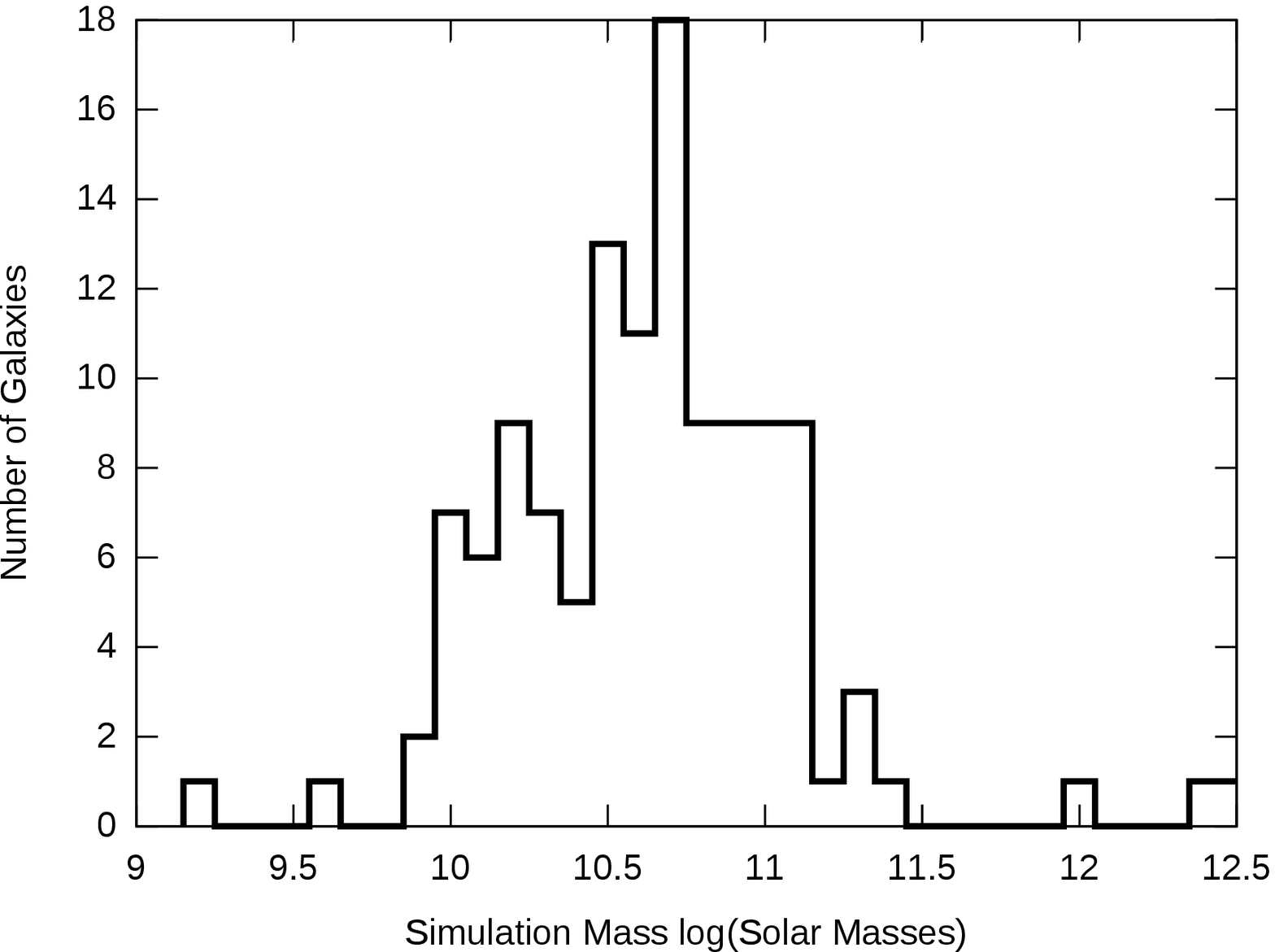}\\
\includegraphics*[width=0.3\textwidth]{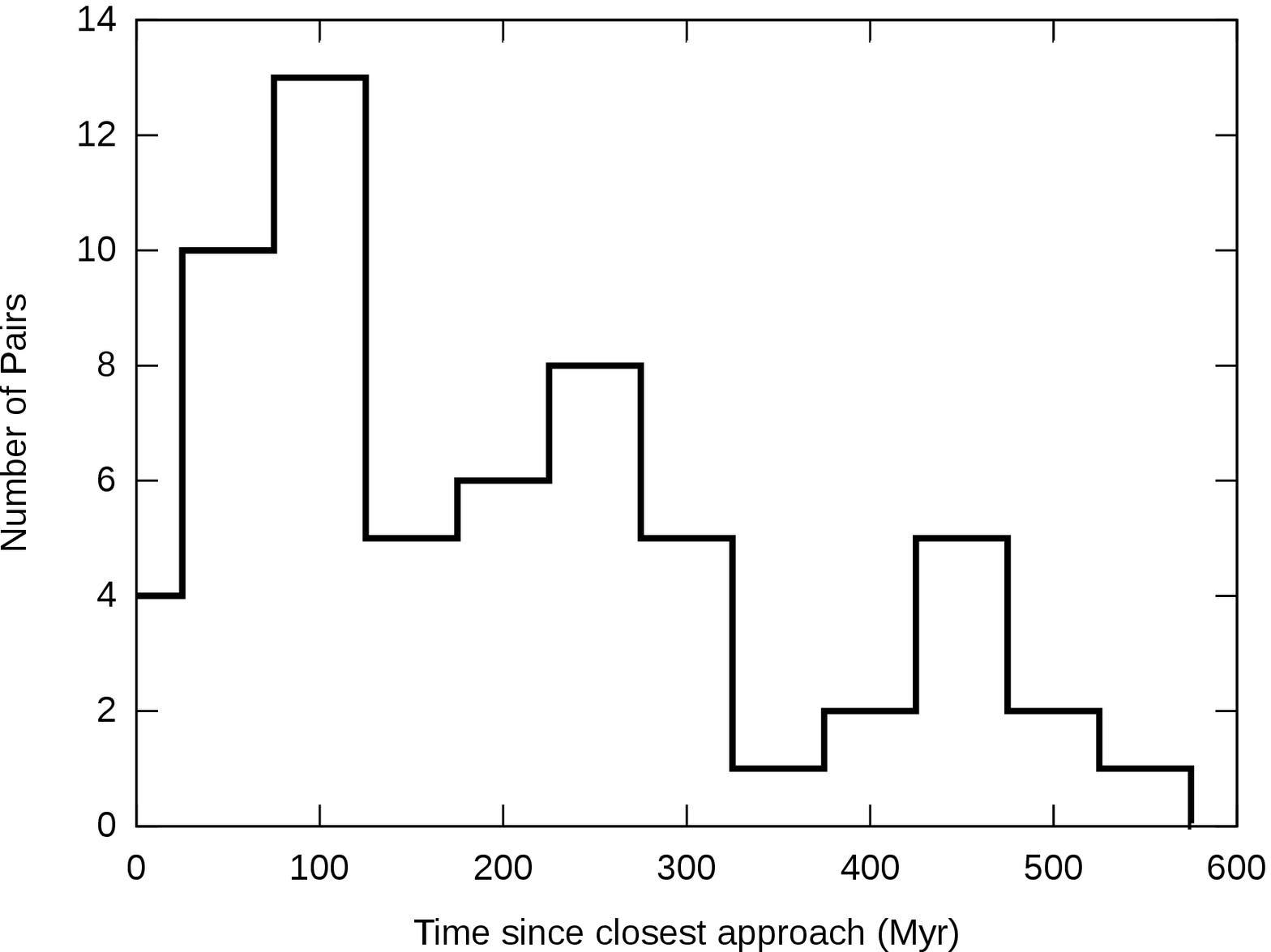}\\
\includegraphics*[width=0.3\textwidth]{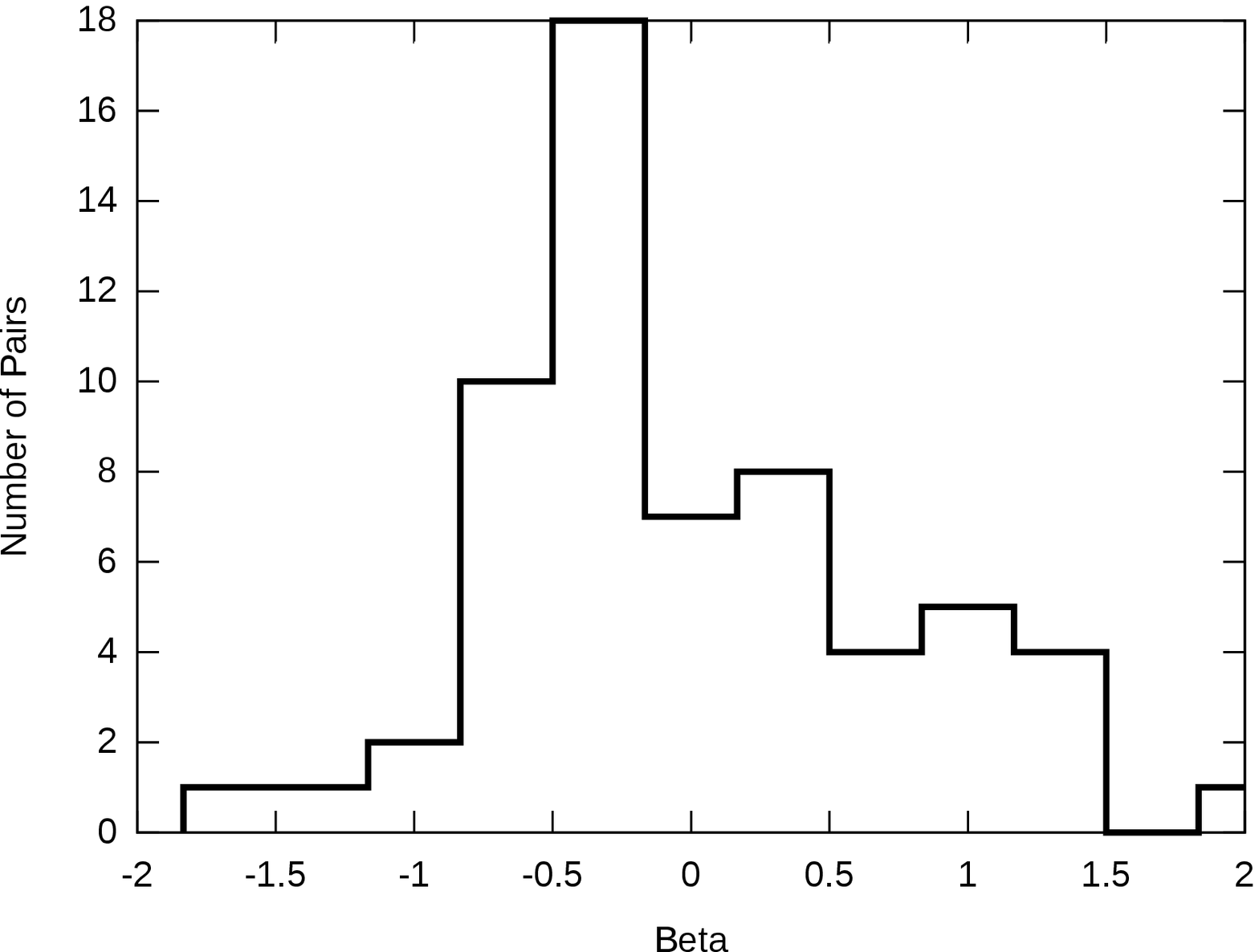}\\
\includegraphics*[width=0.3\textwidth]{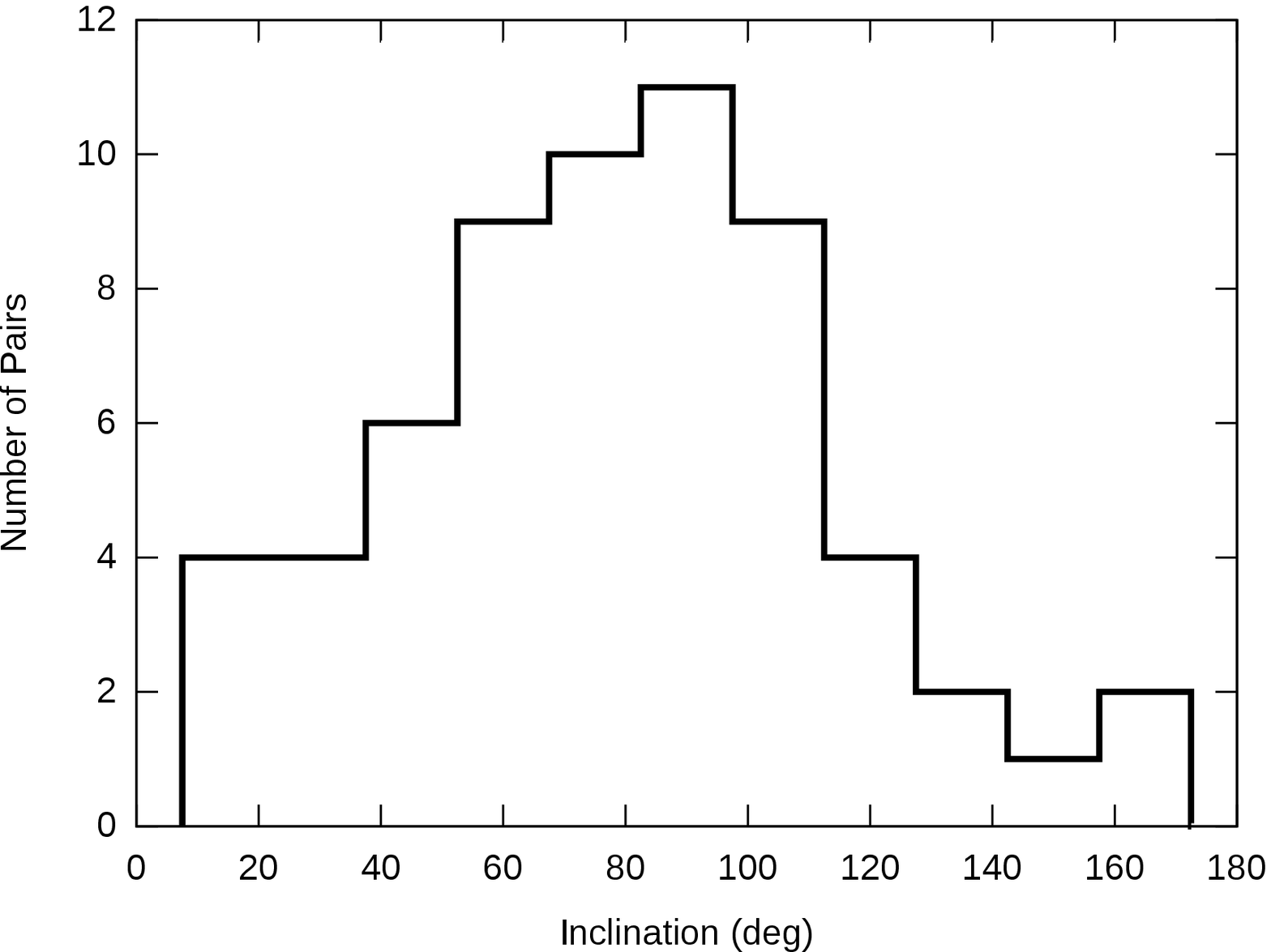}
\end{tabular}
}%
\subfloat{%
\begin{tabular}{c}
\includegraphics*[width=0.3\textwidth]{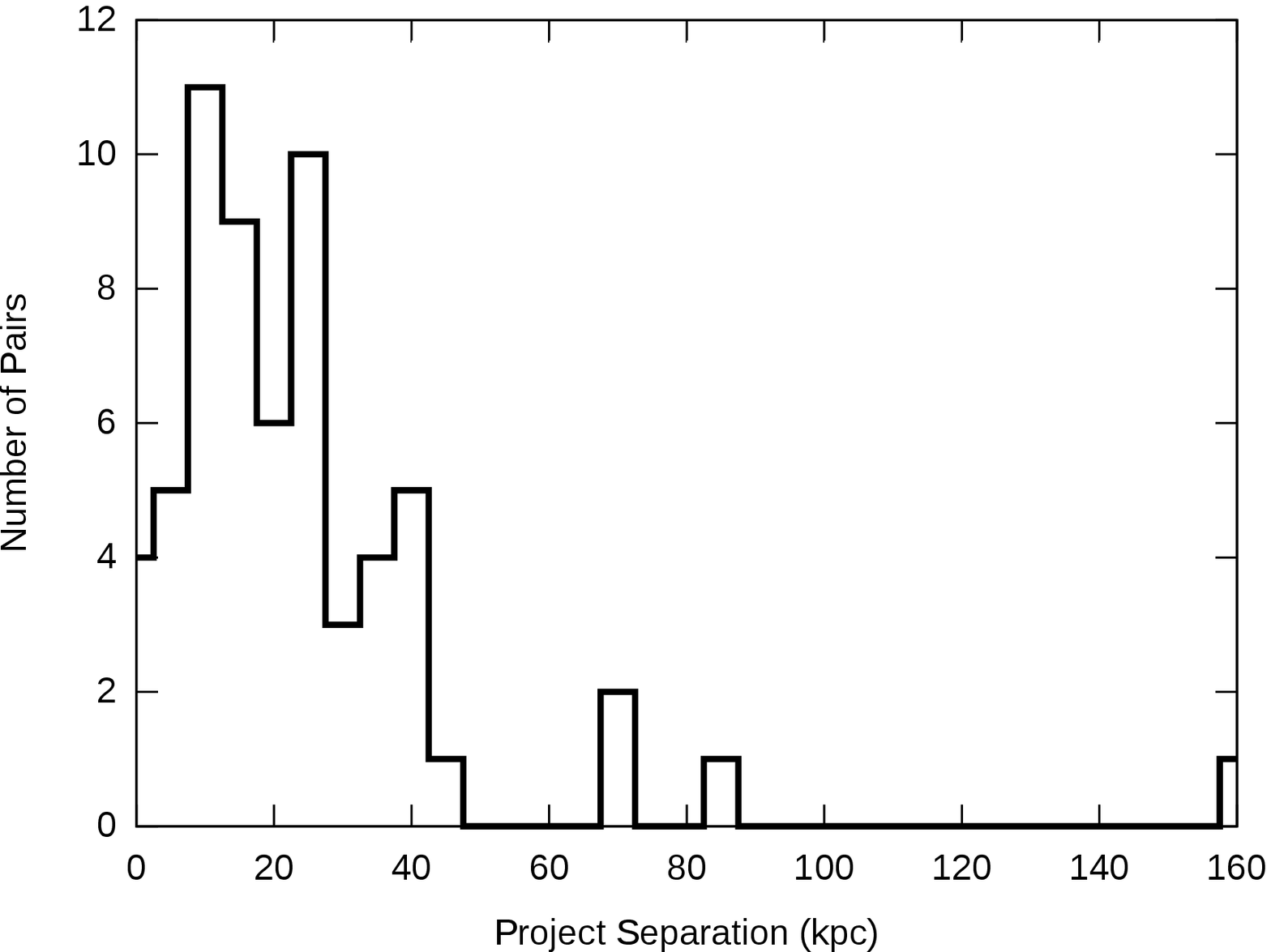}\\
\includegraphics*[width=0.3\textwidth]{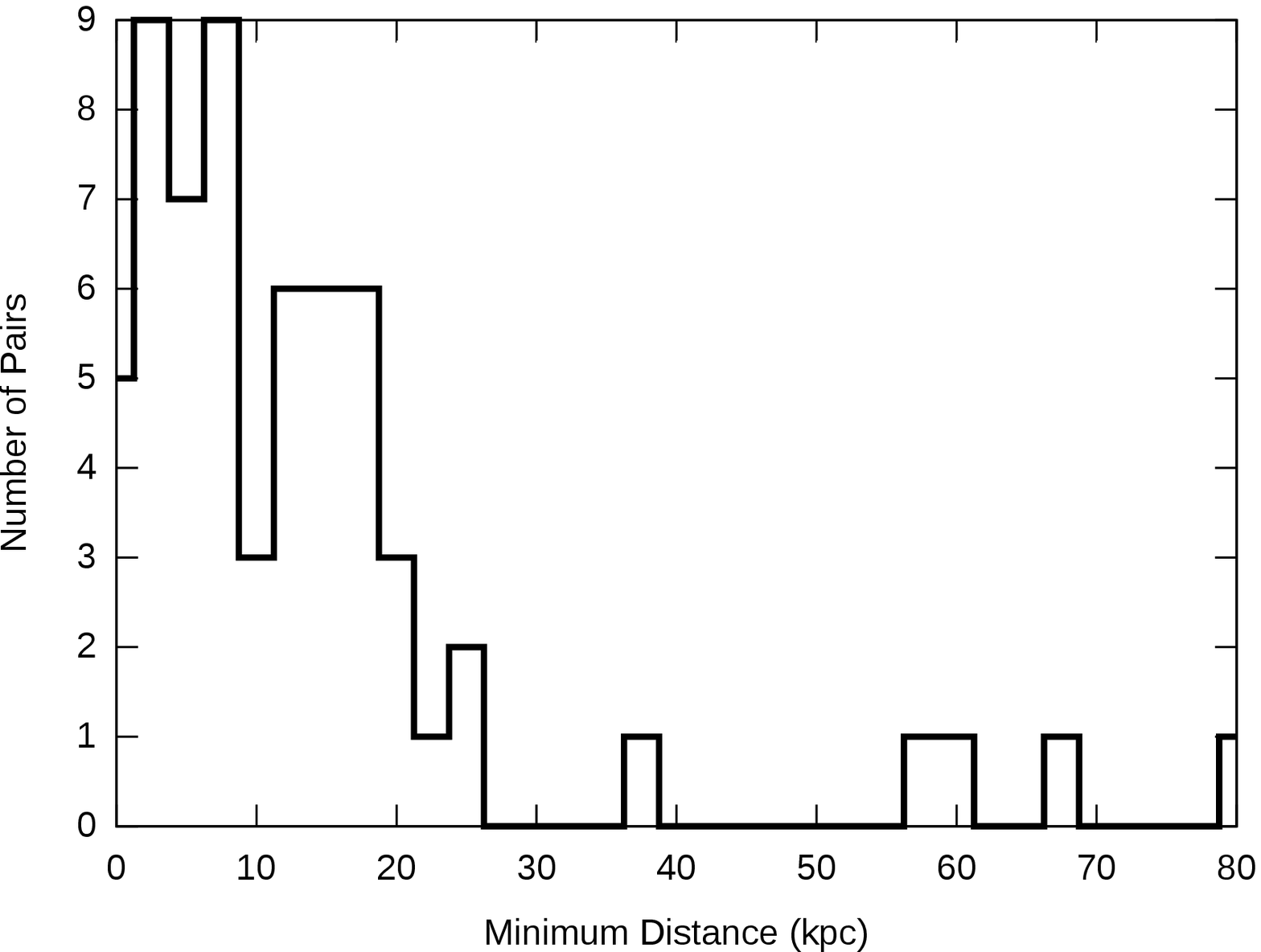}\\
\includegraphics*[width=0.3\textwidth]{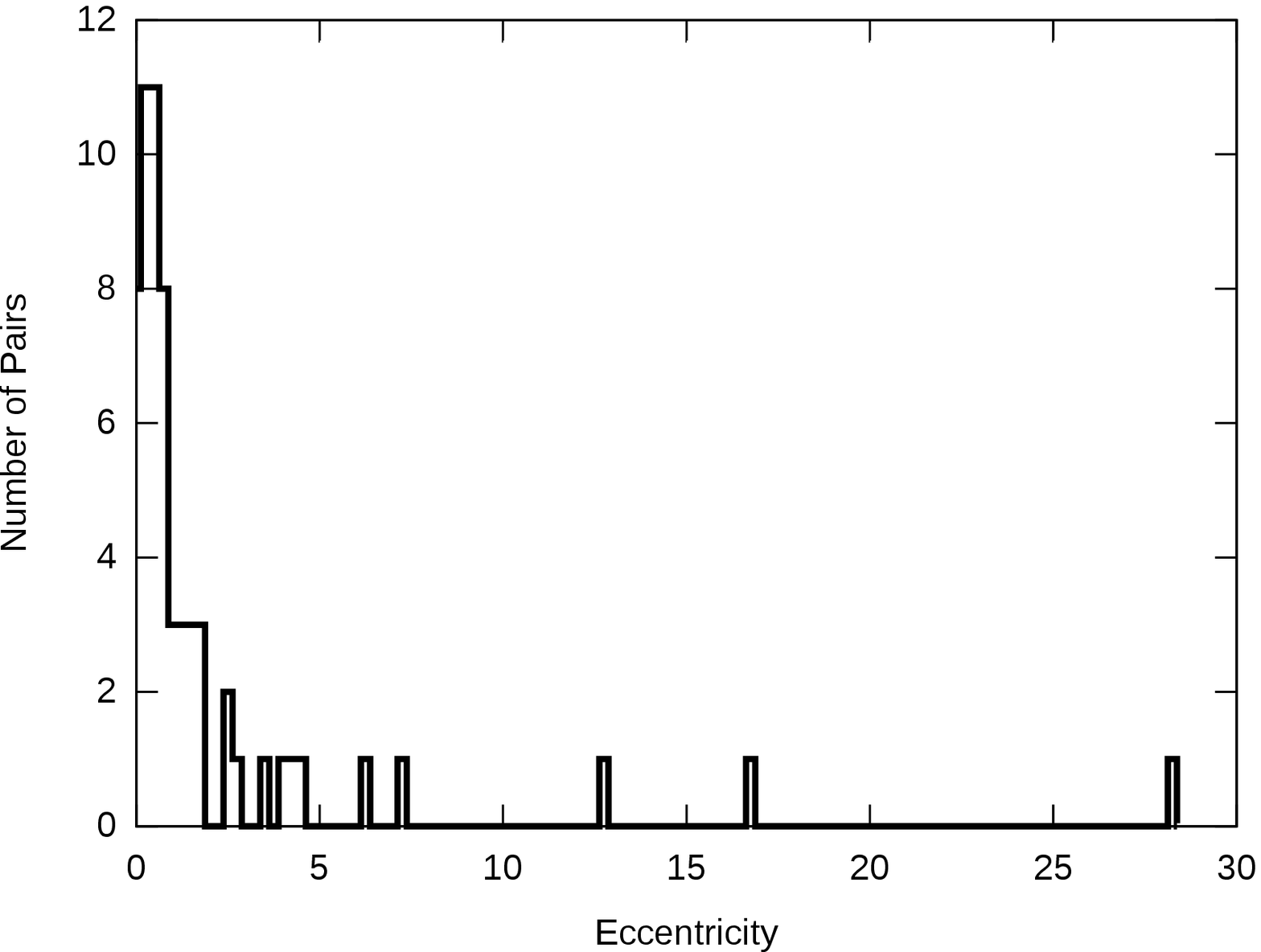}\\
\includegraphics*[width=0.3\textwidth]{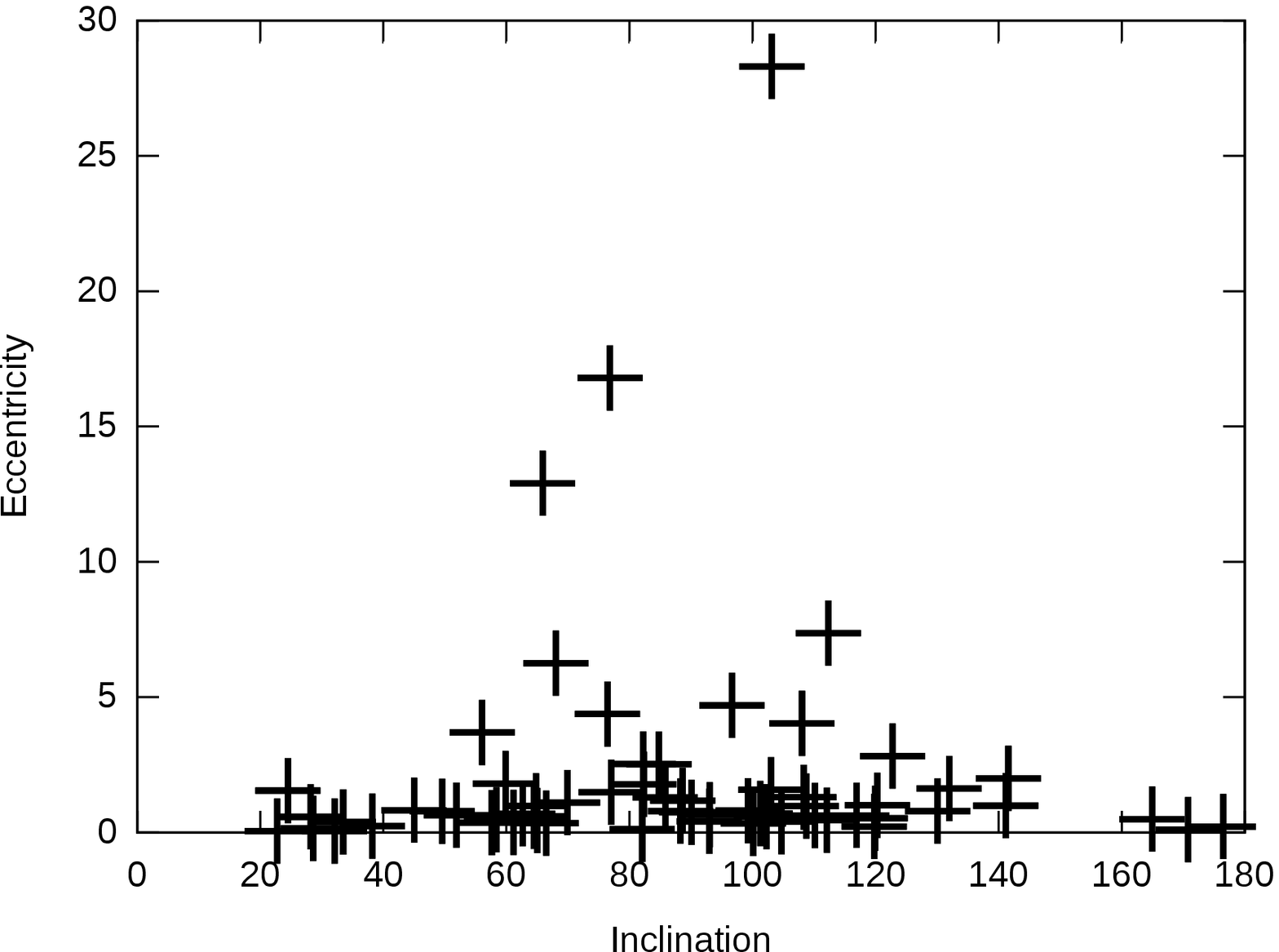}
\end{tabular}
}%

\caption{The distribution of simulation parameters including mass, $t_{min}$, projected separation, $r_{min}$, $\beta$, $ecc$, $inc$, and $ecc$ vs. $inc$. The bottom right plot shows that the highest eccentricities are nearly perpendicular to the plane of the sky.}
\label{figsimdist}
\end{center}
\end{figure*}

Figure \ref{figsimdist} contains seven histograms and one scatter plot. The histograms show the distribution of some important simulation parameters across all 62 pairs.

The values shown are for the best-fit value for each pair. The masses range over more than three orders of magnitude. Most of the projected separations are less than 40 kpc. The $r_{min}$ values tend to be smaller by a factor of two. The $t_{min}$ values range from 0 to 570 Myr. As discussed in section \ref{limitations} and in equation \ref{ml_eqn}, it is important to note that these represent the upper limits of the actual times because we don't have an accurate measurement of the mass-to-light ratio of our systems. However, the majority of the systems have dynamical ages of less than 300 Myr. This suggests that the majority of systems that have prominent tidal features we see have ages of less than 300 Myr. The $\beta$ values range from 0.01 to 100, with all but 5 galaxies having values of greater than 0.1 and less than 30. Most of the eccentricities are between 0 and 2 including a mix of elliptical and nearly parabolic orbits. However a number of orbits are very hyperbolic with eccentricities much greater than 1. The inclination with respect to the sky covers most of the full range. There is a peak in the distribution near 90$^\circ$. The scatter plot in the bottom right of Figure \ref{figsimdist} offers a potential explanation. The orbits with very high eccentricities tend to have inclinations nearly perpendicular to the plane of the sky. This is a result of our selection bias for picking galaxies that have relatively small projected separation distances. A hyperbolic passage would need to be nearly perpendicular to the plane of the sky for us to have a high probability of perceiving the galaxies as still being close to each other.

We can compare our population of orbital eccentricities with cosmologically motivated initial conditions.  \cite{khochfarburkert} studied major mergers of cold dark matter halos within a high-resolution cosmological simulation.  They found that $\sim$ 40\% of mergers have an initial eccentricity close to 1.  Their study found that 95\% of mergers in the simulation had an eccentricity < 1.5.   Our population of 62 models has only 70\% of systems < 1.5. Models of interacting galaxies with high eccentricity can be astrophysically plausible.  For example, \cite{highvelocity} found that the relative velocity of M86 and NGC 4438 exceeded the escape velocity of the system, $\sim$ 1000 \ km~s$^{-1}$. 

However, systems in our sample with very high-eccentricities also have large errors in the best-fit eccentricity.  Figure \ref{figlowskew} shows three pairs of galaxies that, at least at the present epoch, appear to be simple super-positions.  There are no large scale tidal features extending beyond the discs that can be used to constrain the relative orbits.  The eccentricities for these systems are 16.797 $\pm$ 19.740, 7.365 $\pm$ 2.416, and 12.908 $\pm$ 2.523. These systems also had low fitness skewness (and many had negative fitness kurtosis) indicating that overall convergence of the models were poor.  The resolution of the target images and the simulation output likely made it difficult to match internal structures like the spiral arms in Arp 274 and the ring in Arp 148 where the disc was still occupied.  Volunteers tended to focus on tails and bridges and may not have noticed these features. 

There are 13 systems where the uncertainty in eccentricity exceeds 1.  For 8 of those systems, the uncertainty exceeds 2.  If we exclude the systems with poorly constrained eccentricities, then $\sim$ 86\% of our systems have eccentricities < 1.5.  This is closer to the distribution found by \cite{khochfarburkert}, though they found fewer low eccentricity systems than we did.  Our eccentricities are reported as osculating values for the current epoch while their values were calculated one output prior to the merger. If the low eccentricity were treated as an initial condition, rather than a current value, these systems would have to have survived multiple close passages without merging.  We do not claim that.  The eccentricity is an evolving parameter under the influence of dynamical friction.

We have 14 pairs with an eccentricity > 1 and a positive fitness kurtosis indicating potential convergence of the model. For each pair we searched the NED IPAC database and the SDSS galaxy cluster catalog of \cite{berlind} to see if any of these pairs were members of a cluster. High velocity encounters are more likely to occur in clusters than for isolated galaxy pairs.  Only two of our high eccentricity pairs were found to be in clusters.  Further study of the environment of each pair is needed, especially when trying to distinguish between dynamical and environmental effects on star formation history. It is difficult to draw any strong conclusions with this sample about eccentricity, group membership, and tidal interaction. Since the galaxies in our sample were selected because they had obvious tidal features and had not undergone a merger, it is possible that the sample has inadvertently biased towards high eccentricity encounters.  Such encounters would be more likely to result in the well-separated pairs that are typical of our sample.

Our pericentre distances given in Table \ref{restbl2} as $r_{min}$ are relatively close. Over 80\% of our systems have their closest passage within the sum of the two disc radii. This is similar to the result found by \cite{khochfarburkert} where 70\% of all mergers had a first pericentre passage within the virial radius of the larger halo.

Not all of our systems are necessarily mergers.  The interacting pairs in several of our systems could be unbounded.  Furthermore, based on the restricted three-body simulations alone we make no claim that any particular system will fully merge in the future.  Our simulations were run from several hundred Myrs before the current epoch and then stop at the current time.  Our analytic treatment of dynamical friction is expected to alter the eccentricity of the orbit over time, causing parabolic orbits to become elliptical. We do not expect this analytic model to yield full accuracy through the merger process. We plan to use full n-body models to study in more detail the past history and future evolution of these systems.

\subsection{Comparison to Previously Published Models}

Several of our targets, such as Arp 242 (Mice) and Arp 244 (the Antennae), are well studied in the literature including several dynamical models dating back to \cite{tt72}. The paper describing the ``Identikit" software \citep{barnes_identikit_2009} contains a very useful list summarizing papers that contain dynamical models of interacting galaxies. \cite{privon_dynamical_2013} use the ``Identikit" software to model four actual systems, three of which were modelled as part of the {\it Galaxy Zoo:~Mergers} project, and also includes a list of previously published models for Arp 242 and Arp 244. \cite{howard_simulation_1993} present models for Arp 82 and Arp 107. \cite{cullen_hi_2007} present a model for Arp 104. \cite{keel_massive_2003} present a model for Arp 297. \cite{kaufman_interacting_1999} present a model for Arp 84. In Table \ref{prevmodels} we present the previous published values for the mass ratio, closest approach distance, time since closest approach, and eccentricity. For each of these parameters we also present our best-fit values and then calculate the ratio with what was previously published. Most ratios fall between 0.5 and 2 meaning that our model parameters match previously published ones within a factor of two.  

For a few systems such as Arp 82 and Arp 107, there are large differences between our simulation results and those from \cite{howard_simulation_1993}. These models from Howard et al. were taken as best matches for these two systems from a large set of simulations. The N-body method used by \cite{howard_simulation_1993} was a polar particle-mesh code, and there was no fine tuning to make an optimal match for these two systems. The halo-to-disc mass ratios for these simulations was set at 1 instead of the more commonly used values for more recent simulations of $\sim 5$. 

The differences in $t_{min}$  for Arp 84 between our simulations and those by \cite{kaufman_interacting_1999} are also striking. However since this simulation used a restricted three-body code with a softened potential (with SPH), the time scales needed to form these tidal features would be significantly different. For Arp 104, the models used by \cite{cullen_hi_2007} are from an in-house N-body code used to simulate the system based on preliminary simulations using a restricted three-body code. The halo-to-disc mass ratio for the N-body simulations was 1 in this work as well.

The difficulty of comparing simulations of interacting galaxies created by different authors is apparent from these examples. The N-body methods, halo-to-disc masses, and shape of the potential greatly affect the time scales and orbital characteristics of the final models. Although detailed comparisons between the velocity field of a galaxy are needed to set a more precise timescale for our sample, the use of a uniform simulation method, halo-to-disc ratio, and potential shape makes it more uniform than the simulations taken from the wider literature.

\begin{table*}
\caption[Model Comparisons]{\small For mass ratio (MR), $r_{min}$, $t_{min}$, and eccentricity (Ecc) we compare our model to the previously published model.}
\begin{center}
\scalebox{0.70}{
\begin{tabular}{| c | c | c | c | c | c | c | c | c | c | c | c | c | c |}
\hline
Source & Object & MR & GZM MR & MR Ratio & $r_{min}$ (kpc) & GZM $r_{min}$ & $r_{min}$ ratio & $t_{min} (Myr) $ & GZM $t_{min}$ & $t_{min}$ ratio & Ecc & GZM Ecc & Ecc ratio	\\
\hline
Kauffman et al. & Arp 84 & 4 & 4.69 $\pm$ 1.4 & 1.17 & 14.24 & 7.38 $\pm$ 4.74 & 0.52 & 17 & 217.63 $\pm$ 26.59 & 12.8 & 1 & 0.39 $\pm$ 32.08 & 0.39 \\
Howard et al. & Arp 82 & 10 & 1.99 $\pm$ 0.45 & 0.2 & 20 & 6.37 $\pm$ 1.64 & 0.32 & 700 & 131.63 $\pm$ 20.96 & 0.19 & 1 & 0.15 $\pm$ 0.13 & 0.15 \\
Cullen el al. & Arp 104 & 2 & 1.84 $\pm$ 0.56 & 0.92 & 12.5 & 1.07 $\pm$ 0.83 & 0.09 & 330 & 67.55 $\pm$ 26.07 & 0.2 & 1 & 0.64 $\pm$ 0.11 & 0.64 \\
Howard et al. & Arp 107 & 2 & 1.35 $\pm$ 0.47 & 0.68 & 20 & 6.74 $\pm$ 5.66 & 0.34 & 900 & 420.88 $\pm$ 86.79 & 0.47 & 1 & 0.4 $\pm$ 0.25 & 0.4 \\
Privon et al. & Arp 240 & 1 & 1.17 $\pm$ 0.39 & 1.17 & 21.25 & 81.44 $\pm$ 10.21 & 3.83 & 230 & 249.41 $\pm$ 31.5 & 1.08 & 1 & 3.7 $\pm$ 2.35 & 3.7 \\
Privon et al. & Arp 242 & 1 & 0.6 $\pm$ 0.52 & 0.6 & 14.81 & 15.99 $\pm$ 5 & 1.08 & 175 & 429.54 $\pm$ 192.82 & 2.45 & 1 & 0.7 $\pm$ 0.27 & 0.7 \\
Privon et al. & Arp 244 & 1 & 0.76 $\pm$ 0.4 & 0.76 & 4.93 & 1.57 $\pm$ 0.69 & 0.32 & 260 & 55.42 $\pm$ 14.89 & 0.21 & 1 & 0.49 $\pm$ 0.14 & 0.49 \\
Keel $\&$ Borne & Arp 297 & 5 & 2.32 $\pm$ 0.17 & 0.46 & 12.91 & 18.57 $\pm$ 1.28  & 1.44 & 250 $\pm$ 35 & 562.9 $\pm$ 54.04 & 2.25 & 1 & 0.59 $\pm$ 0.03 & 0.59 \\

\hline
\end{tabular}
 \label{prevmodels}
 }
\end{center}
\end{table*}

 \cite{privon_dynamical_2013}  have also gathered a set of previously published model parameters for Arp 242 and Arp 244, with 5 and 8 models respectively. The set of previously published models are treated as a population for which we can compute the mean and standard deviation of each parameter. In Table \ref{prevpopmodels} we compare our best-fit values and uncertainties for mass ratio, $r_{min}$, $t_{min}$, and eccentricity with these values. Because we have uncertainty values for each proposed value, we can do a consistency check. We do this by computing the ratio of the absolute value of the differences to the sum of the uncertainties. For Arp 242, the consistency check is less than 1 for all parameters except  $t_{min}$. Here the value 1.02 indicates that differences in the two values is slightly larger than the sum of uncertainties. For Arp 244, the consistency check is less than 1 for all but $r_{min}$. In this case the value is 1.7 indicating that the difference in values is 70 \% bigger than the sum of uncertainties. 

One of the hardest things to constrain without velocity field measurements of the interacting system is the time scale associated with the interaction. The consequences for this uncertainty and the method we used to set these masses is discussed below in Section \ref{limitations}. However, we often cannot determine how previous authors set their initial masses. The ratio between $r_{min}$ and the outer disc radius of the primary galaxy may also be a more relevant parameter in creating tidal features than just the distance to closest approach. However, we cannot easily determine this parameter in older models as well. This uncertainty in what scales were actually used makes comparing models particularly difficult.

\begin{table*}
\caption[Model Population Comparisons]{\small For mass ratio (MR), $r_{min}$, $t_{min}$, and eccentricity (Ecc) we compare our model values (GZM) to the previously published values (Prev.). Each ``check" value is the absolute value of the difference of the two values divided by the square root of the some of squares of uncertainties.}
\begin{center} 
\scalebox{0.70}{
\begin{tabular}{| c | c | c | c | c | c | c | c | c | c | c | c | c | c |}
\hline
Object & Prev. MR & GZM MR & MR Check & Prev. $r_{min}$ (kpc) & GZM $r_{min}$ & $r_{min}$ Check & Prev. $t_{min} (Myr) $ & GZM $t_{min}$ & $t_{min}$ Check & Prev. Ecc & GZM Ecc & Ecc Check	\\
\hline
Arp 242 & 1 $\pm$ 0 & 0.6 $\pm$ 0.52 & 0.77 & 14.79 $\pm$ 9.53 & 15.99 $\pm$ 5 & 0.08 & 181 $\pm$ 50.30 & 429.54 $\pm$ 192.82 & 1.02 & 0.84 $\pm$ 0.22 & 0.7 $\pm$ 0.27 & 0.28 \\
Arp 244 & 1 $\pm$ 0 & 0.76 $\pm$ 0.4 & 0.61 & 17.22 $\pm$ 8.52 & 1.57 $\pm$ 0.69 & 1.7 & 227 $\pm$ 164 & 55.42 $\pm$ 14.89 & 0.96 & 0.81 $\pm$ 0.26 & 0.49 $\pm$ 0.14 & 0.79 \\
\hline
\end{tabular}
 \label{prevpopmodels}
 }
\end{center}
\end{table*}

\subsection{Separation Distance and Interaction Time}

\begin{figure*}
\begin{center}
\includegraphics[scale=1.5]{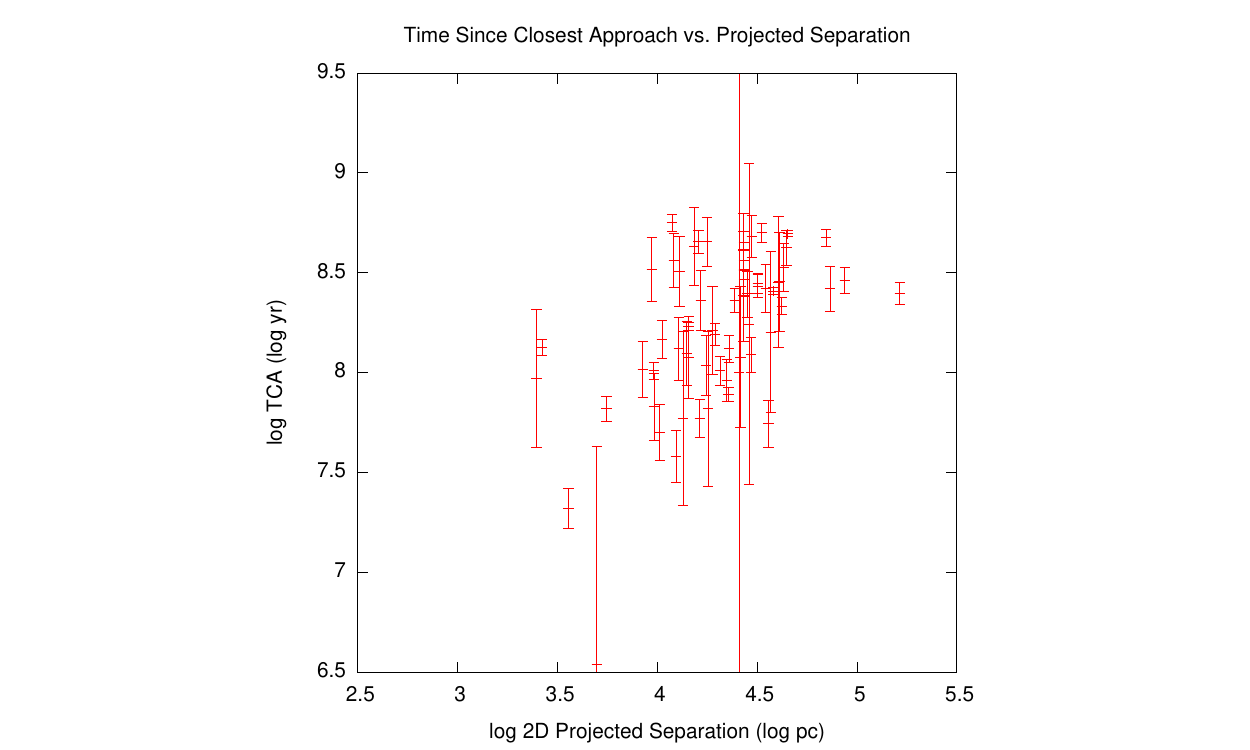}
\caption{Separation Distance and Interaction Time. The time (log yrs) since the point of closest approach is plotted against the 2D projected separation distance (log pc) for the best fit models in our project. The error bars represent the uncertainty analysis described in section \ref{modelparam}. }  
\label{sepdist}
\end{center}
\end{figure*}

Figure \ref{sepdist} shows the relationship between the projected two dimensional separation distance and the derived time since closest approach in our best fit models. In many studies, the projected separation distance is used as a proxy for the time since closest approach. Although the average between these two quantities are clearly correlated, there are a number of systems with the same separation distances which have very different ages. With our systems, the orbits have a wide range of orbital eccentricities. In systems with closed orbits (likely second approaches after dynamical friction), this correlation breaks down.

\subsection{Dynamical Parameters and Star Formation}
Work is on-going to form a consistent set of photometric and spectral measurements of star formation in our targets to be published later. Without presenting star formation measurements for our sample we can still infer how soon after the time of closest approach that tidally induced star formation is triggered. \cite{barton} explore a population of close galaxy pairs and multiples from the CfA2 survey. Using measurements of several key spectral line widths, they are able to fit models of star formation history to their sample. They used starburst models to determine how long ago each galaxy underwent a burst of star formation. They call this time $t_{burst}$. They then plot their population of galaxies with $t_{burst}$ against projected separation distance and line-of-sight velocity difference of the two galaxies.

\begin{figure*}
\begin{center}
\setlength{\tabcolsep}{0mm}
\subfloat{%
\begin{tabular}{c}
\includegraphics*[width=0.45\textwidth]{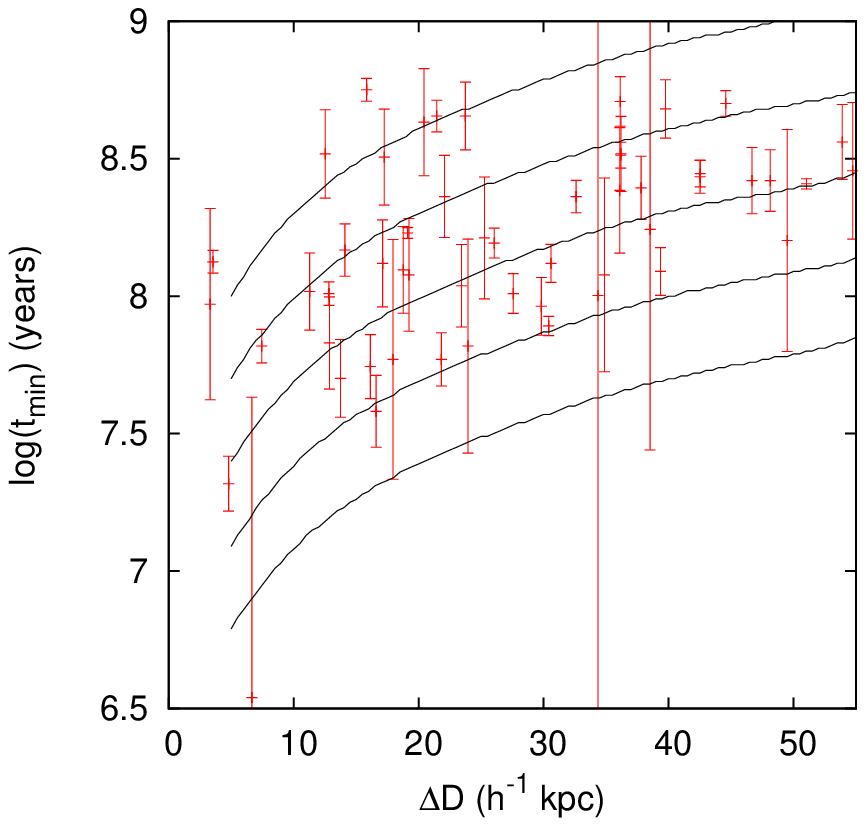} \\
\end{tabular}
}%
\subfloat{%
\begin{tabular}{c}
\includegraphics*[width=0.45\textwidth]{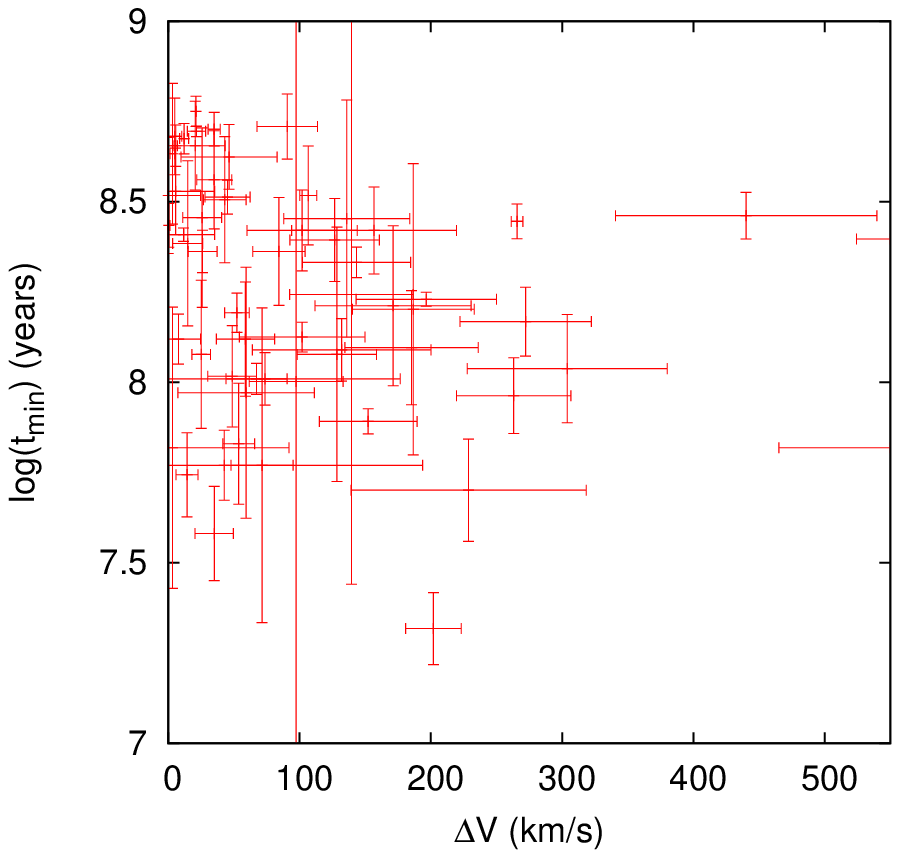} \\
\end{tabular}
}%
\caption{The distributions of $t_{min}$ vs. projected separation and line-of-sight velocity differences for our population are very similar to those of \citet{barton}. See their Fig. 10b and Fig. 18 for comparison. For the left figure, they found interacting pairs with no delay between close-approach time and the triggering of star formation occupied the region on the plot outlined by the contours.}
\label{bartoncomp}
\end{center}
\end{figure*}

Figure \ref{bartoncomp} shows the dynamically derived ages from our models plotted against the projected separation distances in our sample.  
Using the same contour lines as Barton {\it et al} for constant velocity, we demonstrate that our population plotted as $t_{min}$ vs. projected separation distance is very to similar to their measurements of $t_{burst}$ vs projected separation distance (see figure 10b in  Barton {\it et al}).   This suggests that the dynamical ages derived of our models fit the same distribution as the time since the last burst of star formation calculated using starburst models of the observed spectrum in interacting systems.   Although we have different samples of interacting systems, the dynamical clocks in our systems are a good match to the star formation clocks used in these observations. 

\cite{barton} then generate a population of simulated orbits using their simulation code setting a variety of impact parameters and inclinations as initial conditions. They plot time since closest approach, referred to as $t_{pass}$, against projected separation distance and line-of-sight velocity difference of the two galaxies.   With star-formation models of observed galaxies and a population of orbit models, they compare the two distributions. By trying different IMFs and delays in triggered starbursts, they demonstrated that the distribution of points in the $t_{burst}$ vs. projected separation distance plot shifts vertically. They conclude that merger-induced starbursts happen at or soon after ($<$ 50Myr) the time of closest approach.   Our results are consistent with their conclusion that close approaches between galaxies trigger star formation almost immediately.

\section{Conclusions}
\label{conclusions}

A pipeline was developed to allow massively parallel online visual inspection of simulations of pairs of interacting galaxies. The system was able to perform simulations for over 300 million samples of parameter space for 62 pairs of interacting galaxies. The pipeline presented over 3 million simulated images from these parameter sets to Citizen Scientist volunteers who selected 66000 of them for followup evaluation. Those same Citizen Scientists were able to perform 1 million {\it Merger Wars} competitions to assign fitness scores to the selected set of simulations and rank them accordingly. The final result is a plausible set of best-fit, restricted three-body dynamical models of each of our 62 target interacting pairs. We use these models to examine the ages of the tidal features and ranges of orbital parameters in systems of interacting galaxies. Finally we have produced a training set of 66000 human-evaluated simulation results. We expect that these images will be useful as a training set for machine learning and computer vision algorithms.

We provided several methods for visualizing convergence of simulation and orbit parameters. We found a strong correlation between the skewness of the fitness distribution of simulations with the presence of detailed tidal features in the target image. We believe the large sample and convergence information generated by our population will provide guidance to researchers studying similar systems.

\subsection{Limitations to the Models}
\label{limitations}
There may be systems that simply cannot be modelled realistically with restricted three-body approximations or where the morphology may not uniquely constrain the orbit trajectory.  Most of our high-eccentricity systems were poorly converged and lacked distinct tidal tails and bridges.  These models are not likely to be correct. \cite{barnes_identikit_2009} stresses the importance of incorporating velocity information, such as derived from HI observations, to constrain model results. They discuss the history of full N-body models for NGC 7252 where initially the best-fit orbit was thought to be retrograde \citep{borne_merger_1991}, but as improved velocity measurements were made, the best-fit orbit was changed to be a prograde orbit.

To have confidence that these models match real systems, it is important to compare the velocity field of the models to the observed velocity maps of the individual systems. Using our published simulation parameters and software, it is possible to construct simulated velocity maps for researchers who wish to make this comparison. Even without this data, many of the dynamical parameters are well constrained. The types of tidal features and the relative positions and sizes of the galaxies makes some of the models converge relatively well even without the velocity data. Even when velocity information is available, one still needs to model the morphology, and the methods presented in this paper provide a sufficiently accurate and consistent way of doing so. Of particular interest is how some of the parameters are less well constrained than others. For example, the alignment of the rotation axis of the secondary galaxy is often extremely poorly constrained while orbital parameters such as the mass ratio, inclination angle, $r_{min}$, and time since closest approach have been constrained to a relatively small part of phase space. This suggests that the orbits rather than the particular alignment of the discs may be constrained in some cases without velocity fields. Of course, additional comparisons between the galaxies and velocity fields are needed to ensure that these models are correct. In the cases where these don't match, we may have a degenerate solution that is not well constrained by the tidal features or the phase space may not have been explored enough to find the best match.

Another limitation to these models is our use of a restricted three-body code instead of a fully-consistent tree code, such as GADGET \citep{springel_gadget:_2001}. In a separate paper \citep{wallin2013a}, we present a more detailed description of our software \citep{2015ascl.soft11002W} including comparisons to self-consistent models. We are also using the models in this paper as the basis for creating high resolution self-consistent models of these systems. In general, the morphological differences between the restricted three-body code with a realistic mass distribution as used for our project and those from full N-body code are minimal in the regions where tidal features are being formed.

Because there are no measured masses or velocity scales for these simulations other than the mass-to-light ratio we have adopted based on the SDSS colours or {\it B} magnitude, it is likely that there are some systematic errors in velocities and interaction times of the simulations. Fortunately, the mass ratios between the two galaxies in the system are fairly well constrained by the shapes of the tidal feature. The distances and sizes of our galaxies are constrained as well because of their redshifts,  although the accuracy of the redshift measurements does not provide strong constraints on the relative velocity or the orbits of the galaxies during the interaction. The biggest effect of the uncertainty in the masses and the mass-to-light ratio is an offset of the actual times of interaction. In the case of an approximate Keplerian orbit between the two galaxies, if we assume the shape of the orbit and distance scale in our simulations is close to the actual values, we can relate the masses and times as follows:

\begin{equation}
(M_{simulation} )T^2_{simulation} = a^3
\end{equation}

\begin{equation}
(M_{actual})  T^2_{actual} = a^3
\end{equation}

\begin{equation}
T_{actual} =T_{simulation} \sqrt{ \frac{M_{simulation}}{M_{actual}}}
\end{equation}

\noindent For most of our systems, this can be simplified to 
\begin{equation}
T_{actual} =T_{simulation} \sqrt{m_{l}}
\label{ml_eqn}
\end{equation}

\noindent where $m_l$ is the actual mass-to-light ratio of our systems.

\subsection{The Pipeline for Constructing Models}

Applying experience gained by running a website for thousands of Citizen Scientists, we were able to refactor our software tools into a pair of applications that now run independently of {\it Galaxy Zoo:~Mergers}. For interacting pairs similar to those from the Arp catalogue processed by {\it Galaxy Zoo:~Mergers}, the new tool allows a researcher to perform and review several thousand simulations in under an hour. The efficiency of the {\it Merger Wars} algorithm and the meaningful constraints we place on the 14 dimensional simulation parameter space allow us to rapidly model interacting galaxies. The new process could potentially duplicate the convergence of simulation parameters for a single system in only a few hours of work by a single researcher. The {\it Evaluate} task has been eliminated, as well as the Prepare Targets for Merger Wars task. This tool will be used in the future to model even more of the Darg catalogue of mergers identified by {\it Galaxy Zoo} \citep{darg_galaxy_2010}.

\subsection{The {\it Galaxy Zoo:~Mergers} Catalog of Interacting Galaxy Models}

We have created a set of 62, best-fit, dynamical models using a restricted three-body code. The level of convergence for each system varies somewhat and is evaluated visually. The sheer number of simulations reviewed by Citizen Scientists, over 50000, is orders of magnitude more than viewed by current researchers simulating individual systems. It is important to note that current researchers are using the more sophisticated and computationally intensive full N-body codes. The hundred or so simulation runs they review for each target represent an increased level of realism over the restricted three-body simulations run here. However, by reviewing so few simulations, they are unable to estimate the uniqueness of their final models. The restricted three-body portion of our multi-model process is currently the best mechanism available for exploring a wide volume of parameter space in order to achieve an estimate of uncertainty in final simulation parameters.

\subsection{Machine Learning Training Set}
\label{fitness}

The 66000 images with fitness scores can be used by other researchers to develop a better automated fitness function. Our initial attempts at performing computer vision analysis on this data set has identified Zernike moments as potentially useful image characteristics. We expect to be able to make use of these images to produce an automated fitness function that will allow genetic algorithms to work nearly as well as the Citizen Science pipeline. The creation of an automated fitness function has only been made possible because of the contributions of our volunteers and our new ability to examine the human derived fitness scores of tens of thousands of models of interacting galaxies. The data for this project are archived at \url{http://data.galaxyzoo.org/mergers.html} and available for analysis in other projects. The simulation software for the project is available through the Astrophysics Source Code Library at \url{http://ascl.net/1511.002}.  The specific version of the software used in the applet run by the volunteers can be accessed at \url{https://github.com/jfwallin/JSPAM} in the {\it archive} folder of the repository.

\section*{Acknowledgements}

The authors would like to thank an anonymous reviewer for useful comments throughout and helpful references for astrophysical plausibility.

The data in this paper are the result of the efforts of the {\it Galaxy Zoo:~Mergers} volunteers, without whom none of this work would be possible. Their efforts are individually acknowledged at \url{http://data.galaxyzoo.org/galaxy-zoo-mergers/authors.html}. The version of the interface displayed on the Merger Zoo site was the result of many rounds of feedback. For their help in refining the tool and the backend software the authors thank the following people: Philip Murray, Geza Gyuk, Mark SubbaRao, Nancy Ross Dribin, Jordan Raddick, and Rebecca Ericson. 

The development of {\it Galaxy Zoo:~Mergers} was supported by the US National Science Foundation under grant DRL-0941610.

Target images and objects from this study were taken from the Sloan Digital Sky Survey.

Funding for the SDSS and SDSS-II has been provided by the Alfred P. Sloan Foundation, the Participating Institutions, the National Science Foundation, the U.S. Department of Energy, the National Aeronautics and Space Administration, the Japanese Monbukagakusho, the Max Planck Society, and the Higher Education Funding Council for England. The SDSS Web Site is http://www.sdss.org/.

The SDSS is managed by the Astrophysical Research Consortium for the Participating Institutions. The Participating Institutions are the American Museum of Natural History, Astrophysical Institute Potsdam, University of Basel, University of Cambridge, Case Western Reserve University, University of Chicago, Drexel University, Fermilab, the Institute for Advanced Study, the Japan Participation Group, Johns Hopkins University, the Joint Institute for Nuclear Astrophysics, the Kavli Institute for Particle Astrophysics and Cosmology, the Korean Scientist Group, the Chinese Academy of Sciences (LAMOST), Los Alamos National Laboratory, the Max-Planck-Institute for Astronomy (MPIA), the Max-Planck-Institute for Astrophysics (MPA), New Mexico State University, Ohio State University, University of Pittsburgh, University of Portsmouth, Princeton University, the United States Naval Observatory, and the University of Washington.

This research has made use of the NASA/IPAC Extragalactic Database (NED) which is operated by the Jet Propulsion Laboratory, California Institute of Technology, under contract with the National Aeronautics and Space Administration.


\bibliographystyle{apj}
\bibliography{bibfile}


\bsp

\label{lastpage}

\end{document}